\documentclass[fleqn, usenatbib]{mnras}

\usepackage{newtxtext}

\usepackage[T1]{fontenc}
\usepackage{ae,aecompl}

\usepackage{graphicx}	
\usepackage{amsmath}	
\usepackage{amssymb}	
\usepackage{subfig}
\usepackage[font=footnotesize]{caption}
\usepackage{wrapfig}
\usepackage{gensymb}
\usepackage{multirow}
\usepackage{tabularx}
\usepackage{float}
\usepackage{stackengine}
\usepackage[dvipsnames]{xcolor}
\usepackage{tablefootnote}
\usepackage{enumitem}
\usepackage{longtable}
\usepackage{threeparttablex}
\usepackage{hhline}
\usepackage{bm}		
\usepackage{pdflscape}	

\def\maxififth{MAXI~J1535--571}
\def\maxithirt{MAXI~J1348--630}
\def\maxieight{MAXI~J1820$+$070}
\def\xte{XTE~J1550--564}
\def\hh{H1743--322}
\def\gx{GX~339--4}

\newcommand{\tu}{\textup}

\title[2019/2020 outburst of \maxithirt{}]{The black hole transient \maxithirt{}: evolution of the compact and transient jets during its 2019/2020 outburst}

\author[F. Carotenuto et al.]{F. Carotenuto,$^{1}$\thanks{E-mail: francesco.carotenuto@cea.fr}
S. Corbel,$^{1,2}$
E. Tremou,$^{1,3}$
T. D. Russell,$^{4,5}$
A. Tzioumis,$^{6}$
R.P. Fender,$^{7,8}$
\newauthor
P.A. Woudt,$^{8}$
S.E. Motta,$^{7}$
J.C.A. Miller-Jones,$^{9}$
J. Chauhan,$^{9}$
A. J. Tetarenko,$^{10}$
\newauthor
G. R. Sivakoff,$^{11}$
I. Heywood,$^{7,12}$
A. Horesh,$^{13}$
A. J. van der Horst,$^{14,15}$
E. Koerding,$^{16}$
\newauthor
K. P. Mooley$^{17,18,19}$
\\ 
$^{1}$AIM, CEA, CNRS, Universit\'{e} de Paris, Universit\'{e} Paris-Saclay, F-91191 Gif-sur-Yvette, France\\
$^{2}$Station de Radioastronomie de Nan\c cay, Observatoire de Paris, PSL Research University, CNRS, Univ. Orl\'eans, 18330 Nan\c cay, France\\
$^{3}$LESIA, Observatoire de Paris, CNRS, PSL Research University, Sorbonne Universit\'{e}, Universit\'{e} de Paris, Meudon, France\\
$^{4}$Anton Pannekoek Institute for Astronomy, University of Amsterdam, Science Park 904, NL-1098 XH Amsterdam, The Netherlands\\
$^{5}$INAF, Istituto di Astrofisica Spaziale e Fisica Cosmica, Via U. La Malfa 153, I-90146 Palermo, Italy\\
$^{6}$Australia Telescope National Facility, CSIRO, PO Box 76, Epping, New South Wales 1710, Australia\\
$^{7}$Astrophysics, Department of Physics, University of Oxford, Keble Road, Oxford OX1 3RH, UK\\
$^{8}$Inter-University Institute for Data-Intensive Astronomy, Department of Astronomy, University of Cape Town, Private Bag X3, Rondebosch  7701, South Africa\\
$^{9}$International Centre for Radio Astronomy Research, Curtin University, GPO Box U1987, Perth, WA 6845, Australia\\
$^{10}$East Asian Observatory, 660 N. A'\!oh$\bar{o}$k$\bar{u}$ Place, University Park, Hilo, Hawaii 96720, USA\\
$^{11}$Department of Physics, University of Alberta, CCIS 4-181, Edmonton, AB T6G 2E1, Canada\\
$^{12}$Department of Physics and Electronics, Rhodes University, PO Box 94, Grahamstown 6140, South Africa\\
$^{13}$Racah Institute of Physics, The Hebrew University of Jerusalem, Jerusalem 91904, Israel\\
$^{14}$Department of Physics, The George Washington University, 725 21st Street NW, Washington, DC 20052, USA\\
$^{15}$Astronomy, Physics and Statistics Institute of Sciences (APSIS), 725 21st Street NW, Washington, DC 20052, USA\\
$^{16}$Department of Astrophysics/IMAPP, Radboud University Nijmegen, P.O. Box 9010, 6500 GL Nijmegen, The Netherlands\\
$^{17}$Department of Physics, University of Oxford, Keble Road, Oxford OX1 3RH, UK\\
$^{18}$National Radio Astronomy Observatory, Socorro, NM 87801, USA\\
$^{19}$Caltech, 1200 E. California Blvd. MC 249-17, Pasadena, CA 91125, USA
}
\date{Accepted XXX. Received YYY; in original form ZZZ}

\pubyear{2021}

\begin{document}
\label{firstpage}
\pagerange{\pageref{firstpage}--\pageref{lastpage}}
\maketitle

\begin{abstract}
\noindent
We present the radio and X-ray monitoring campaign of the 2019/2020 outburst of \maxithirt{}, a new black hole X-ray binary (XRB) discovered in 2019 January. We observed \maxithirt{} for $\sim$14 months in the radio band with MeerKAT and the Australia Telescope Compact Array (ATCA), and in the X-rays with MAXI and \emph{Swift}/XRT. Throughout the outburst we detected and tracked the evolution of the compact and transient jets. Following the main outburst, the system underwent at least 4 hard-state-only re-flares, during which compact jets were again detected. For the major outburst, we observed the rise, quenching, and re-activation of the compact jets, as well as two single-sided discrete ejecta, launched $\sim$2 months apart and travelling away from the black hole. These ejecta displayed the highest proper motion ($\gtrsim$100 mas day$^{-1}$) ever measured for an accreting black hole binary. From the jet motion, we constrain the ejecta inclination and speed to be $\leq$46\degree and $\geq$0.69 $c$, and the opening angle and transverse expansion speed of the first component to be $\leq$6\degree and $\leq$0.05 $c$. We also infer that the first ejection happened at the hard-to-soft state transition, before a strong radio flare, while the second ejection was launched during a short excursion from the soft to the intermediate state. After traveling with constant speed, the first component underwent a strong deceleration, which was covered with unprecedented detail and suggested that \maxithirt{} could be located inside a low-density cavity in the interstellar medium, as already proposed for \xte{} and \hh{}.
\end{abstract}

\begin{keywords}
accretion, accretion discs -- black holes physics  -- X-rays: individual:~\maxithirt{} -- ISM: jets and outflows -- radio continuum: stars -- X-rays: binaries
\end{keywords}



\section{Introduction}
\label{Intro}
Relativistic jets are outflows powered by accreting compact objects over a wide range of physical scales, including stellar-mass black holes (BHs) and their supermassive counterparts. In black hole low mass X-ray binaries (BH LMXBs), the companion star fills its Roche lobe and feeds a hot accretion disk that surrounds the compact object. In this configuration, the system is able to channel a fraction of the accreted mass and energy into powerful jets that can have different forms on very different scales: from compact jets at small scales ($\sim$tens of AU), to moving, discrete plasmons or hot-spots and persistent radio lobes at pc scales (e.g. \citealt{Fender2006}).
While BH LMXBs spend most of their existence in a low-luminosity quiescent state, they occasionally enter outburst phases that typically last from months to years (e.g. \citealt{Remillard_xrb, Watchdog, Fender_balance}). Sharing many similarities with active galactic nuclei (AGN), BH LMXBs provide suitable laboratories to study scale-invariant properties of black holes, specifically those linked to the connection between accretion and ejection, as they evolve through different regimes of accretion on human timescales (e.g. \citealt{Kording_2005}). 

During an outburst, the system cycles through different accretion states that are characterized by distinct spectral and timing signatures (e.g. \citealt{Homan_belloni, Remillard_xrb, Belloni_2010, Belloni_Motta2016, Ingram_2019}).
As the outburst begins, these systems are typically observed in a rising hard X-ray state with high rms variability, during which the X-ray spectrum is dominated by a non-thermal power law that is thought to be the result of inverse-Compton scattering by a corona of hot electrons close to the compact object (e.g. \citealt{Zdziarski_corona, Poutanen_2014}). In addition, subdominant thermal emission, likely from a truncated, optically thin accretion disk, may also be present. The accretion rate increases as the outburst progresses, and the system transitions to the intermediate state (IMS), which could be (conveniently) divided into the hard (HIMS) and soft (SIMS) intermediate states, during which the X-ray spectrum becomes progressively softer as the thermal emission from the accretion disk increases and the rms variability decreases. The system subsequently enters the soft state, characterised by a drop in the rms variability and in which the X-ray spectrum is completely dominated by a $k_{\tu{B}}T \sim 1$ keV thermal emission from an optically thick, geometrically thin accretion disk, whose inner radius is thought to reach the BH Innermost Stable Circular Orbit (ISCO, e.g. \citealt{Steiner_2010}).
Some weeks to months later, the luminosity decreases and, at some point, the system re-enters a lower luminosity IMS, before a transition back to the hard state, after which the outburst ends and the system slowly approaches quiescence.

During the outburst evolution, the jets produced by the system also evolve significantly. In the hard state, collimated and continuously replenished compact jets are typically observed in the radio band \citep{Corbel_2000, Fender_2001}, emitting self-absorbed synchrotron radiation. This results in a flat or slightly inverted radio spectrum ($\alpha \gtrsim 0$, where the radio flux density follows $S_{\nu} \sim \nu^{\alpha}$) up to a break frequency $\nu_{\tu{break}}$ located in the IR \citep{Markoff_2001, Corbel2002, Russell_2013b, Russell_2014, Russell_2020_break_frequency}, after which the jet becomes optically thin. Synchrotron radiation from compact jets can also contribute to the hard X-ray emission (e.g. \citealt{Markoff_corona}). There are indications that matter in compact jets could be accelerated up to bulk Lorentz factors $\Gamma_{\tu{j}} \lesssim 3.5$ (e.g. \citealt{Ribo, Russell_2015, Tetarenko_2019, Saikia, Peault_2019}).
No radio emission is usually detected during the soft state, as compact jets are strongly quenched \citep{Fender1999, Corbel_2000}, by at least 3.5 orders of magnitude \citep{Russell_1535, Maccarone_2020}. Compact jets are then gradually re-activated as the systems transition back to the final hard state, which precedes quiescence \citep{Miller-Jones_h1743, Kalemci, Corbel2013_IR, Russell_2014}.

One of the most interesting properties of BH LMXBs is the capability of launching transient jets, which are bipolar, discrete blobs of plasma ejected from the system in opposite directions. These transient jets have been observed so far in a small fraction of the known BH LMXB population (e.g. \citealt{Mirabel1994, GRO1655, Fender1999, Mioduszewski, Gallo_2004, Yang2010, Rushton, Russell_1535, Miller-Jones2019, Bright}). As the plasmons move away from the core at relativistic speeds, they often display apparent superluminal motion and emit optically thin radio emission ($\alpha \sim-$ 0.6, e.g. \citealt{Corbel2002_xte}). The ejected components can travel unseen for months to years, before they are detected again at large ($\sim$pc) scales, with the emission coming from their interaction with the surrounding interstellar medium (ISM). 
Notably, discrete jets have been simultaneously detected at large scales in radio and X-rays for \xte{} \citep{Corbel2002_xte, Tomsick_2003, Kaaret_2003, Migliori2017}, \hh{} \citep{Corbel2005_h17} and \maxieight{} \citep{Espinasse_xray}, with a broadband power law spectrum suggesting the synchrotron nature of the emission coming from the shock region. 

Despite the progress made so far, the mechanism responsible for the jet launching has not yet been identified. 
In many systems, strong radio flares are observed to happen close to the transition between the hard and the soft state and are believed to be the signature of the ejection of transient jets \citep{Corbel2004, Fender_belloni_gallo}, even in the cases for which the presence of extended jets cannot be confirmed by imaging. Discrete ejections could originate within compact jets due to internal shocks produced by a change in the accretion flow and a subsequent sharp increase in the Lorentz factor \citep{Kaiser, Fender_belloni_gallo}. In alternative, the transient jet could be the result of the ejection of the coronal material at state transitions \citep{Rodriguez_2003, Vadawale_2003}, although this scenario could imply an abrupt change in the coronal spectral properties that is not observed. It is of first importance to accurately model the jet motion. With such models, we can precisely infer its ejection date and possibly identify, along with other observational features, the causal relation that leads to a discrete ejection. It was proposed that transient jets are associated with the change in the X-ray variability and with presence of Type-B Quasi Periodic Oscillations (QPOs; \citealt{Soleri2008, Fender_2009}). While a possible connection has been identified in \maxieight{} \citep{Homan_qpo}, there is no clear evidence of a causal relation for other sources (e.g. \citealt{Fender_2009, Miller-Jones_h1743, Russell_1535}).

\subsection{\maxithirt{}} 
\label{sec:source_presentation}

\maxithirt{} was discovered as a bright X-ray transient on 2019 January 26 \citep{Yatabe2019} by the Monitor of All-sky X-ray Image (MAXI) onboard the International Space Station \citep{Matsuoka_maxi} and, thanks to immediate and intense  multi-wavelength follow-up observations, it was subsequently identified as a black hole candidate \citep{Kennea_atel, Nicer_atel, HXMT_atel, Denisenko_atel, Nesci_atel, DRussell_atel_opt_1, TRussell_atel}. \maxithirt{} displayed a typical outburst that was followed by four reported subsequent hard state re-flares \citep{DRussell_atel_opt_2, Negoro_atel, Yazeedi_atel_2, Pirbhoy_atel, Carotenuto_atel_2, Zhang_atel, Shimomukai_atel}. While currently no information is available on the black hole inclination, mass, spin and companion star, it has been possible to measure a source distance $D = {2.2}_{-0.6}^{+0.5}$ kpc \citep{Chauhan2020} from observations of H\textsc{i} absorption carried out with the Australian Square Kilometre Array Pathfinder (ASKAP) and with MeerKAT. From the analysis of the MAXI data for the first part of the outburst, \cite{Tominaga_1348} estimated a black hole mass $M_{\tu{BH}} \simeq 16 (D/5$ kpc$) \ M_{\odot}$, assuming a face-on system with a non-spinning BH (with the $M_{\tu{BH}}$ estimate increasing with the BH spin and inclination). At the known distance of \maxithirt{}, this corresponds to $M_{\tu{BH}} \sim$ 7 $M_{\odot}$, which is fully consistent with measured mass values among the known population of Galactic BHs (e.g. \citealt{Watchdog}). 

In this paper we present a dense and comprehensive radio and X-ray monitoring of \maxithirt{} during its discovery outburst. In Section \ref{sec:Observations} we describe the observations and the data processing. In Section \ref{sec:results} we present the results of the monitoring from the X-ray point of view and we describe the evolution of the compact jets and the motion of the discrete ejections from our radio observations, while in Section \ref{sec:discussion} we discuss our findings. Our conclusions are summarized in Section \ref{sec:conclusion}.
The study of the radio/X-ray correlation will be presented in a forthcoming paper.

\section{OBSERVATIONS} 
\label{sec:Observations}

\subsection{MeerKAT radio observations} 
\label{sec:MeerKAT}
We observed \maxithirt{} with MeerKAT \citep{Jonas2016, Camilo2018} as part of the ThunderKAT Large Survey Programme \citep{ThunderKAT}. MeerKAT is a new radio-interferometer and a precursor of the SKA, which is located in the Karoo desert in South Africa. MeerKAT consists of 64 antennas of 13.5 m diameter, currently equipped with L-band receivers (0.86--1.71 GHz). Characterised by a dense core and with a longest baseline of 8 km, the array offers a very good snap-shot \textit{uv}-coverage and a large field of view (1.69 deg$^{2}$), achieving a resolution of $\sim$5 arcsec in L-band. \maxithirt{} was monitored with an approximately weekly cadence from 2019 January 29 (MJD 58512) to 2020 March 03 (MJD 58910) for a total of 48 epochs. Each observation consisted of a single 15 min long scan on source, except for the first observation (MJD 58512), which was 20 mins of on-source time separated into two 10 min scans. The primary calibrator PKS~1934--638 was observed either at the beginning or at the end of each observing block, and it was used for flux, bandpass, and complex gain calibration.
All observations were taken with 8 seconds of integration times, at a central frequency of 1.28 GHz and with a total bandwidth of 860 MHz, divided into 4096 equivalent 209 kHz channels. 

The data were first flagged using AOFlagger \citep{Offringa} and then averaged in frequency to reduce the number of channels (and the data size) by a factor 8. The calibration was subsequently performed using standard practices with the Common Astronomy Software Application (\textsc{casa} version 4.7, \citealt{CASA}), while imaging was carried out with the wide-band, wide-field imager DDFacet \citep{Tasse}, adopting a uniform weighting scheme to achieve the best possible angular resolution. To mitigate the presence of artifacts from bright sources close to the target, the images were then self-calibrated in amplitude and phase using the calibration software \textsc{killMS} \citep{Tasse} and \textsc{DDFacet}, with a solution interval of 5 min.
Between the observations of 2019 July 22 and November 01, the sky coordinates of PKS~1934--638 used by MeerKAT were slightly offset from the VLBI ones, therefore we artificially rotated the visibilities to the correct position using the {\tt chgcentre} task, which is part of the \textsc{wsclean} package \citep{WSCLEAN}. 
For each epoch, the positions of the target and the jet knots were corrected using a stable background point source close to \maxithirt{}, both in Right Ascension and Declination, while its reference position was determined from higher resolution ATCA observations (that were part of the monitoring and are presented in Section \ref{sec:ATCA}). The corrected global positional offsets were $\lesssim 1$ arcsec. The level of rms noise reached ranged from 30 to 60 $\mu$Jy beam$^{-1}$, depending on the observing conditions of the telescope and variable levels of radio-frequency interference (RFI). 
To obtain the radio flux density $S_{\nu}$, we fit point sources in the image plane with the \textsc{casa} task {\tt imfit}. All flux densities are reported in Tables \ref{tab:core_flux_table}, \ref{tab:RK2_flux_table}. Positions were extracted before self-calibration, while flux densities were extracted afterwards.

\subsection{ATCA radio observations} 
\label{sec:ATCA}
\maxithirt{} was also monitored with the Australia Telescope Compact Array (ATCA) for 31 epochs in total, from January to December 2019 (project codes C1199 and CX423). The first part of the monitoring consisted of a dense coverage at the beginning of the outburst, with observations performed roughly every two days, from 2019 January 26 (MJD 58509) to 2019 February 17 (MJD 58531), and a total observing time that ranged from 8 minutes to 5 hours, depending on the expected flux density of the target. The second part of the monitoring started on 2019 March 31 (MJD 58573) and ended on 2019 December 14 (MJD 58789), with observations taken every 1--2 or 3 weeks and longer exposure times of up to 12 hours on MJD 58689\footnote{We do not consider the 5.5 and 9 GHz ATCA observation performed on MJD 58816 (2019 November 29), as we noticed significant discrepancies in the flux densities of background check sources at both frequencies, which are likely of instrumental origin.}.
All of the observations were taken simultaneously at central frequencies of 5.5 and 9 GHz, while for some epochs (specified in Table \ref{tab:core_flux_table}) we added a second pair of bands, and the central frequencies were either 17.0 and 19.0 GHz, or 16.7 and 21.2 GHz.
For each central frequency, the total bandwidth was 2 GHz. We also included two epochs (MJD 58515 and 58608), during which ATCA was part of the Long Baseline Array (LBA) as a phased-array element for VLBI observations (project code V447). In this configuration, the central frequency was 8.4 GHz, with a total bandwidth of 128 MHz. In this paper we report results of the ATCA-only correlated data. We used either PKS~1934--638 (preferred) or PKS~0823--500 as primary calibrators, depending on the visibility of the calibrator at the time of the observation, while PKS~1352--63 was used for gain calibration. The data were first flagged and then calibrated using standard procedures in \textsc{casa}. Imaging was carried out with the standard {\tt tclean} algorithm in \textsc{casa} with a Briggs weighting scheme. We chose a robust parameter of 0 for an optimal trade-off between angular resolution and sensitivity. 
The level of rms reached 6 $\mu$Jy beam$^{-1}$ for the longest duration observations. To obtain the radio flux density $S_{\nu}$, we again fitted a point source in the image plane with the \textsc{casa} tasks {\tt imfit}, except for several short observations, in which the poor \emph{uv}-coverage did not allow a proper deconvolution: in that case the flux densities were obtained by directly fitting the visibility data, using \textsc{uvmultifit} \citep{uvmultifit}.
All flux densities, measured positions and telescope configurations are reported in Tables \ref{tab:core_flux_table}, \ref{tab:RK2_flux_table}.

\subsection{\emph{Swift}/XRT observations}
\label{sec:Swift/XRT observations}
To monitor the evolution of the X-ray emission, \maxithirt{} was observed with the Neil Gehrels \emph{Swift} Observatory \citep{Gehrels}, as part of a dedicated monitoring associated with the ThunderKAT Large Survey Program on X-ray transients, for a total of 85 epochs since its discovery (target ID: 00011107).
The on-board X-ray Telescope (XRT, \citealt{Burrows_xrt}) was operated in Windowed Timing (WT) mode for the first part of the outburst (from MJD 58509 to 58720), during which \maxithirt{} became extremely bright, up to a count rate of $\sim$1700 counts s$^{-1}$. The Photon Counting (PC) mode was used for the rest of the monitoring, when the count rate was $\lesssim 1$ count s$^{-1}$.
We used the HEASOFT package (version 6.25) and the related calibration files to reduce the publicly available raw XRT data. We ran the {\tt xrtpipeline} task to first reprocess the data and create exposure maps for each observation. The source spectra were then manually extracted using {\tt xselect}. For observations in WT mode, a circular region of 20 pixels (1 pixel = 2.35 arcsec) radius centered on the target position was used to extract the source spectrum when the source count rate was lower than 100 counts s$^{-1}$. To mitigate for the well-known pile-up problem affecting observations of bright sources, for higher count rates we filtered grade 0 events and used an annulus of variable (increasing with the count rate) inner and outer radii. The radii values (listed in Table \ref{tab:xray_1348}) were chosen to have $\lesssim 100$ counts s$^{-1}$ in the extraction region and to avoid the presence of calibration residuals in the source spectra. For the annulus extraction region, the inner radius ranged between 1 and 33 pixels and the outer radius ranged between 30 and 50 pixels. To extract the source spectrum in observations carried out in PC mode, we used a circular region with a radius of 20 pixels centered on the target, except for the epochs on MJD 58816, 58821 and 58889. During these three epochs we used an annulus of fixed 30 pixels outer radius and variable inner radius (listed in Table \ref{tab:xray_1348}) to ensure $\lesssim 0.5$ counts s$^{-1}$ in the extraction region and avoid pile-up. The background spectra for all observations were extracted using an annulus region centred on the source with a 50 pixels inner radius and 100 pixels outer radius.
We created ancillary response files with {\tt xrtmkarf} and used the up-to-date response matrix files from the High Energy Astrophysics Science Archive Research Center (HEASARC) calibration database (CALDB). When possible, we used {\tt grppha} to group the spectra in bins of 20 counts each to enable the use of $\chi^2$ statistics for the spectral fitting. For low-counts observations we used Cash statistics {\tt cstat} \citep{cstat} instead.

Spectral analysis was carried out in the 0.7--10 keV energy range with XSPEC \citep{Arnaud_xspec}. Our main goal with the X-ray spectral analysis was to estimate the unabsorbed flux in the 1--10 keV band. Therefore, we used simple models to achieve acceptable fits. For hard state observations, we fit the spectra with a simple absorbed power law model ({\tt tbabs$\times$powerlaw}), where in {\tt tbabs} the interstellar absorption is modelled with an equivalent hydrogen column density ($N_{\textup{H}}$) by using {\tt wilm} abundances \citep{Wilms} and {\tt vern} cross-sections \citep{Verner1996}. For the soft state, we used an absorbed multi-color disk blackbody to which we add a power law component to take into account residual high energy tails, {\tt tbabs}$\times$({\tt diskbb}$+${\tt powerlaw}). Since in the latter cases the power law photon index $\Gamma$ could not be constrained, for soft state observations we fixed $\Gamma = 2.4$. We first fitted all the epochs while leaving the $N_{\textup{H}}$ free to vary, and then, assuming no variation of the hydrogen column density due to the source activity, we froze the $N_{\textup{H}}$ to the average value of $0.86\times10^{22}$ cm$^{-2}$. We then calculated the 1--10 keV X-ray unabsorbed flux using the XSPEC convolution model {\tt cflux}. All the results of the spectral analysis are reported in Table \ref{tab:xray_1348}. In case of hard-state observations close to quiescence, for which the photon index could not be constrained, we fixed $\Gamma = 2.2$, as measured for BH LMXBs approaching quiescence \citep{Corbel2006, Plotkin2013}.

\subsection{MAXI/GSC Observations}
\label{sec:MAXI}
The Monitor of All-sky X-ray Image (MAXI, \citealt{Matsuoka_maxi}) observed \maxithirt{} with the Gas Slit Camera (GSC) throughout the whole outburst and all the data are publicly available. We downloaded the data from the MAXI on demand website\footnote{\url{http://maxi.riken.jp/mxondem/}} and we primarily used them to follow the outburst evolution and to obtain the Hardness-Intensity Diagram (HID), which is shown in Figure \ref{fig:maxi_hid}. See \cite{Tominaga_1348} for details on the MAXI data.

\section{RESULTS} 
\label{sec:results}

With our MeerKAT, ATCA, \emph{Swift}, and MAXI observations, we monitored the entire 2019/2020 outburst of \maxithirt{}.
During the discovery outburst, \maxithirt{} was detected early during the rise and subsequently transitioned through all the typical BH XRB accretion states, displaying a complex and rich behaviour in terms of X-ray and radio emission, as can be seen from the MAXI HID in Figure \ref{fig:maxi_hid}. 

\subsection{X-ray spectral and temporal evolution of the outburst}
\label{sec:X-ray emission and spectral evolution}

\begin{center}
\begin{figure*}

\includegraphics[width=\textwidth]{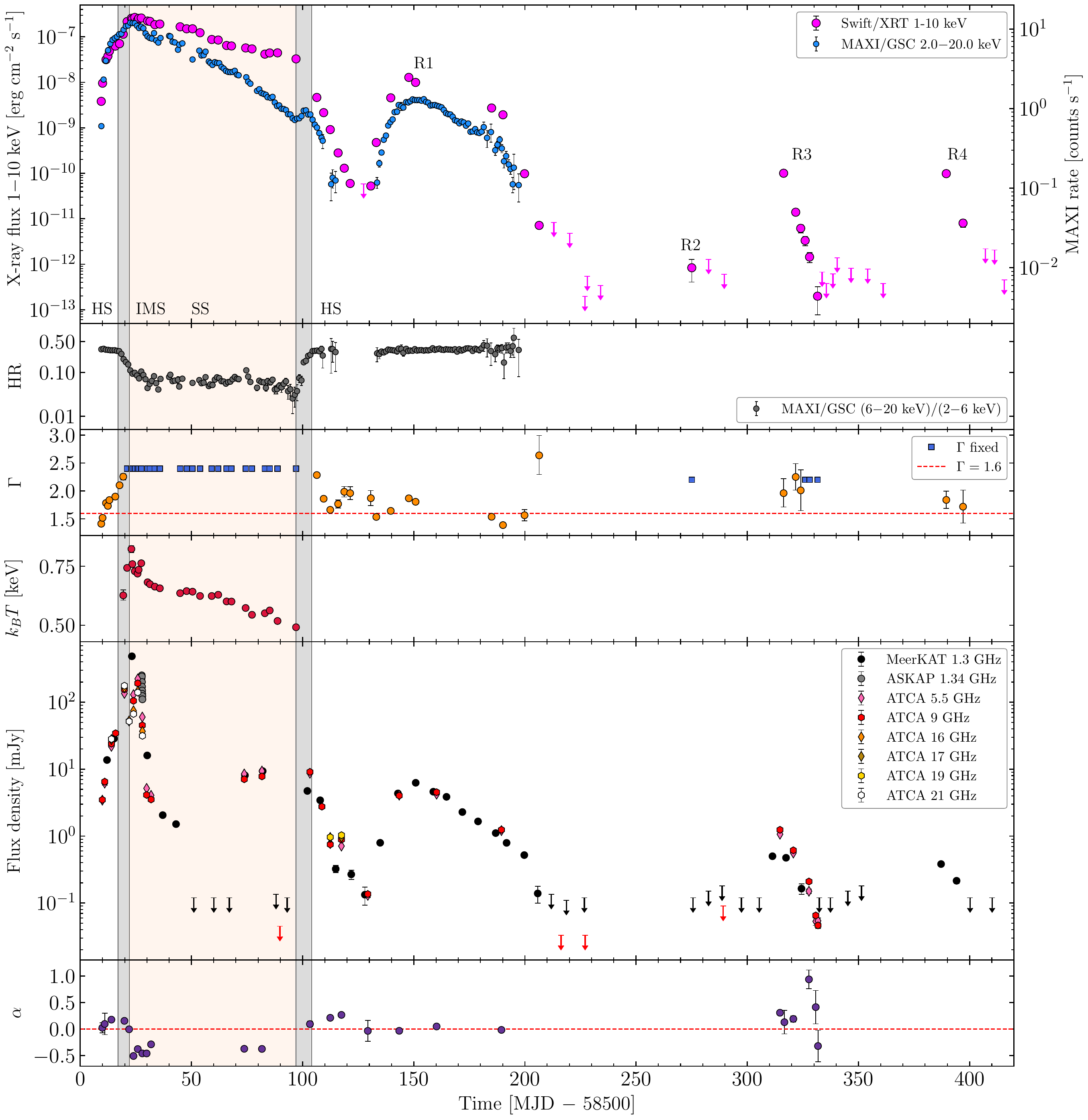}
\caption{X-ray and radio light curves of \maxithirt{} during its 2019/2020 outburst. The un-shaded regions mark the periods in which the source was in the hard state, while the light grey shaded regions indicate the IMS and the light pink regions indicate the soft state (see Section \ref{sec:X-ray emission and spectral evolution} and Table \ref{tab:state_transitions} for the timing of the spectral state transitions, which are obtained from \citealt{Zhang2020}). \emph{First (top) panel}: Unabsorbed X-ray flux light curve of \maxithirt{} from \emph{Swift}/XRT observations in the 1--10 keV energy range (left \textit{y-axis}). The four re-flares that followed the main outburst are marked on the figure with R1,2,3,4.  We also show the MAXI/GSC daily count rate (right \textit{y-axis}). For clarity, we only show the first part of the outburst, in the MJD interval 58505--58710, and select daily epochs with a count rate above 0.1 counts s$^{-1}$. \emph{Second panel}: MAXI/GSC daily hardness ratio, defined as the ratio between the counts in the two adjacent \emph{hard} and \emph{soft} bands of 6--20 keV and 2--6 keV, respectively.
\emph{Third panel}: X-ray photon index $\Gamma$ obtained from \emph{Swift}/XRT spectra. Epochs for which the value of the photon index was fixed are shown with blue square points. \emph{Fourth panel}: Temperature of the inner edge of the accretion disk obtained from \emph{Swift}/XRT spectra fitted with {\tt diskbb}. \emph{Fifth panel}: Radio light curve of \maxithirt{} at the core location (excluding large scale jets). Black dots are for 1.28 GHz MeerKAT observations, while the others are for multi-frequency ATCA observations. ASKAP 1.34 GHz data taken on MJD 58527 from \protect\cite{Chauhan2020} are also shown. \emph{Sixth (bottom) panel}: Radio spectral index of \maxithirt{} obtained from ATCA simultaneous multi-frequency observations, as reported in Table \ref{tab:core_flux_table}.}
\label{fig:main_lc}
\end{figure*}
\end{center}

A global view on the outburst can first be obtained from the MAXI count-rate light curve and from the \emph{Swift}/XRT flux light curve, both shown in Figure \ref{fig:main_lc}. We describe the outburst evolution using our observations, but we rely on the Neutron Star Interior Composition Explorer (NICER) X-ray data \citep{Zhang2020} for the dates of the state transitions, which are obtained with high accuracy from spectral and timing analysis, and are reported in Table \ref{tab:state_transitions}.

\setlength{\extrarowheight}{.2em}
\setlength{\tabcolsep}{18pt}
\begin{table}
\caption{Summary of the X-ray spectral state evolution of \maxithirt{} during the 2019/2020 major outburst. Dates are obtained from \protect\cite{Zhang2020}, using spectral and timing NICER data. We note that frequent excursions to the SIMS were observed during the first part of the soft state (until MJD 58542, see \protect\citealt{Zhang2020}).}
\centering
\begin{tabular}{l l}
\hline
\hline
X-ray state & MJD interval\\
\hline
Hard state & 58509 -- 58517\\
HIMS & 58517 -- 58522.6\\
Soft state & 58522.6 -- 58597\\
IMS & 58597 -- 58604\\
Hard state & 58604 -- 58910\\
\hline
\end{tabular}
\label{tab:state_transitions}
\end{table}

\begin{figure}
\begin{center}
\includegraphics[width=\columnwidth]{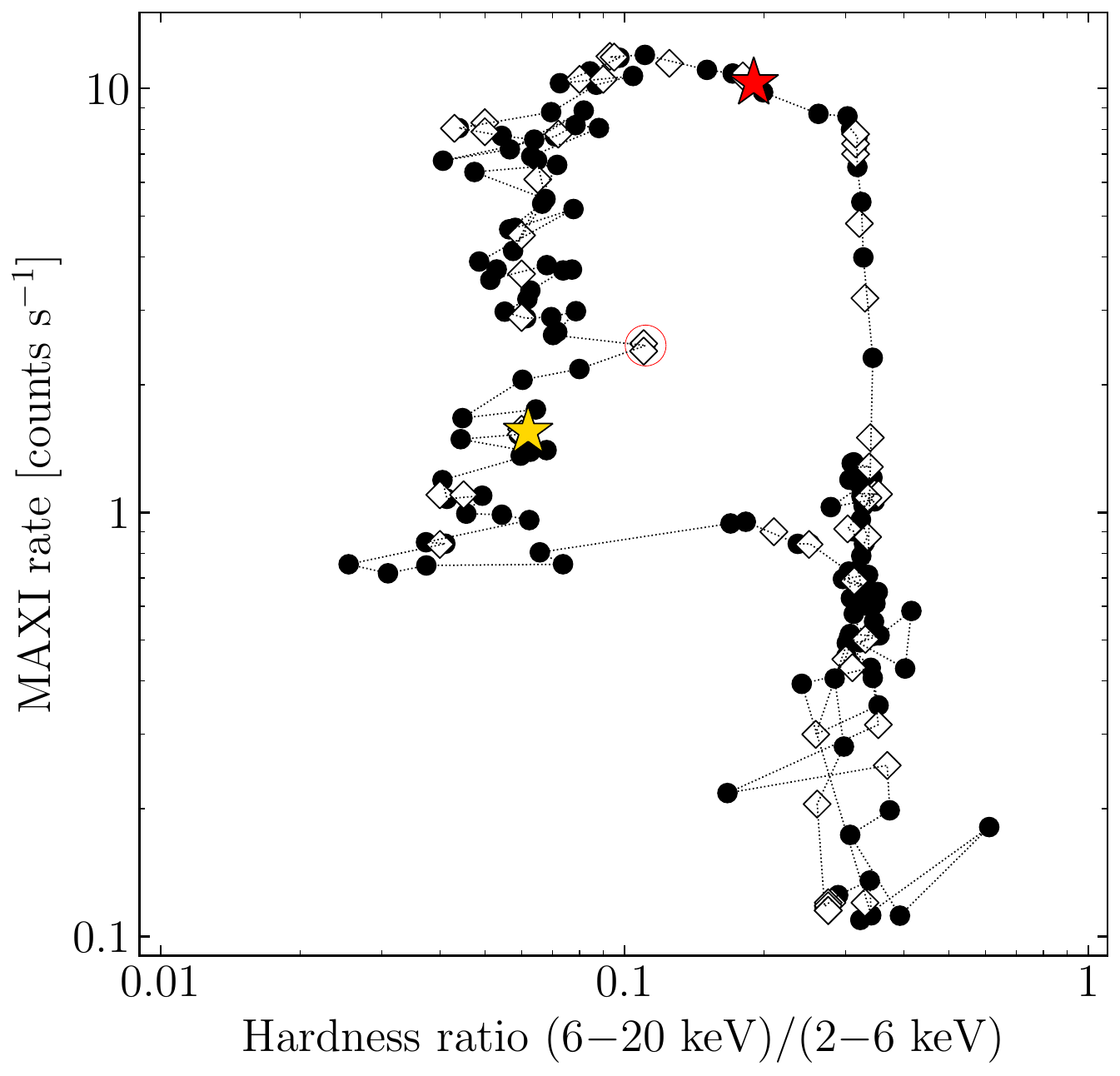}
\caption{Hardness-Intensity Diagram (HID) of \maxithirt{} during its major outburst, where the hardness ratio (HR) was determined from MAXI data. The source performs a typical cycle in the HID during the major outburst. For clarity, we only show MAXI observations above 0.1 counts s$^{-1}$ in the MJD interval 58505--58710, to include the brightest part of the outburst that covered an entire cycle in the HID. Times of radio observations are marked on the HID as white diamonds. Radio observations for the second part of the outburst (MJD $> 58710$) lie on the bottom part of the hard line (HR $\sim$ 0.4). The inferred ejection dates for RK1 and RK2 are shown, respectively, as red and yellow stars (see Sections \ref{sec:multiple_radio_flares} and \ref{sec:ejection_RK2}). The hardness spike on MJD 58574, believed to be associated with a short excursion to the SIMS (see Section \ref{sec:ejection_RK2}), is highlighted with a red circle.}
\label{fig:maxi_hid}
\end{center}
\end{figure}

The outburst started with a rapid rise that was first detected by MAXI and \emph{Swift} on MJD 58509 (2019 January 26, \citealt{Yatabe2019, Kennea_atel}). During the rise, the system was in the hard state, as can be initially seen from Figure \ref{fig:maxi_hid} and \ref{fig:main_lc}. In this state the X-ray spectrum can be well fitted with a single absorbed power law, with a photon index starting from $\Gamma \sim$1.4 and becoming softer as the system became brighter. \maxithirt{} entered the IMS on MJD 58517 (2019 February 03), during which the flux continued increasing and the spectrum started to be dominated by the thermal accretion disk emission at $\sim$keV energies.

On MJD 58522 (2019 February 08) the system subsequently entered the soft state, approaching the outburst peak on MJD 58524. During the soft state the disk temperature evolution clearly tracked the light curve, reaching $\sim$0.8 keV during the brightest part, a low value among BH XRBs \citep{Tominaga_1348}.
Then the system exponentially decayed for two months, during which the spectrum stayed soft (see Figure \ref{fig:maxi_hid}). The system transitioned back to the IMS on MJD 58597 and the low hard state started on MJD 58604. During the first weeks of the low hard state the source exhibited an exponential decay. As the decay progressed, the spectrum initially became steeper, but, after MJD 58612, we observed a progressive spectral softening, as typically observed when BH LMXBs get closer to their quiescent state \citep{Corbel2006, Plotkin2013}. Then, \maxithirt{} was not detected on MJD 58627\footnote{The source was below the detection threshold of \emph{Swift}/XRT in WT mode, during an observation characterised by an unusually high background level.}, but during the following weeks we observed a rise in flux (hereafter R1, the first re-flare, as marked on Figure \ref{fig:main_lc}). A smooth decay then followed, the first half of which was not densely covered by \emph{Swift}, but was monitored with MAXI. Again, in the decay phase of R1, between MJD 58699 and 58706 (August 2019), the spectrum became softer as the system approached quiescence. \maxithirt{} was then detected again in X-rays on MJD 58775 (2019 October 19, R2), after the optical re-brightening reported by \cite{Yazeedi_atel_2}, but went quickly back to quiescence.

The last part of the outburst is characterised by two additional short hard-state re-flares, in November-December 2019 (R3, MJD interval 58789--58831) and the fourth in February 2020 (R4), after which \maxithirt{} faded back to quiescence. 
We note that after the main outburst and R1, from MJD 58706 the majority of the observations had to be fitted with Cash statistics due to low total counts, except for epochs on MJD 58816 and MJD 58889, for which the number of counts was high enough to allow $\chi^2$ statistics to be used.
Moreover, the photon index $\Gamma$ had to be frozen for the hard-state epochs on MJD 58720, 58775, 58826, 58828 and 58831, since the low number of counts did not allow the power law to be constrained. Their values are listed on Table \ref{tab:xray_1348}.

\begin{center}
\begin{figure*}

\includegraphics[width=\textwidth]{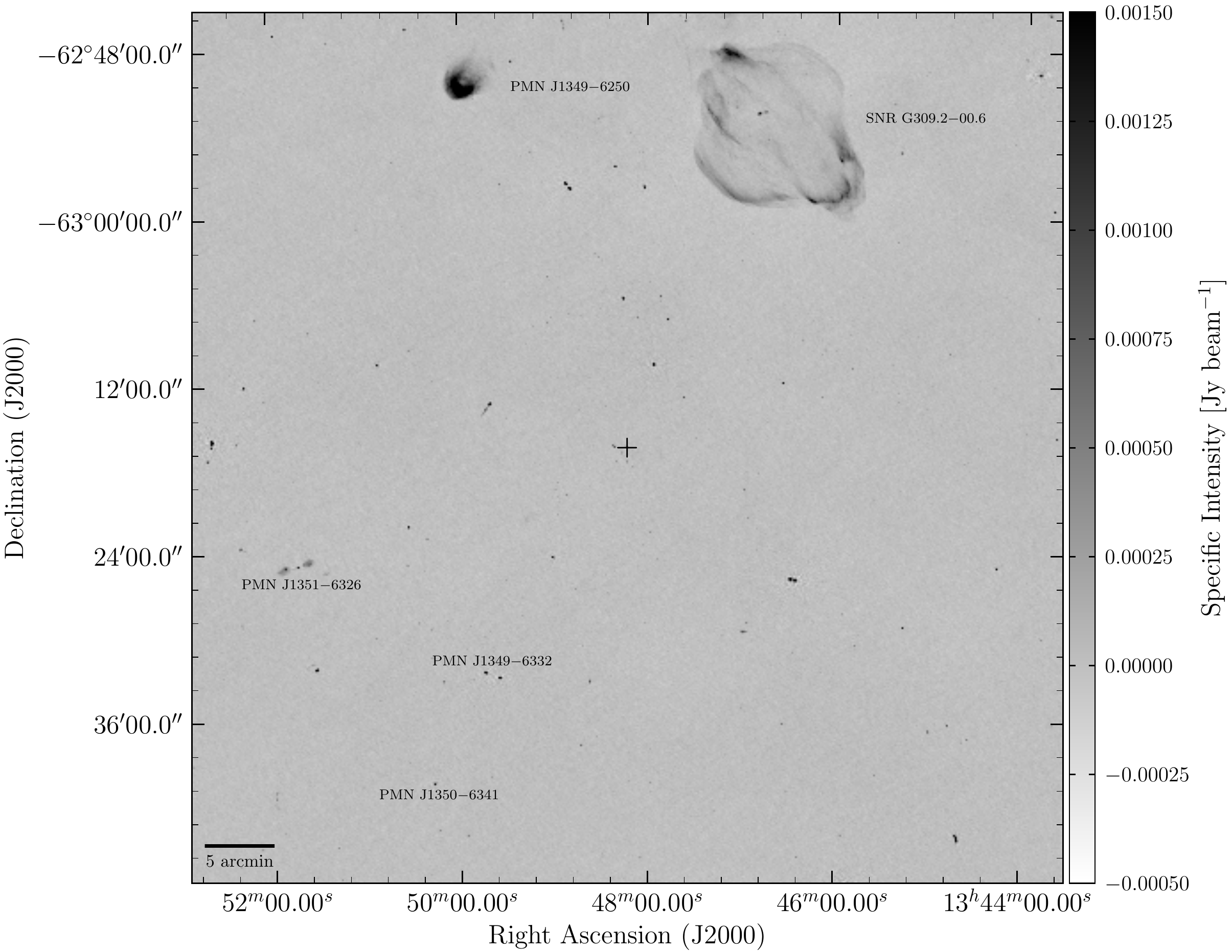}
\caption{MeerKAT large scale radio image at 1.28 GHz of the \maxithirt{} field. The image is obtained by concatenating the individual observations of July 14, 22, 27 and August 04 2019, corresponding to 1 hour of total observing time, and in which the \maxithirt{} core was reasonably stable in radio brightness (see Table \ref{tab:core_flux_table}). 
North is up and East is left. The colour bar shows the radio brightness in units of Jy beam$^{-1}$ and the black cross marks the position of \maxithirt{}. The rms is $\sim$30 $\mu$Jy beam$^{-1}$. The size of the image is approximately 1 deg$^2$, containing a field largely dominated by point sources, with the presence of several jetted radio galaxies. The two prominent sources to the North are the SNR G309.2--00.6 in the West \citep{SNR} and the HII region PMN J1349--6250 in the East.}
\label{fig:large_image}

\end{figure*}
\end{center}

\subsection{Radio emission from the \maxithirt{} core position}
\label{sec:radio_core_results}

With the start of the radio monitoring immediately after the X-ray detection of \maxithirt{}, our initial ATCA and MeerKAT observations clearly detected the radio counterpart of the source \citep{TRussell_atel}, consistent with the X-ray position. The radio light curve and the evolution of the radio spectral index for the entire monitoring are reported in Table \ref{tab:core_flux_table} and shown in the two lower panels of Figure \ref{fig:main_lc}, while a wide field image of the \maxithirt{} field observed with MeerKAT is shown in Figure \ref{fig:large_image}.

We first report our best measured position of \maxithirt{}, obtained from the weighted average of all the ATCA hard-state compact jet detections at 9 GHz, which are optimal in terms of resolution and signal-to-noise ratio. The Right Ascension (R.A.) and Declination (Dec) of the radio counterpart of \maxithirt{} are:
\begin{equation*}
 \begin{aligned}
  &\textup{R.A. (J2000)} = 13^{\textup{h}}48^{\textup{m}}12.79^{\textup{s}} \pm 0.01^{\textup{s}} \\   
  &\textup{Dec (J2000)} = -63\degree16\arcmin28.6\arcsec \pm 0.2\arcsec ,\\
 \end{aligned}
\end{equation*}
which is consistent with the \textit{Swift} localization \citep{Kennea_atel}. The associated errors are obtained from the combination in quadrature of the statistical and systematic errors that take into account the distance between the source and the phase calibrator and the accuracy on the position of the phase calibrator itself.

\begin{figure}
\begin{center}
\includegraphics[width=\columnwidth]{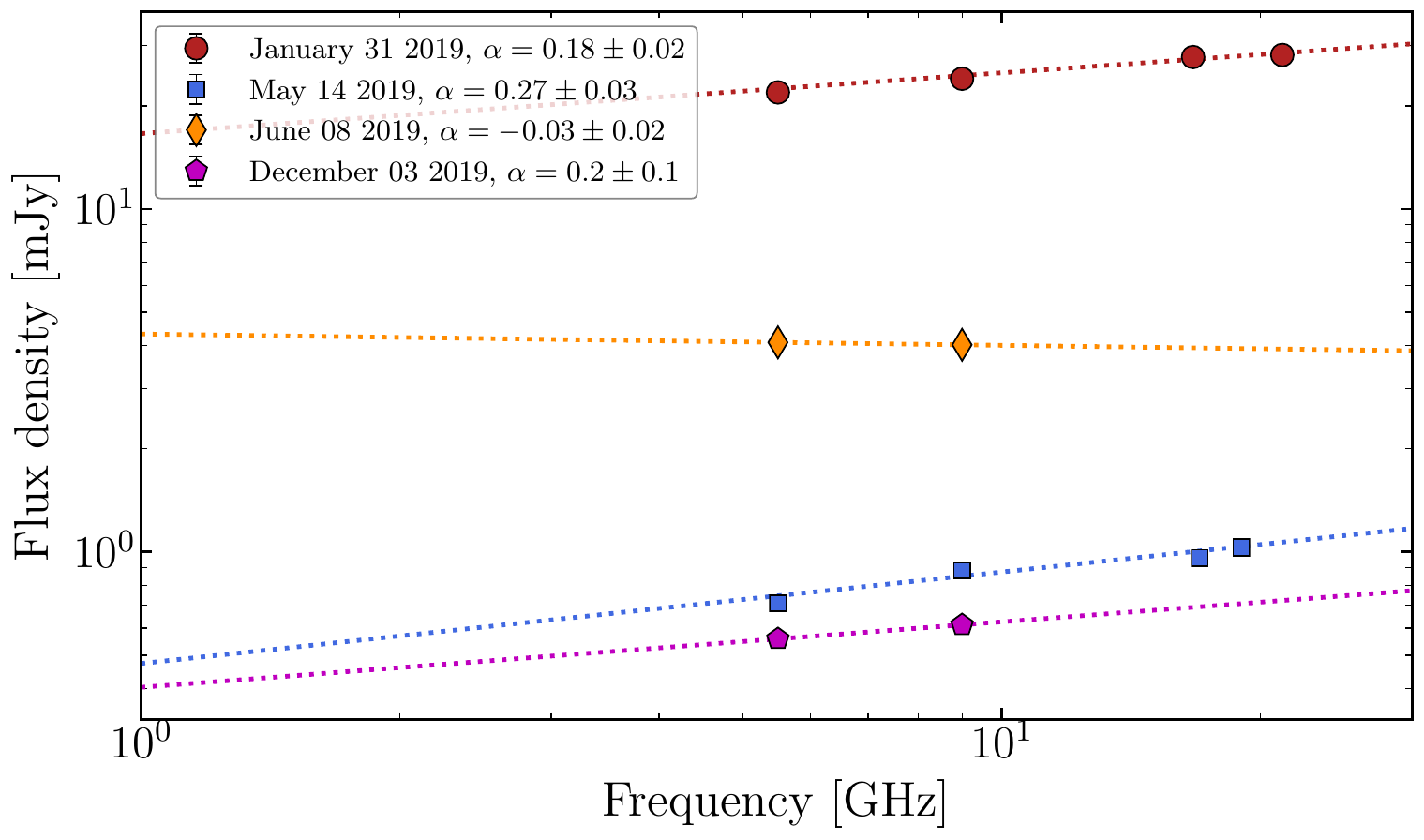}
\caption{Radio band spectra of \maxithirt{} of four epochs in the hard state, obtained with ATCA observations. Respectively in red, blue, yellow and purple we show data from MJD 58514, 58617, 58643, 58820, between January and December 2019. The fitted power law spectrum is shown as a dotted line for each plotted observation. Due to the fast evolution of radio emission, for the spectra we only use simultaneous ATCA multi-frequency observations. The flat-to-inverted spectra observed throughout the whole outburst are a signature that the radio emission is coming from self-absorbed compact jets.}
 \label{fig:spectrum}
\end{center}
\end{figure}

\begin{figure*}
\begin{center}
\includegraphics[width=\textwidth]{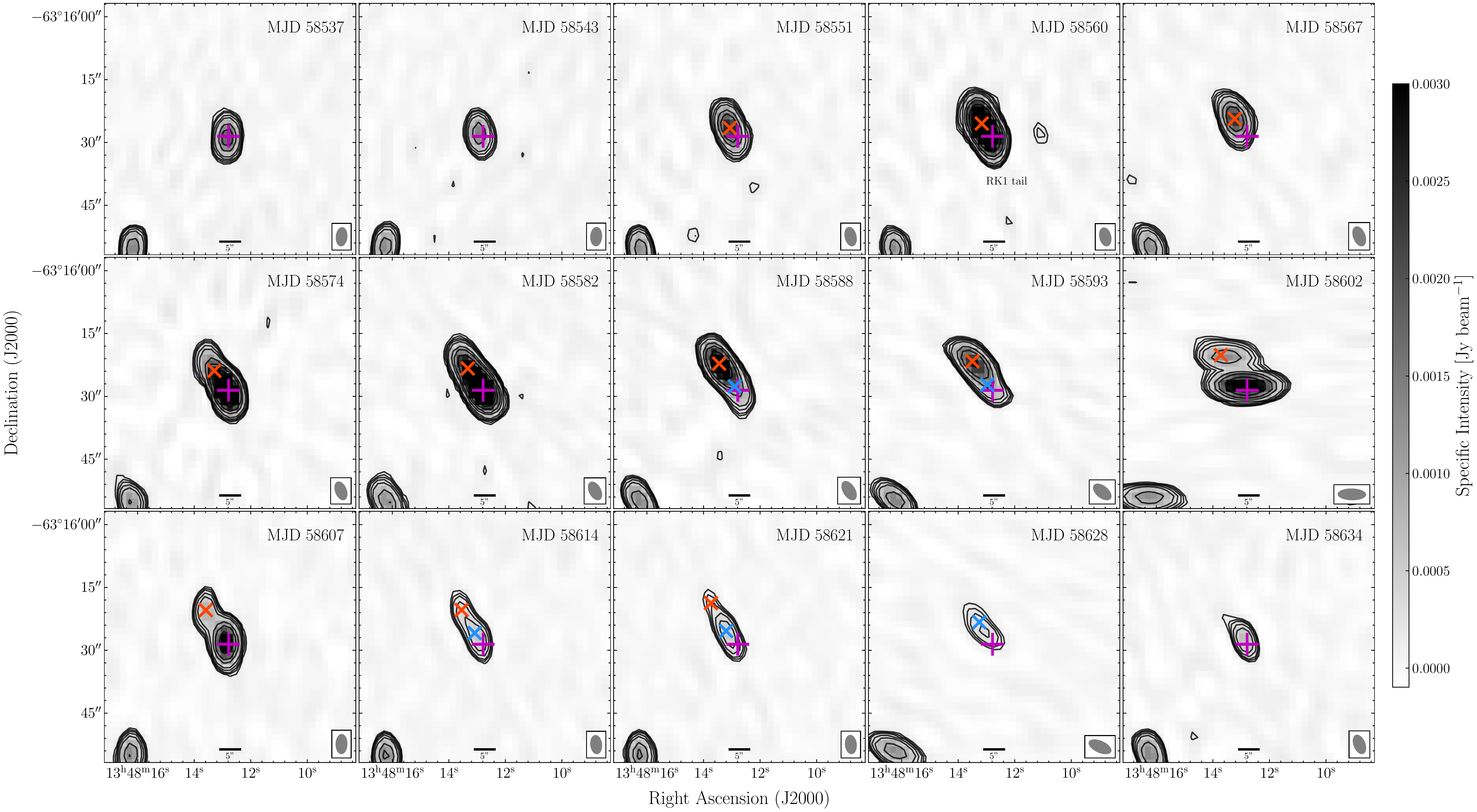}
\caption{A sample of the 1.28 GHz MeerKAT monitoring of \maxithirt{}, including 15 images that cover a period between MJD 58537 (2019 February 23) and MJD 58634 (2019 May 31). Contours start at three times the rms, with the rms ranging between $\sim$35 and $\sim$60 $\mu$Jy beam$^{-1}$. The \maxithirt{} position is marked with a magenta cross, while red and blue crosses are used, respectively, for the fitted positions of RK1 and RK2. We show the first part of the evolution of RK1, as it moved away from the core at a position angle of $33.2\degree \pm 1.4\degree$ East of North, and we also show the detections of RK2, which was never completely resolved with MeerKAT. The re-activated compact jets are present in the panels starting from MJD 58602. We show on MJD 58560 a resolved radio source that we fitted with two components: a jet \textit{head} and a jet receding \textit{tail} (see Section \ref{sec:RK1}). The radio emission at the core position detected on MJD 58574 and 58582 is discussed in Section \ref{sec:ejection_RK2}. The stable, background, point source (possibly consistent with the IR source 2MASS J13481645--6316501) in the bottom left corner is used to correct the astrometry for all the MeerKAT observations.}
\label{fig:meerkat_images_first_part}
\end{center}
\end{figure*}

\begin{figure*}
\begin{center}
\includegraphics[width=\textwidth]{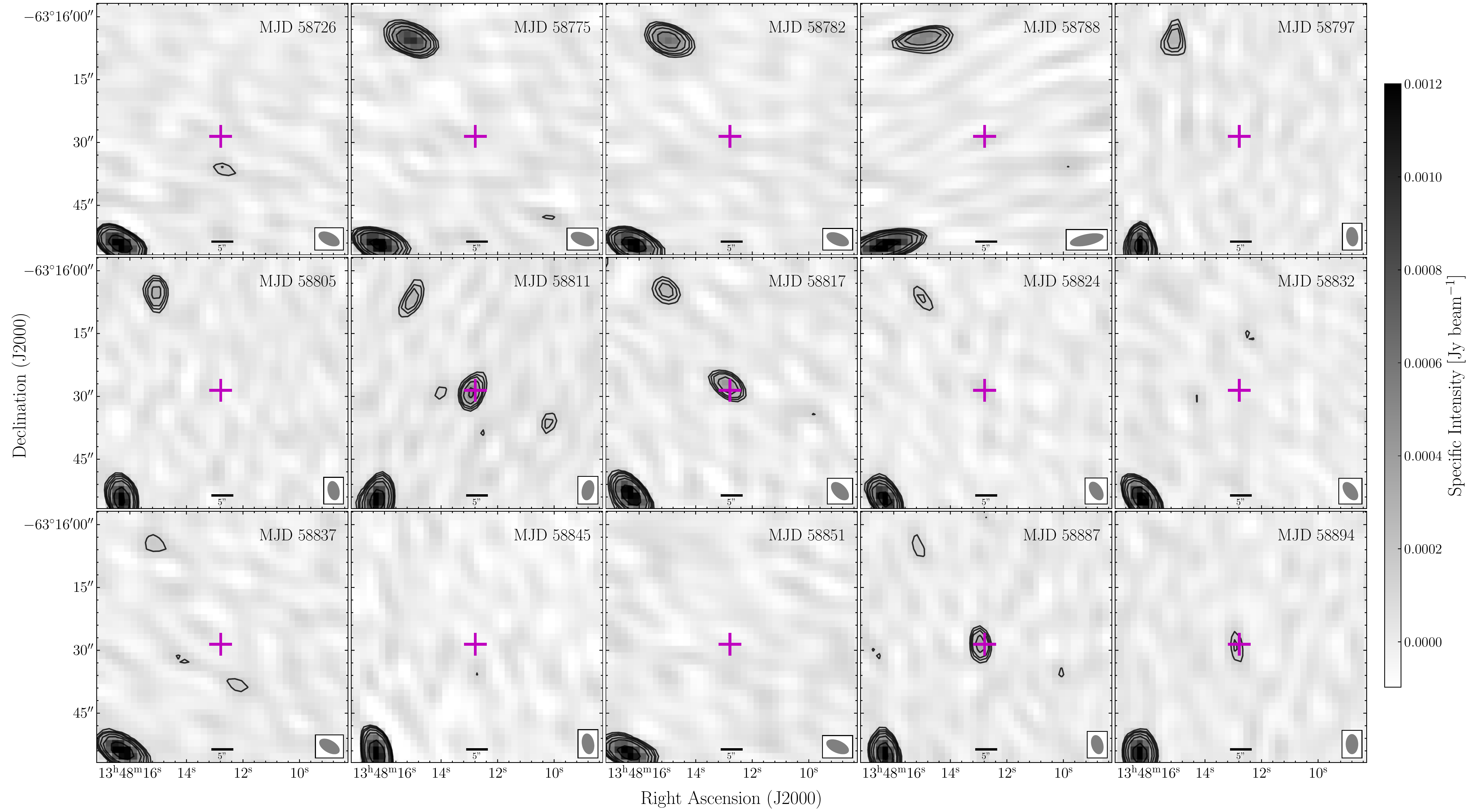}
\caption{A sample of the 1.28 GHz MeerKAT monitoring of \maxithirt{}, including 15 images that cover a period between MJD 58726 (2019 August
31) and MJD 58894 (2020 February 15). Contours start at three times the rms, with the rms ranging between $\sim$35 and $\sim$60 $\mu$Jy beam$^{-1}$. The \maxithirt{} position is marked with a magenta cross. We show the late-time evolution of RK1, as it re-brightened $\sim$8 months after the ejection at a distance of $\sim$27 arcsec from the core position, undergoing a strong deceleration (see Figure \ref{fig:first_jet_angsep}). Compact jets re-activated and decayed during the third and fourth re-flares. The stable, background, point source (possibly consistent with the IR source 2MASS J13481645–6316501) in the bottom left corner is used to correct the astrometry for all the MeerKAT observations.}
\label{fig:meerkat_images_late_times}
\end{center}
\end{figure*}

Radio emission was immediately detected at the beginning of the outburst, at a level of $\sim$3 mJy with ATCA, as the first epoch is on the same day as the first MAXI detection (MJD 58509 --- 2019 January 26). As the system became brighter in the hard state, the emission quickly rose, reaching $\sim$30 mJy with ATCA and MeerKAT on MJD 58515, before the transition to the IMS. In this phase the rising radio emission is consistent with the presence of powerful compact jets. This is confirmed by the positive spectral index, as the flat-to-inverted spectrum is a signature of self-absorbed synchrotron emission, typical of compact radio jets \citep{Blandford_Konigl, Falcke_Biermann_1996, Heinz_Sunyaev_2003}.
Spectral indices are only computed using strictly simultaneous multi-frequency ATCA observations. We detected a sharp drop in the radio flux density between the ATCA observations of MJD 58519 and 58521, when the 5.5 GHz flux density decreased from $\sim$135 mJy to $\sim$52 mJy, before rising again to $\sim$130 mJy on MJD 58523. A strong radio flare was observed on MJD 58523, when the system reached the peak flux density of $\sim$486 mJy at 1.28 GHz \citep{Carotenuto_atel}. To understand if the flare peaked before or after our observation, we divided the 15 minutes long scan of MJD 58523 into three chunks of equal length, and we imaged them separately. We found a decreasing flux density ($\sim$1\%), which implies that the first flare peak was located before MJD 58523.214 (see Table \ref{tab:core_flux_table}). A second, shorter flare ($\sim$150 mJy) was observed some days later with ATCA and ASKAP \citep{Chauhan2020}. We note that the radio flare was accompanied by a significant evolution of the spectral index towards a steep spectrum (from $\alpha \gtrsim 0$ to $\alpha < 0$). The details of the transition are discussed in Section \ref{sec:compact_jets}.
During the weeks following the flare, the emission displayed a steep spectrum and quickly decayed until no radio source was detected on MJD 58551, with a 3$\sigma$ upper limit of 120 $\mu$Jy beam$^{-1}$. This is not surprising, as generally no compact jet radio emission is expected for BH XRBs in the soft state. Interestingly, optically thin radio emission was again detected between MJD 58573 and 58582, before the transition back to the IMS. The emission faded from $\sim$10 mJy to less than 1 mJy, and its origin is discussed more in detail in Section \ref{sec:transient_jets}.

From MJD 58597, the system re-entered the IMS and then transitioned back to the hard state. During this period, we started to detect again optically thick ($\alpha \gtrsim 0$) radio emission, a clear signature of the re-activation of compact jets (see Figure \ref{fig:spectrum}), and the radio and X-ray light curves followed the same smooth evolution. We observed a first decay from MJD 58603 to 58628, in which the flux density decreased from $\sim$5 mJy to $\sim$0.13 mJy (at 1.28 GHz, MeerKAT) and the spectrum became more inverted, a behaviour already observed in several BH XRBs (e.g. \citealt{Fender_2001, Corbel2013_IR, Russell_2015}). We then tracked the compact jet in a first re-flare, during which the jet flux reached $\sim$6 mJy at 1.28 GHz on MJD 58650, and then slowly faded for $\sim$3 months, while maintaining a flat spectrum, until non-detection on MJD 58711 (3$\sigma$ upper limit of 135 $\mu$Jy beam$^{-1}$ with MeerKAT).
During this period of temporary quiescence, the monitoring was interrupted, and it was subsequently resumed in October 2019 after the optical re-brightening reported by \cite{Yazeedi_atel_2}. However, the radio core was not detected during the second, short re-flare. The radio core was detected with MeerKAT and ATCA during the third re-flare (November-December 2019). During the third re-flare the spectrum was consistently flat-to-inverted, when the compact jet re-activated at the $\sim$mJy level, at the same time as the hard-state X-ray re-flare (Figure \ref{fig:main_lc}), before fading below our detection threshold on MJD 58832. 
At the faintest level, we clearly detected \maxithirt{} at 50 $\mu$Jy with ATCA at 5.5 and 9 GHz on MJD 58830 and 58831 (2019 December 13--14) with two 12 hours long observations.
A fourth, fainter radio flare was observed with MeerKAT in February 2020 (MJD 58887 and 58894), until \maxithirt{} went back to quiescence. Our last MeerKAT observations yielded no detection, with a 3$\sigma$ upper limit of 120 $\mu$Jy beam$^{-1}$.

\subsection{First discrete ejection - RK1}
\label{sec:RK1}

\begin{figure}
\begin{center}
\includegraphics[width=\columnwidth]{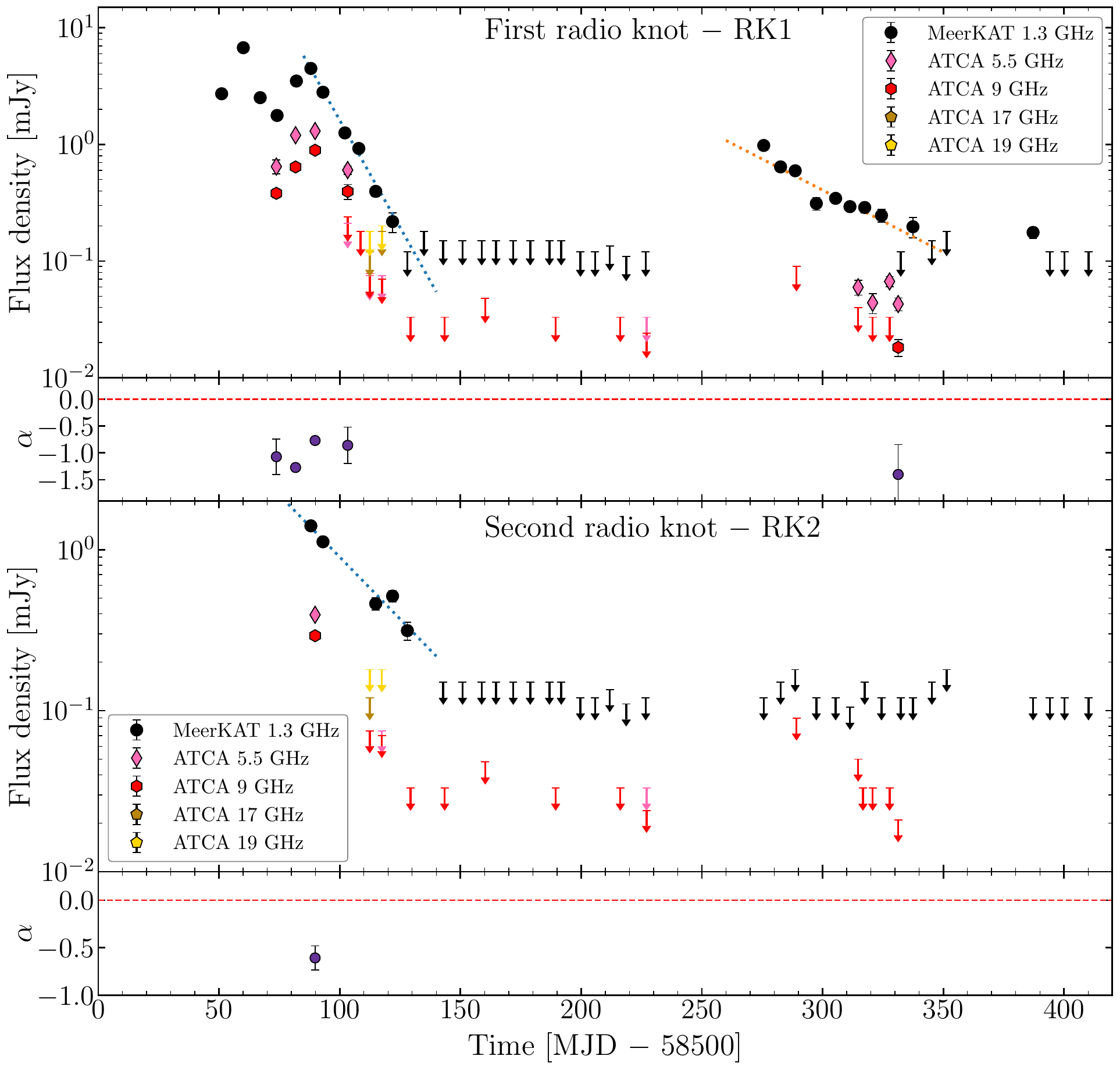}
\caption{Radio monitoring of RK1 and RK2: the two radio knots launched by \maxithirt{} during its 2019/2020 outburst. \emph{First (top) panel}: Radio light curve of RK1 obtained with MeerKAT and ATCA observations. Exponential fits to the decay phases (only MeerKAT) are represented with dashed lines. \emph{Second panel}: Radio spectral index of RK1 obtained from ATCA simultaneous multi-frequency observations, as reported in Table \ref{tab:core_flux_table}. \emph{Third}: Radio light curve of RK2 obtained with MeerKAT and ATCA observations.  An exponential fit to the decay (only MeerKAT) is represented with a dashed line. \emph{Fourth (bottom) panel}: Radio spectral index of RK2 obtained from ATCA simultaneous multi-frequency observations, as reported in Table \ref{tab:RK2_flux_table}.}
\label{fig:twojets_lc}
\end{center}
\end{figure}

We first detected radio emission slightly displaced from the core position ($\sim$2.5 arcsec) on the MeerKAT observation of MJD 58551 (2019 March 09), while \maxithirt{} was in the soft state. The source displayed a flux density of $\sim$3 mJy and appeared to be moving in the North-East direction. We followed the motion of this component for the rest of the outburst, and hereafter we refer to it as RK1. It is likely that the emission of RK1 was initially confused with the fading core emission before MJD 58551. With the proper motion and ejection date obtained for RK1 (discussed in Section \ref{sec:Motion of the first transient jet}), we deduce that RK1 would have reached an angular distance from the core of $\sim$2 arcsec ($\sim$25--40\% of the MeerKAT synthesized beam) on MJD 58538. However, we do not detect such displacement on MJD 58538 and 58545, during which we still observe decaying emission at the core position. This could be due to RK1 being still too close to the core for its emission to be resolved, similarly to \cite{Russell_1535}, or being too faint with respect to the fading core, as the emission appeared to be rising during its first detection (see Figure \ref{fig:twojets_lc}).
MeerKAT radio images from the first part of the outburst are shown in Figure \ref{fig:meerkat_images_first_part} and they illustrate the motion of RK1 during the first phase, while the detections of RK1 at large scale with MeerKAT are shown in Figure \ref{fig:meerkat_images_late_times}. The radio light curve is shown in Figure \ref{fig:twojets_lc} and radio data are reported in Table \ref{tab:core_flux_table} and \ref{tab:first_jet_angsep}. We observed RK1 moving away from \maxithirt{} with a linear trajectory, at a position angle of $33.2\degree \pm 1.4\degree$ East of North.

\subsubsection{Evolution of the major radio flare and associated transient jet}
\label{sec:Evolution of the major radio flare and associated transient jet}

The first part of the RK1 light curve is characterized by some flux variability between MJD 58551 and MJD 58588, as can be seen in Figure \ref{fig:twojets_lc}.
On the MeerKAT observation of MJD 58560 (2019 March 18), the source appeared to be resolved along the jet axis, as it is shown on Figure \ref{fig:meerkat_images_first_part} (fourth panel on the top row). We simultaneously fitted the extended emission with two point sources, interpreting the component located further from the core position as the jet \emph{head}, and obtaining a second component consistent with the core position (at a distance of 0.5 $\pm$ 0.5 arcsec from the core). In the study of the motion of RK1, which is discussed in Section \ref{sec:Motion of the first transient jet}, we use the coordinates of the jet \emph{head}. For the second component, the emission is unlikely to be produced by the core. On MJD 58560 \maxithirt{} was in the soft state, and we do not detect any quick excursion to the SIMS around that time. While we cannot exclude a short radio flare from the core, signature of an additional ejection on very short timescales (e.g. \citealt{Tetarenko2017}), we expect that such flare would have been accompanied by a change in X-ray spectral state, and that such ejection would have been detected in the following epochs.
Therefore, our adopted scenario is that the second component is a receding \emph{tail} associated with RK1. Discrete knot tails have been already observed in X-rays for \xte{} \citep{Migliori2017}, where electrons could be accelerated by a reverse shock and undergo a fast adiabatic cooling. This rapid cooling could explain why the tail is only detected in this epoch.

This phase is followed by a smooth exponential decay that lasted $\sim$6 weeks, from $\sim$4.4 mJy on MJD 58588 until non-detection on MJD 58628, with a 3$\sigma$ upper limit of 120 $\mu$Jy beam$^{-1}$ with MeerKAT. The decay can be well-fitted with an exponential function, which yields a characteristic timescale of 11.8 $\pm$ 0.2 days (e-folding time).
At this point the ejected jet knot appeared to have travelled $\sim$10 arcsec on the plane of the sky in less than 3 months, displaying an extremely high proper motion, which can be seen in Figure \ref{fig:meerkat_images_first_part} and that we discuss in Section \ref{sec:Motion of the first transient jet}.
On MJD 58628, the jet knot decayed below our detection threshold and remained undetected until MJD 58775 (2019 October 19), when additional MeerKAT observations were performed in response to the core optical detection during the second re-flare (R2, \citealt{Yazeedi_atel_2}). On MJD 58775 we detected emission from RK1, this time at an angular distance of $\sim$27 arcsec from \maxithirt{} and at a flux level of $\sim$1 mJy. The re-brightening of RK1 was then followed by a 9 week-long smooth decay, with the exception of the non detection of RK1 on MJD 58832. MeerKAT images of the detection at large scale of RK1 are shown in Figure \ref{fig:meerkat_images_late_times}.
RK1 was also detected with ATCA during the re-flare phase, but only in four epochs, when the length of the observations was large enough to achieve the required $\sim$10 $\mu$Jy sensitivity, and mostly at 5.5 GHz. For the last ATCA detection of RK1, we stacked the two observations taken on December 13 and 14 2019 (MJD 58830 and 58831). While the two epochs were considered separately for the core, as it was quickly decaying in this phase, we do not have hints of variation of RK1 between the two epochs, and therefore the combination of the two observations led to an rms noise of $\sim$6 $\mu$Jy and a detection of RK1 both at 5.5 and 9 GHz, as reported in Table \ref{tab:core_flux_table}. During the re-brightening, RK1 was moving extremely slowly, or not moving at all, as its angular separation from \maxithirt{} ranged between $\sim$26.4 and $\sim$28 arcsec (see Figure \ref{fig:first_jet_angsep}, discussed in Section \ref{sec:Motion of the first transient jet}).
Following the reports of as the re-brightening in February 2020 \citep{Pirbhoy_atel, Shimomukai_atel}, we obtained our last detection of RK1 with MeerKAT at $\sim$170 $\mu$Jy on MJD 58887, roughly 11 months after the first detection.

The radio spectrum of RK1 was steep for all of the five double-frequency ATCA detections, with a spectral index that ranged from $-0.7$ to $-1.4$.
This explains why RK1 was mostly detected with MeerKAT, despite the lower rms achieved in the several $\sim$10 hours observations performed with ATCA during the monitoring. In addition, in case of a diffuse knot, we expect ATCA to resolve out some of the emission due to the higher frequencies and to the smaller number of short baselines compared to MeerKAT. The second decay, without including the last detection on MJD 58887, can also be well-fitted by an exponential function, yielding a much longer characteristic timescale of 41 $\pm$ 4 days, with respect to the first decay. 

Given the light curve evolution, the radio spectrum and the proper motion, RK1 is qualitatively consistent with a typical discrete ejection from an X-ray binary, consisting in synchrotron-emitting plasma bubbles that first expand adiabatically (e.g. \citealt{vanderlaan}) and then re-brighten due to later interactions with the interstellar medium at large scales (e.g. \citealt{Corbel2002_xte, Migliori2017}).

\subsubsection{Motion of the first transient jet}
\label{sec:Motion of the first transient jet}

\begin{figure*}
\begin{center}
\includegraphics[width=\textwidth]{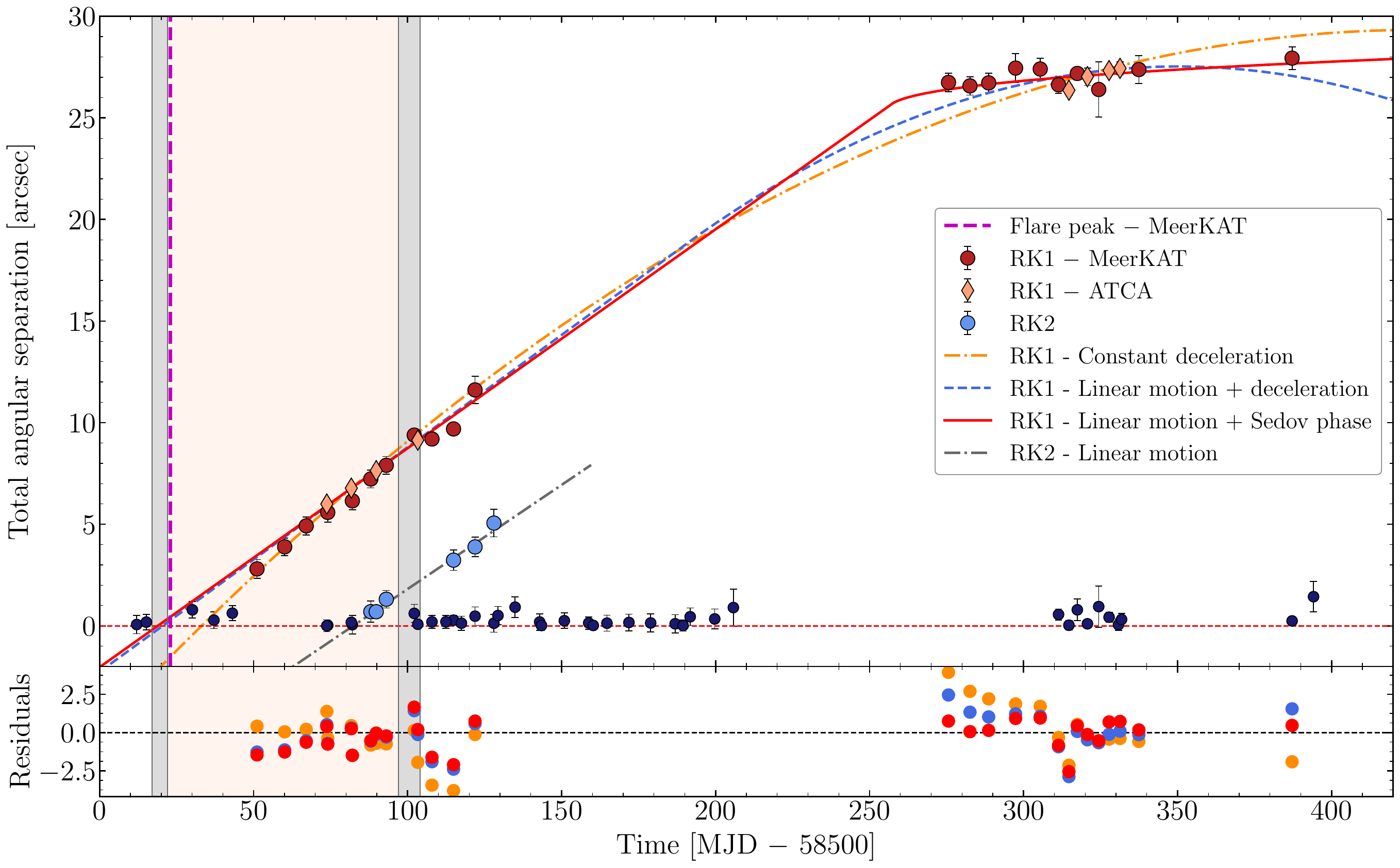}
\caption{Angular separation in arcsec between RK1 and \maxithirt{}. We use here MeerKAT 1.28 GHz and ATCA 5 GHz observations, shown respectively as red and orange points. Detections of the core are shown as blue points. The un-shaded, grey and pink regions mark periods in which \maxithirt{} was, respectively, in the hard, intermediate and soft state. The data are listed in Table \ref{tab:first_jet_angsep}. The astrometry has been corrected using a stable background point source. The red horizontal line represents the zero separation from the core, while the time of the major flare observed with MeerKAT (MJD 58523) is represented with a magenta vertical line. As discussed in Section \ref{sec:RK1}, we fitted the RK1 motion with a constant deceleration model and with two models consisting of a combination of a linear motion with either a simple deceleration, or with a Sedov phase \citep{Wang_model}. Residuals ([data -- model]/uncertainties) are reported for each model in the bottom panel with corresponding colors. They show the better agreement of the linear+Sedov model with the data. Our best estimate for the RK1 ejection date is $t_{\tu{ej}} =$ MJD 58518.9 $\pm$ 2.4. We also show the angular separation of the second discrete ejection RK2, which is discussed in Section \ref{sec:RK2}. A simple linear motion was used to fit the data  of RK2, and our best estimate for the RK2 ejection date is $t_{\tu{ej}} =$ MJD 58582.3 $\pm$ 2.5. All fit results are shown in Table \ref{tab:fit_params_jets}.}
\label{fig:first_jet_angsep}
\end{center}
\end{figure*}

The proper motion of RK1 is shown in Figure \ref{fig:first_jet_angsep} and the data are reported in Table \ref{tab:first_jet_angsep}, for which we used data from MeerKAT and from ATCA at 5.5 GHz that have a higher signal-to-noise ratio with respect to the 9 GHz ones. The offset is computed as the great circle distance between the radio knot and the \maxithirt{} position, either the fitted one in case of core detection on the same epoch, or the core position reported in Section \ref{sec:radio_core_results} in case of core non-detection. In the first case we are not affected by global systematics in the position error estimation.

The motion of RK1 can be divided in two parts: a first phase in which the jet appeared to travel ballistically with a high, constant proper motion and a second part in which the jet underwent a strong deceleration, after the re-brightening on MJD 58775. We discuss the possible reasons for the deceleration in Section \ref{sec:transient_jets}. We start with a simple linear motion model:
\begin{equation}
\alpha(t) = \mu(t-t_{\tu{ej, lin}})
\label{eq:linear}
\end{equation}
where $\alpha(t)$ is the total angular separation from \maxithirt{}, $\mu$ is the proper motion of the jet and $t_{\tu{ej}}$ is the time on which the plasmon is launched. This simple model is clearly not adequate for fitting the whole data set. If we assume instead a constant deceleration throughout out its evolution, we can use the following model:
\begin{equation}
\alpha(t) = \frac{1}{2}\dot{\mu}(t-t_{\tu{ej, decel}})^2 +  \mu(t-t_{\tu{ej, decel}})       
\label{eq:decel}
\end{equation}
A preliminary fit of the data with a simple constant deceleration shows that the model does not seem to describe well the later part of the jet motion, with a reduced $\chi^2 = 3.09 \ (80.3/26)$. The fit yielded an extremely high launching proper motion of $\mu = 148 \pm 4$ mas day$^{-1}$ and a deceleration of $\dot{\mu} = -0.38 \pm 0.03$ mas day$^{-2}$, while the inferred ejection date is $t_{\tu{ej, decel}} =$ MJD 58533.1 $\pm$ 1.3, $\sim$10 days after the radio flare. A better (statistical and physical) result can be obtained by combining the two motions mentioned above in Equation \ref{eq:linear} and \ref{eq:decel}, assuming a jet that starts with constant speed and then decelerates after a \emph{change} time $t_{\tu{c}}$, after eventually having swept up enough mass on its path. By fitting the data with the combined model, we obtain a lower reduced $\chi^2 = 1.69 \ (42.3/25)$. For the linear part we obtain a constant proper motion of $\mu = 111 \pm 4$ mas day$^{-1}$. Then the motion changes at an inferred $t_{\tu{c}} =$ MJD 58689 $\pm$ 17, to continue with a constant deceleration of $\dot{\mu} = -0.68 \pm 0.08$ mas day$^{-2}$. We note that the model predicts that the jet completely stops at MJD 58850.
The fit yields an ejection at earlier times, $t_{\tu{ej, lin+decel}} =$ MJD 58520.4 $\pm$ 2.3.

We then follow the approach of \cite{Miller_jones_sedov} by applying the external shock dynamic model from \cite{Wang_model}, in analogy with models developed for GRB afterglows \citep{Rees_1992, Meszaros_1997}. In the context of such model, the relativistic motion of the ejecta should produce a shock wave that propagates into the ISM and that decelerates as material is swept up along the plasmon trajectory. As a result, the late time behaviour of the adiabatically expanding jet obeys the known Sedov solution $\alpha(t) \propto t^{\frac{2}{5}}$ that in our case reads:
\begin{equation}
\alpha(t) = \alpha_0 + k(t-t_0)^{\frac{2}{5}}     
\label{eq:sedov}
\end{equation}
where $t_0$ is the time at which the jet is at an angular separation $\alpha_0$ from the core position. In this model, the motion is first linear, with a proper motion $\mu$, until a \textit{change} time $t_{\tu{c}}$, after which it follows Equation \ref{eq:sedov}. This fit yields a better reduced $\chi^2 = 1.17 \ (29.3/25)$, and we can also see from Figure \ref{fig:first_jet_angsep} that this model is more adequate to describe the jet motion. For the first, ballistic part we still obtain a high proper motion $\mu = 108 \pm 4$ mas day$^{-1}$. The jet motion later approaches the Sedov phase on $t_{\tu{c}} =$ MJD 58757 $\pm$ 11, with a $t_0 =$ MJD 58755 $\pm$ 12, two days earlier than $t_{\tu{c}}$. Our best estimate for the jet ejection date is $t_{\tu{ej, lin+Sedov}} =$ MJD 58518.9 $\pm$ 2.4, a result very similar to the one obtained with the linear+deceleration model. We find $\alpha_0 = 25.2 \pm 3.8$ arcsec and a normalization parameter $k = 0.3_{-0.3}^{+3.7}$ arcsec day$^{-0.4}$, which is not well constrained.
Given that from our fit $k$ could also be 0, we have to consider the possibility that the jet simply stopped at $\alpha(t) = \alpha_0$ before MJD 58750. Therefore, we tried to fit the data with a combination of a linear motion (Equation \ref{eq:linear}) and a constant position. The result is a reduced $\chi^2 = 1.39$, which is slightly worse than the one obtained from the linear+Sedov model, implying that the scenario of a moving jet is still favored with respect to that of an arrested jet. All the fits for the different models are shown in Figure \ref{fig:first_jet_angsep} and the parameters obtained are listed in Table \ref{tab:fit_params_jets}.

While our data seem to favour the scenario of a linear motion followed by a Sedov phase, we can also consider the information provided by the X-ray and radio monitoring to discriminate between diferent models, with an approach very similar to the one adopted for the single radio knot of \maxififth{} \citep{Russell_1535}. In Section \ref{sec:X-ray emission and spectral evolution} we report that the system transitions from the hard state to the IMS on MJD 58517, and then enters in the soft state on MJD 58522.6 (see Table \ref{tab:state_transitions}). Moreover, we observe with MeerKAT the peak of the strong radio flare on MJD 58523. 
It is generally accepted that radio flares in XRBs are produced by the adiabatic expansion of ejected plasma blobs, hence these components should be launched before strong radio flares are observed (e.g. \citealt{Fender_belloni_gallo, Miller-Jones_h1743, Fender_2019_equipartition}).
On that basis, we can discard the scenario of constant deceleration, since its inferred ejection date $t_{\tu{ej, decel}} =$ MJD 58533.1 $\pm$ 1.3 is $\sim$10 days after the bright radio flare, when the radio emission from the core had been constantly fading, until reaching non detection on MJD 58551. However, we note that, although unlikely, we cannot in principle rule out a totally unrelated radio flare. On the other hand, the two other models both start with a linear motion and they yield very similar results regarding the inferred ejection date, with the $t_{\tu{ej}}$ ranging between MJD 58518 and 58520 (see Table \ref{tab:fit_params_jets}), which are more plausible as they lie before the MeerKAT radio flare, when the system was still in the IMS. We note that, since these ejection dates are obtained from composite models, our estimates for $t_{\tu{ej}}$ are based for the most part on the first half of the data (RK1 detections before MJD 58640).

Assuming that our best estimate for the proper motion is the one obtained by the linear+Sedov model, our measurement of $\mu = 108 \pm 4$ mas day$^{-1}$ during the ballistic part of the motion corresponds to a superluminal, apparent transverse speed of $1.37 \pm 0.05$ $c$, assuming a distance of 2.2 kpc, while it is still superluminal ($\sim c$) for the closest acceptable distance (1.6 kpc).  We can then conclude that RK1 is almost certainly the approaching component and that it is intrinsically relativistic for the first part of its motion. \maxithirt{} is, therefore, a new source displaying discrete ejecta with apparent superluminal motion.

\setlength{\tabcolsep}{8pt}
\setlength{\extrarowheight}{.1em}
\begin{table*}
\caption{Parameters obtained from the fitting of the motion of RK1 and RK2, discussed in Sections \ref{sec:RK1} and \ref{sec:RK2}. We show the proper motion $\mu$, the acceleration $\dot{\mu}$, and, for the composite models, the time $t_{\tu{c}}$ at which the deceleration starts. For composite models, the parameter $\mu$ refers only to the initial, linear part of the motion. For all the models we present the best estimate for the ejection date $t_{\tu{ej}}$ and the reduced $\chi^2$.}
\label{tab:fit_params_jets}
\begin{tabular}{*{7}{c}}
\hline
\hline
Component & Model & $\mu$ & $\dot{\mu}$ & $t_{\tu{c}}$ & $t_{\tu{ej}}$ & Reduced\\
& & [mas day$^{-1}$] & [mas day$^{-2}$] & [MJD] & [MJD] & $\chi^2$\\
\hline
RK1 & Constant deceleration & 148 $\pm$ 4 & $-0.38 \pm 0.03$ & & 58533.1 $\pm$ 1.3 & 3.09 (80.3/26)\\
RK1 & Linear+Deceleration & 111 $\pm$ 4 & $-0.68 \pm 0.08$ & 58689 $\pm$ 17 & 58520.4 $\pm$ 2.3 & 1.69 (42.3/25)\\
RK1 & Linear+Sedov & 108 $\pm$ 4 & & 58757 $\pm$ 11 & 58518.9 $\pm$ 2.4 & 1.17 (29.3/25)\\
\hline
RK2 & Linear motion & 103 $\pm$ 12 & & & 58582.3 $\pm$ 2.5 & 0.21 (0.84/4)\\

\hline
\end{tabular}
\end{table*}

\subsection{Second discrete ejection - RK2}
\label{sec:RK2}

\begin{figure}
\begin{center}
\includegraphics[width=\columnwidth]{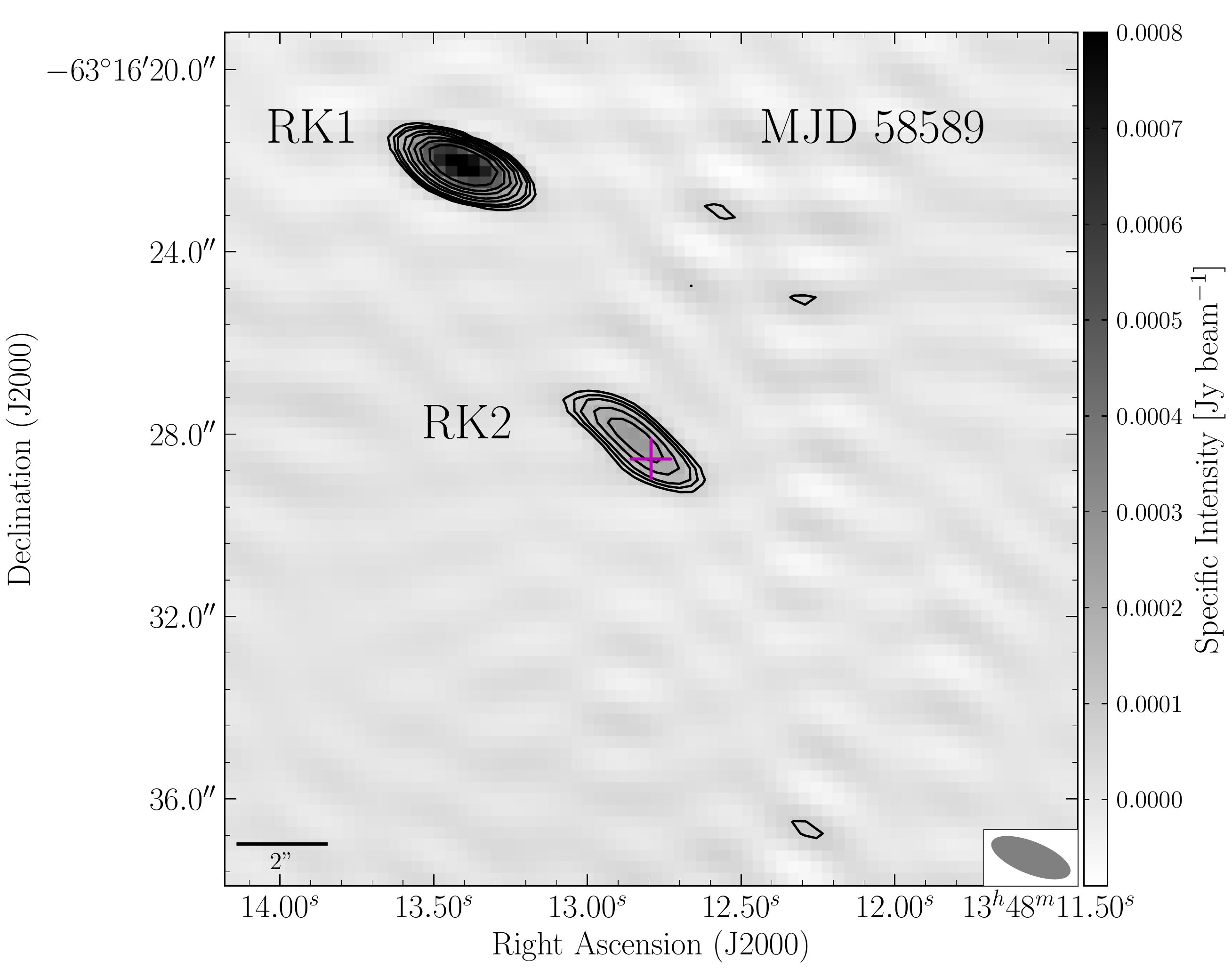}
\caption{ATCA 9 GHz image of the 2019 April 16 observation (MJD 58589), showing the motion of the discrete ejecta RK1 and RK2 as they moved away from the core at position angles of, respectively, $33.2\degree \pm 1.4\degree$ and $33\degree \pm 11\degree$ East of North. Contours start a 3 times the rms ($\sim$20 $\mu$Jy beam$^{-1}$). The \maxithirt{} position is marked with a magenta cross.  RK1 detected an angular separation of $\sim$7.7 arcsec from the core, while RK2 at this stage is at the beginning of its motion, at an angular distance of $\sim$0.7 arcsec from the core.}
\label{fig:atca_images}
\end{center}
\end{figure}

A second radio knot launched by \maxithirt{} was detected approximately one month after the first detection of RK1, when on MJD 58589 (2019 April 16) the radio emission detected with ATCA at 5.5 GHz and 9 GHz was incompatible with the location of the core (see Figure \ref{fig:atca_images}).
We detected it three more times with MeerKAT in May 2019, as it traveled further from \maxithirt{} at a position angle of $33\degree \pm 11\degree$ East of North, which is fully consistent with the measured sky direction of RK1. We can therefore exclude a significant projected precession of the jet axis of the system\footnote{As instead observed for V404 Cygni \citep{Miller-Jones2019}.} between February and April 2019. Hereafter we refer to the second detected radio knot as RK2. The MeerKAT images in which RK2 is detected are shown in  Figure \ref{fig:meerkat_images_first_part}.

The light curve of RK2 is shown in the bottom panel of Figure \ref{fig:twojets_lc}, the radio data are listed on Table \ref{tab:RK2_flux_table} and its motion is displayed in Figure \ref{fig:first_jet_angsep}.
Lower resolution MeerKAT observations on April 15 and 20 2019 (MJD 58588 and 58593) showed a point source slightly offset from the core position in the direction of the RK2 motion.
While the lower resolution MeerKAT data alone for these two epochs would not have allowed us to firmly conclude on a possible displacement (offsets of less than 20\% of the synthesised beam, see Figure \ref{fig:meerkat_images_first_part}, third and fourth panels on the central row), the detection of RK2 with ATCA (see Figure \ref{fig:atca_images}) in between the two MeerKAT observations allows us to associate also the MeerKAT radio emission to RK2.
From the ATCA detection, we measured $\alpha = -0.6$ $\pm$ 0.1. This index is consistent with optically thin synchrotron emission from a second transient jet launched by \maxithirt{} during its 2019/2020 outburst, in the same direction as the previous radio knot. Again we detect no counter-jet in the opposite direction. 
Due to the steep spectrum and radio emission fainter than RK1, RK2 was mostly detected with MeerKAT, and its evolution was tracked for roughly 1 month. The detections are characterised by lower positional accuracy and lower signal-to-noise ratio with respect to RK1, since, for the last three MeerKAT detections, RK2 was never completely resolved from the re-activated compact jet in the second hard state of \maxithirt{}. With the first radio detection, on MJD 58588, RK2 shows a moderate flux density of $\sim$1.4 mJy at 1.3 GHz, while on the following weeks the emission faded to $\sim$400 $\mu$Jy at 1.3 GHz with a $28 \pm 2$ days decay timescale. After the last detection on MJD 58628, RK2 faded below our detection limits for the rest of the whole outburst.

To fit the motion of RK2 we adopt a simple linear, ballistic motion from Equation \ref{eq:linear}. Fitting the data with such model yields a reduced $\chi^2$ of 0.21 (0.84/4), with the low value likely to be coming from overestimated uncertainties on the data points. The parameters obtained from the fits are listed in Table \ref{tab:fit_params_jets}. We obtain again a very high proper motion of $\mu = 103 \pm 12$ mas day$^{-1}$, similar to RK1. The ejection date, $t_{\tu{ej, linear}} =$ MJD 58582.3 $\pm$ 2.5, points towards an unusual ejection in the middle of the soft state; this is discussed in Section \ref{sec:ejection_RK2}. 
For RK2, the proper motion derived from the linear fit corresponds to an apparent transverse speed of $1.30 \pm 0.15$ $c$ at 2.2 kpc, and the jet appears to be superluminal for all but the closest distance (transverse speed $\sim$0.95 $c$ at 1.6 kpc). Hence, RK2 is likely to be the approaching component of a second, intrinsically relativistic, discrete ejection.

\section{DISCUSSION} 
\label{sec:discussion}

\maxithirt{} is a new BH XRB that displayed a rather classical outburst, which we followed with our X-ray and radio monitoring. After completing a whole cycle in the HID, it exhibited four subsequent hard state-only re-brightenings. Our multi-frequency radio observations probed the evolution of compact jets through their initial brightening in the hard state, their quenching as the system transitioned to the soft state and their re-activation in the following hard state re-flares.
Two single-sided, approaching discrete ejections have been detected by their motion away from the core. They were launched in the same direction $\sim$2 months apart, and displayed the highest proper motion ($\gtrsim$ 100 mas day$^{-1}$) observed so far for plasmons ejected from XRBs.

\subsection{Overview of the outburst}
\label{sec:overview}

Thanks to the dense coverage of MAXI and \emph{Swift}/XRT, the X-ray evolution of \maxithirt{} was tracked over almost 6 orders of magnitude in X-ray luminosity, with $L_X$ ranging from 10$^{32}$ to 10$^{38}$ erg s$^{-1}$ in the 1--10 keV energy range, a luminosity interval typical for BH XRBs in outburst (e.g. \citealt{Corral_santana}). The mass of the BH is not known, but if we assume a 7 $M_{\odot}$ black hole (see Section \ref{sec:source_presentation}), this corresponds to a range from $\sim$10$^{-7}$ to $\sim$10$^{-1}$ in units of Eddington luminosity\footnote{The Eddington luminosity is $L_{\tu{Edd}} \sim$ 1.3 $\times 10^{38} \ (M_{\tu{BH}}/M_{\odot}$) erg s$^{-1}$ for a BH accreting hydrogen.}.
During the first part of the outburst, \maxithirt{} performed a whole cycle in the Hardness-Intensity diagram, tracing the complete well-known Q-shape, as can be seen from Figure \ref{fig:maxi_hid}. The appearance in succession of a fast rise in X-ray flux, the transition between the hard, intermediate and soft state, a flare episode and an exponential decay has been already observed in a large number of sources, including, for example, \gx{} (e.g \citealt{Belloni_2005, Tomsick_2008}, Tremou et al. \textit{in prep.}), \maxififth{} \citep{Russell_1535}, XTE~J1752--223 \citep{Brocksopp}, \xte{} \citep{Sobczak_2000} and \hh{} \citep{McClintock_2009, Miller-Jones_h1743}. 
The first state transition from the hard state to the IMS happened at $\sim$3 $(M_{\tu{BH}}/10 M_{\odot})$ \% $L_{\tu{Edd}}$, while the following transition to the soft state was at $\sim$10 $(M_{\tu{BH}}/10 M_{\odot})$ \% $L_{\tu{Edd}}$. The reverse transition to the IMS instead took place at $\sim$1.5 $(M_{\tu{BH}}/10 M_{\odot})$ \% $L_{\tu{Edd}}$, while the system went back to the hard state at $\sim$0.4 $(M_{\tu{BH}}/10 M_{\odot})$ \% $L_{\tu{Edd}}$, consistent with the hysteresis observed in the majority of XRBs undergoing a canonical outburst \citep{Maccarone, Dunn_2010, Kalemci, Vahdat}. 

The following evolution is instead marked by four smaller re-flares in the hard state, interleaved with periods of quiescence, as shown in Figure \ref{fig:main_lc}. Re-flares have already been observed in several BH LMXBs (e.g. \citealt{Chen_1997, Tomsick_2004, Homan_2013, Parikh}), but they are not easily explained by the commonly accepted Disk Instability Model (e.g. \citealt{Lasota2001}) and the mechanism producing them is not yet understood. This behaviour might be due to a smaller amount of available mass in the disk that was not fully replenished after the first flare. Irradiation of the outer accretion disk might also be important in this case. X-ray irradiation illuminates the disk and increases the duration of the outburst by contributing to the disk ionization \citep{Dubus2001}. 
Once in the soft state, the system transitions between a viscous decay and an irradiation-controlled decay, and these two phases determine different profiles in the X-ray light curve \citep{Tetarenko_B_2018}. In this context, an irradiation geometry that varies in space or in time could possibly explain the re-flaring episodes observed after the main outburst \citep{Tetarenko_B_2018, Tetarenko_B_2020}.
Alternatively, re-flares might also be triggered by increased irradiation from the companion star \citep{Hameury2000}. During the re-flares, the source did not exhibit again an hysteresis pattern in the HID, in a similar manner as the 2017/2018 outburst of \maxififth{} \citep{Russell_1535, Parikh}.

It is interesting to note how the X-ray spectrum becomes softer as the source approaches phases of quiescence. This behaviour has been observed in the majority of BH LMXBs, with the photon index that appears to saturate to $\Gamma \sim$ 2 in the quiescent state \citep{Corbel2006, Corbel2008, Plotkin2013, Liu_index}. The anti-correlation between the luminosity and $\Gamma$ in the low luminosity part of the outburst is predicted in the context of Advection-Dominated Accretion Flow (ADAF) models, which are used to describe radiatively inefficient accretion flows \citep{Esin}. The spectral softening is explained as a result of the decrease in efficiency at low luminosities of the inverse Compton scattering that is responsible for the hard X-ray emission. Alternatively, compact jets could start to dominate the X-ray emission at low luminosities and synchrotron cooling from the jet, due to increased radiative losses, would produce a steeper X-ray spectrum \citep{yuan_cui_2005}, although this scenario has been ruled out for V404 Cyg \citep{Plotkin_2017_v404}. Moreover, synchrotron self-Compton from thermal particles at the base of the jet has also been invoked as a possible explanation for the X-ray emission at low luminosities \citep{Markoff_corona, Poutanen_2014, Plotkin2015}.

\subsection{Evolution of the compact jets}
\label{sec:compact_jets}

\begin{figure}
\begin{center}
\includegraphics[width=\columnwidth]{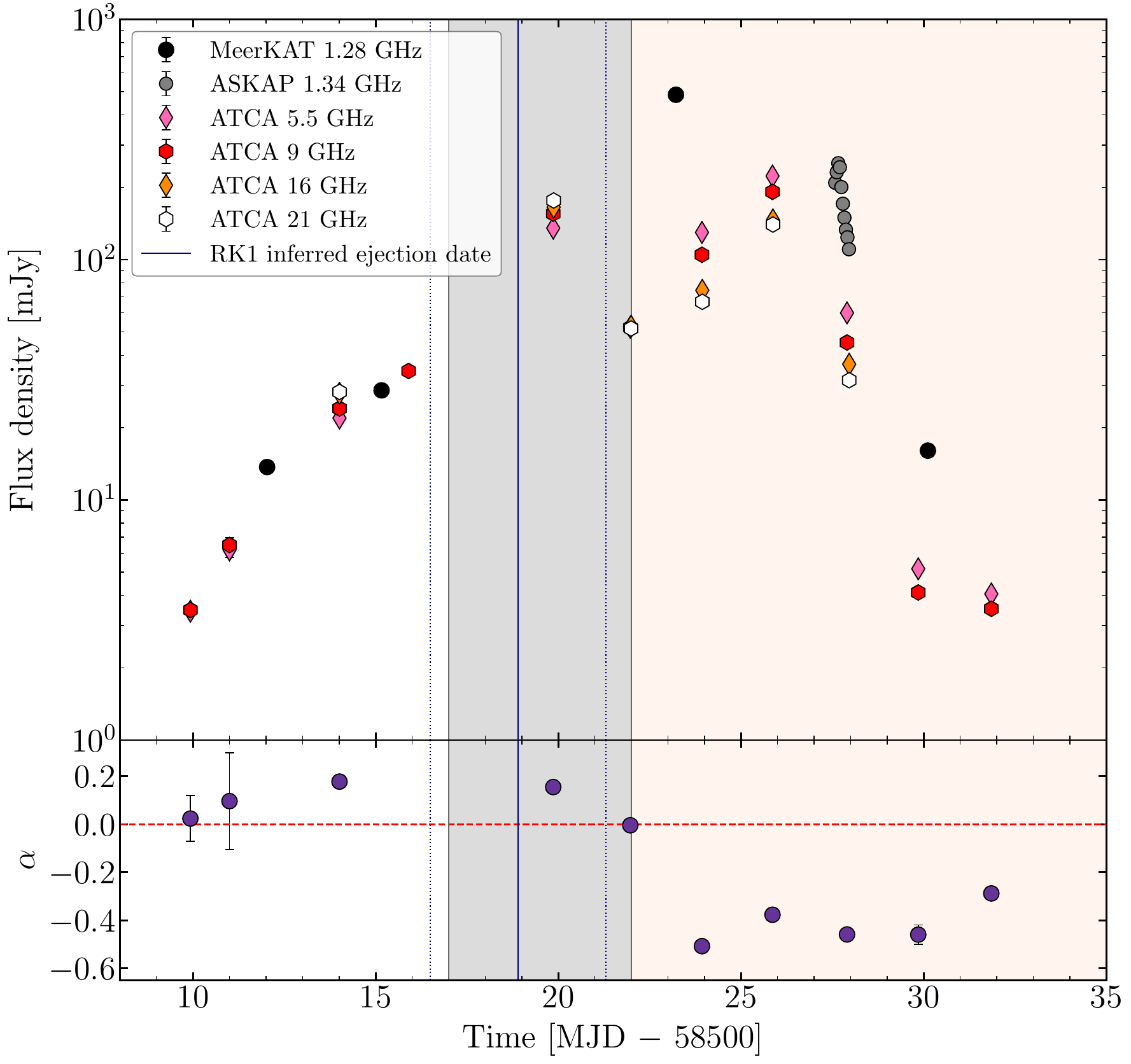}
\caption{First part of the radio light curve at the core location of \maxithirt{}, as shown in the last two panels of Figure \ref{fig:main_lc}. We choose here the MJD interval 58508--58535 to clearly follow the evolution of the light curve at different frequencies and of the spectral index. The un-shaded regions mark the periods in which the source was in a hard state, while the light grey shaded regions indicate the IMS and the light pink regions indicate the soft state. The dates of the state transitions are obtained from \protect\cite{Zhang2020}. The blue vertical line marks the inferred ejection date of RK1: $t_{\tu{ej}} =$ MJD 58518.9 $\pm$ 2.4, where the two blue dashed lines representes the upper and lower bounds of the $t_{\tu{ej}}$ confidence interval.}
 \label{fig:flare_zoom}
\end{center}
\end{figure}

The activation and evolution of compact jets in the 2019/2020 outburst of \maxithirt{} has been detected and tracked with the radio monitoring performed with MeerKAT and ATCA. According to the standard model, compact jets emit self-absorbed synchrotron radiation \citep{Blandford_Konigl}. The resulting spectrum is flat or slightly inverted at radio frequencies, with a turnover frequency observed in the near infrared band \citep{Corbel2002, Russell2013}. Our observations followed the rise of the compact radio jets during the initial hard state and followed it through the transition to the IMS and then to the soft state.

To better visualize the jet behaviour in this phase, we show in Figure \ref{fig:flare_zoom} the radio light curve at the core location for the first 25 days of the outburst, taken from Figure \ref{fig:main_lc}. The compact jet emission peaks on MJD 58520 and then starts to decay on MJD 58521, one day before the system enters in the soft state and two days after the inferred RK1 ejection date (see Section \ref{sec:RK1}). This whole evolution is accompanied by a smooth transition between the optically thick and the optically thin regimes of synchrotron radio emission. Compact jets are usually observed to quench at the transition from the hard to soft state (e.g. \citealt{Fender1999, Corbel_2000}). This phenomenon is observed to start at higher frequencies (where the emission is produced closer to the compact object) and terminate as the jet break evolves through the radio band \citep{Russell_2013b, Russell_2014, Russell2020}. It is not yet clear if compact jets switch off before or during the launch of discrete ejections, and how (or if) the two events are linked \citep{Russell_2020_break_frequency}. We do not have a MeerKAT observation on MJD 58521, but, as the flat spectrum obtained with ATCA suggests, the flux density would have been at the $\sim$50 mJy level. The following observations show instead a quick rise in flux density and a subsequent decay until MJD 58531, when the source is steadily in the soft state. With our data, we cannot conclude on the origin of the radio emission on MJD 58521, which could be produced either from the compact jet that is quenching (with no evidence of the spectral break in the radio band), or by the first self-absorbed part of the radio flare observed to peak on MJD 58523. The latter scenario implies that the jet significantly quenched in less than two days, a shorter timescale with respect to what observed for other sources (e.g. \citealt{Russell_2013b, Russell_2020_break_frequency}). Therefore, we are not able to precisely order in time the ejection of RK1 and the quenching of the compact jets, and thus we cannot draw conclusions on a potential link between the two events.

During the soft state, we detect no radio emission from the compact jets, hence we can place constraints on the jet quenching factor by using the highest flux density compatible with the compact jet in hard state and the lowest upper limit on the non-detection during the following soft state. We find a $\sim$3.5 orders of magnitude quenching factor, which is, so far, one of the strongest constraints on the suppression of compact jets during the soft state of a BH LMXB \citep{Coriat, Russell_1535, Maccarone_2020}.

\subsection{Multiple radio flares}
\label{sec:multiple_radio_flares}
The strong radio flare observed on MJD 58523 is likely to be associated with the ejection of RK1, but the mechanism capable of launching and accelerating these relativistic plasma knots is still not understood. The current picture links the launch of the jets with the transition from the HS to the SS, i.e. during the IMS, where usually ejections take place prior to strong radio flares (e.g. \citealt{Fender2006, Miller-Jones_h1743}) and are marked by changes in the X-ray timing and spectral properties \citep{Belloni_2005}. RK1 fits very well in this picture, as the inferred ejection date 
$t_{\tu{ej, lin+Sedov}} =$ MJD 58518.9 $\pm$ 2.4 (see Section \ref{sec:RK1}) is only 4 days before the strong radio flare observed with MeerKAT, and when \maxithirt{} was in the IMS. 
The presence of Type-B QPOs (tentatively linked to discrete ejections, e.g. \citealt{Soleri2008, Fender_2009, Miller-Jones_h1743, Homan_qpo}) has been detected with NICER close to the first hard-to-soft state transition during the main outburst of \maxithirt{} \citep{Belloni_1348, Zhang2020}. In particular, Type-B QPOs started to be detected on MJD 58522.6 \citep{Zhang2020}. This is $\gtrsim 3$ days after our inferred ejection date, similar to what was already observed for \hh{} \citep{Miller-Jones_h1743} and \maxififth{} \citep{Russell_1535}. However, a detailed discussion of the NICER timing results in relation to the inferred ejection dates will be discussed in a forthcoming paper. The $\sim$5 arcsec MeerKAT resolution does not allow us to resolve the two components soon after the ejection, while the ATCA observations were short (providing a non optimal \emph{uv}-coverage). As the synchrotron-emitting component is launched and expands, it is predicted to have an optically thick rising phase due to an increasing surface area, followed by an optically thin decay produced by adiabatic expansion losses \citep{vanderlaan}. 
Flares produced by discrete ejecta are observed to rise on very different timescales (from minutes or hours to days, e.g. \citealt{Brocksopp_2007, Tetarenko2017, Bright}), depending on the size and energy of the ejected component.
It is possible that the flare had a fast rise, and thus we missed the optically thick phase with our coverage. Another possibility is that the flare was entirely optically thin, as already observed for some sources  (e.g. \citealt{Fender1997}). The ATCA observation taken on MJD 58524 and the decreasing flux observed on MJD 58523 (see Section \ref{sec:radio_core_results}) seem to confirm that we are observing the optically thin decay of the transient jet (see Figure \ref{fig:flare_zoom}).

The following ATCA observations clearly show a second, longer, optically thin flare taking place after the first one, and the emission detected between MJD 58526 and 58532 is possibly resulting from the superposition of the decaying first flare with the rise of the second flare. This is also confirmed by the 1.34 GHz ASKAP observation of MJD 58527, showing a low frequency emission rising, peaking ($\sim$250 mJy) and decaying within the same epoch \citep{Chauhan2020}. In this context, it is not clear what could be producing the second flare. We might be observing the signature of a second ejection, as multiple ejections have been already associated to multiple radio flares on short timescales (e.g. \citealt{Tetarenko2017, Miller-Jones2019}). This would also be consistent with the fact that lower frequencies (ASKAP) appear to peak after the higher ones (ATCA), as predicted for ejected plasmons \citep{vanderlaan}. However, we caution that we might not have sampled the flare adequately, and we cannot infer precisely the location in time of the high frequency peaks. 
Moreover, we do not detect any additional discrete component leaving the system. A possibility is that the emission linked to RK1, observed at later times, could arise from the collision of two ejecta launched in coincidence with the two radio flares. As an alternative, pairs of flares close in time could be produced by the approaching and receding components of a single ejection event, as for instance reported in \cite{Tetarenko2017} for V404 Cyg. However, in V404 Cyg the timescales for the detection of both flares were of the orders of minutes to hours, while in our case there are at least three days between the two flare peaks, hence it seems unlikely that they are produced by a single ejection event.

Our data do not allow us to draw more precise conclusions regarding the origin of the second flare. Obtaining a more accurate estimation of the RK1 ejection date would be useful to get a better understanding of the observations, and a denser coverage, or higher-resolution VLBI observations, could have helped us to discriminate between different scenarios. Nevertheless, this highlights the importance of radio monitorings during flares and state transitions.

\subsection{Two relativistic discrete ejections}
\label{sec:transient_jets}

We detected and tracked the motion of the two discrete ejecta RK1 and RK2 for more than 300 days. Discrete ejections have only been detected so far in a limited subset of the known BH XRBs population (e.g. \citealt{Mirabel1994, GRO1655, Corbel2002_xte, Yang2010, Rushton, Russell_1535, Bright}). The two ejecta both displayed initial proper motions larger than 100 mas day$^{-1}$, which are the highest ever measured for BH XRBs ejections, and, more in general, for objects outside the Solar System. This can be likely ascribed to the relative proximity (2.2 kpc) of \maxithirt{} in comparison to other XRBs.

\begin{figure}
\begin{center}
\includegraphics[width=\columnwidth]{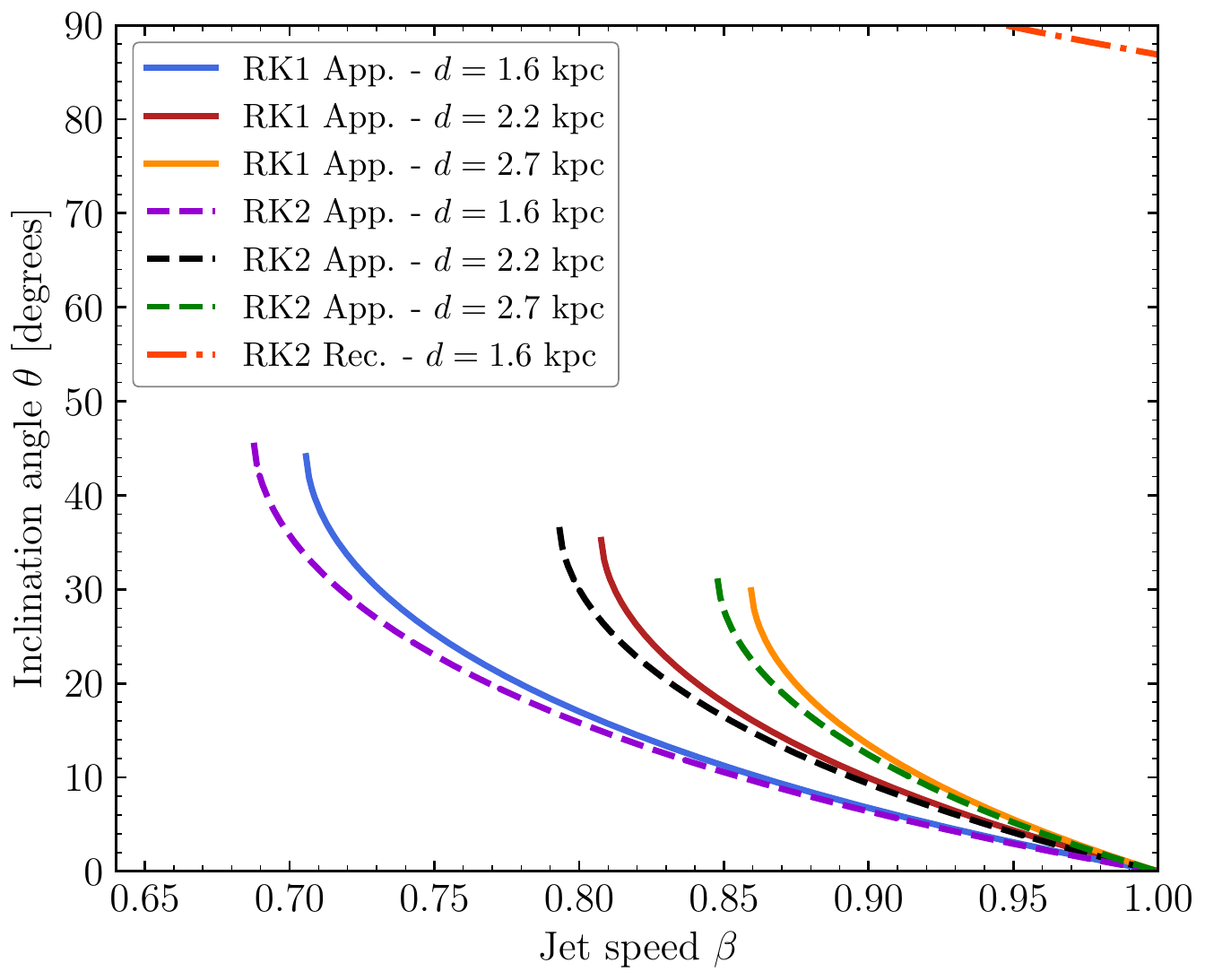}
\caption{Constraints on the jet speed and inclination angle to the line of sight for RK1 (continuous lines) and RK2 (dashed lines). The constraints are derived from Equation \ref{eq:theta_beta}, by using the initial measured proper motions and the whole range of possible distances (2.2$_{-0.6}^{+0.5}$ kpc, \protect\citealt{Chauhan2020}). RK1 has only solutions as the approaching component. RK2, while being almost certainly another approaching jet, does have a small subset of solutions as a receding component for the closest possible distance (top right corner of the bottom panel, dot-dashed line), but we deem them unlikely, as discussed in the text.}
 \label{fig:theta_beta}
\end{center}
\end{figure}

We can put constraints on the true speed $\beta$ of the knots and their inclination angle $\theta$ to the line of sight at the time of ejection, under the assumption of intrinsically symmetric jets. For unpaired components we can only solve for $\beta \cos{\theta}$ in the following equation, once the distance is known (e.g. \citealt{Rees_1966, Mirabel1994}):
\begin{equation}
        \mu_{\tu{app}} = \frac{\beta \sin\theta}{1 - \beta \cos{\theta}} \frac{c}{D} \ \ \ \ \ \ \ \ \ \mu_{\tu{rec}} = \frac{\beta \sin\theta}{1 + \beta \cos{\theta}} \frac{c}{D}
        \label{eq:theta_beta}
\end{equation}
where $\mu$ is the proper motion for the approaching or receding component and $D$ is the source distance. The results for RK1 and RK2 are shown in Figure \ref{fig:theta_beta}, where we plot the existing solutions for the best distance estimate and for its upper and lower bounds. For RK1, Equation \ref{eq:theta_beta} has solutions only for the jet being the approaching component at all possible distances, in agreement with its apparent superluminal motion. We obtain $\beta_1 \cos{\theta_1} \geq 0.51$,  which implies that $\beta_1 \geq 0.70$ and that $\theta_1 \leq 44\degree$ (accepting only solutions for which $\beta$ decreases as $\theta$ increases). A low inclination angle is consistent with RK1 appearing brighter due to Doppler boosting, and at the same time it is consistent with the non detection of the receding paired component, as strongly Doppler de-boosted. Moreover, if we assume a jet axis perpendicular to the accretion disk\footnote{This does not appear to be universal among BH XRBs (e.g. \citealt{Maccarone_2002}).}, our constraints point to a relatively face-on configuration of the system, which can in turn suggest a significant Doppler boosting of the compact jet itself.

Regarding RK2, we can see from Figure \ref{fig:theta_beta} that the detected knot is almost certainly the approaching component of the second ejection. Similarly to RK1, we obtain $\beta_2 \cos{\theta_2} \geq 0.48$, leading to $\beta_2 \geq 0.69$ and $\theta_2 \leq 46\degree$. However, Equation \ref{eq:theta_beta} has also a small number of solutions with RK2 being the receding component, only at the closest possible distance of 1.6 kpc, implying an extremely fast jet with $\beta$ close to 1 and an inclination angle close to 90\degree. This scenario is highly unlikely for two main reasons. First, with such an intrinsically fast jet we would expect to detect also the approaching component, which would be highly Doppler boosted and should be even brighter than the receding component. Second, the results would suggest an almost edge-on disk configuration, which we can discard on the basis of the fact that we do not detect any modulation of the X-ray light curve that would eventually be due to the occultation by the companion star. Moreover, these constraints would be totally inconsistent with the ones obtained for RK1. Hence, we can safely affirm that also RK2 is the approaching component, which is possibly traveling with the same trajectory of RK1, given that we do not find any evidence for a varying jet speed or inclination angle between the two ejections.

\subsubsection{Ejection of RK2}
\label{sec:ejection_RK2}

Considering RK2, we obtained the ejection date $t_{\tu{ej}} =$ MJD $58582.6 \pm 2.3$, implying an ejection during an interval in which the system appeared to be in the soft state, a behaviour that has never been clearly observed before in BH XRBs. We discuss here that the ejection rather happened during a short excursion from the soft state to the IMS. While the sequence of events leading to a discrete ejection from a BH XRB has not been understood yet, these events are classically linked to the hard-to-soft state transitions (e.g. \citealt{Fender_belloni_gallo, Corbel2004, Fender2006}). Optically-thin radio emission from the core position was detected with MeerKAT and ATCA between MJD 58573 and 58582, reaching almost 10 mJy at 1.3 GHz on MJD 58582, the same day as our inferred ejection date. The emission then faded below our detection threshold on the following epoch. While the observed flux density is one order of magnitude lower than the one measured at the beginning of the outburst (on MJD 58523), it may still indicate the presence of a radio flare in coincidence with our inferred ejection date. 
Considering the information provided by the X-ray monitoring, a possible additional hint can be found in the MAXI hardness-ratio evolution: while the system was in the soft state and the X-ray emission was completely dominated by the accretion disk (see Section \ref{sec:X-ray emission and spectral evolution}), we detected a fast spike in the hardness, with the source getting harder on MJD 58574 (see Figures \ref{fig:maxi_hid} and \ref{fig:main_lc}). This may suggest a very quick return of the source to a particularly short-lived IMS, a spectral state that is strongly connected to discrete ejections (e.g. \citealt{Corbel2004,Fender_belloni_gallo}).
While the spike in hardness happens at least $\sim$5 days prior to our inferred $t_{\tu{ej}}$, the sudden change in spectral state of the source is likely associated to the ejection of RK2.

This behaviour is not new among the known population of BH XRBs. Sources have already been observed to rapidly oscillate between the soft and the hard state (e.g. \citealt{Homan_2001, Fender_2009}) and multiple radio flares and discrete ejections have already been observed as as the source moved back and forth across the top branch of the HID (e.g. \citealt{Brocksopp_2001, Tetarenko2017}). An example is the 1999 outburst of XTE~J1859$+$226, during which several short excursions to the hard-intermediate state (or the so-called Steep Power Law state, given the spectral hardness and the presence of high-frequency QPOs), were clearly correlated with radio flares, signature of multiple ejections \citep{Brocksopp_2002, Fender_2009}. In our case, however, the quick return to the IMS did not happen on the top branch of the HID, but at later times, when the source was in the middle of the soft state, descending the left branch on the diagram (see Figure \ref{fig:maxi_hid}).

\subsection{Late time behaviour of the discrete ejecta}
\label{sec:late_time}

The motion of RK1 was monitored for $\sim$300 days and, after a first phase in which it traveled at constant speed, it appeared to re-brighten while being strongly decelerated in the second phase of its motion. Decelerating discrete ejecta have been observed in a number of other sources, namely \xte{} \citep{Corbel2002_xte}, \hh{} \citep{Corbel2005_h17}, XTE~J1752--223 \citep{Yang2010, Miller_jones_sedov}, \maxififth{} \citep{Russell_1535} and \maxieight{} \citep{Bright, Espinasse_xray}. For some of them the discrete knots were detected both in radio and X-rays, with the \emph{Chandra} observatory.
The current picture is that the plasma blob decelerates via interactions with the ISM, causing constant \emph{in-situ} particle acceleration (up to TeV energies) and producing broadband, optically thin synchrotron radiation \citep{Corbel2002_xte, Migliori2017, Espinasse_xray}.

\subsubsection{Is \maxithirt{} in a low density cavity?}
\label{sec:low density cavity?}

After a first part of ballistic, high-speed motion, the deceleration of RK1 was rather abrupt, which is something not observed in the majority of discrete ejecta from BH XRBs (e.g \citealt{Mirabel1994, Fender1999, Miller-Jones_h1743}).
This scenario is consistent a jet that travels first at constant speed in a low-density region of the ISM, which constitutes a large scale cavity around the system, before hitting the higher-density wall of the cavity itself, as already proposed in \cite{Hao}. Radio emission at late times would be produced by the external shock between the plasma blob and the ISM cavity wall, in analogy with GRB afterglows \citep{Wang_model}. Those cavities have been suggested to exist at $\sim$pc scales at least for \xte{} and \hh{} \citep{Hao, Steiner_xte, Steiner_h17, Migliori2017}.
We obtained the best modeling for the RK1 motion by a combination of a linear motion and a Sedov phase (see Section \ref{sec:RK1}), achieved at late times due to the jet sweeping up ISM material on its path, in a similar way as \cite{Miller_jones_sedov}, which derived it from \cite{Wang_model}. We therefore suggest that \maxithirt{} is located in a similar cavity, possibly carved by previous jet activity, or resulting from the action of accretion disk winds. However, so far such winds have not been detected for \maxithirt{}. If we take our estimate of $\alpha_0 \sim$ 25 arcsec, the upper limit on the inclination angle of the discrete ejection of $\theta_1 \sim$ 44\degree and the lowest acceptable distance of 1.6 kpc, we obtain that RK1 traveled at least $\sim$0.3 pc before reaching the angular distance $\alpha_0$, which is consistent with what has been observed for other discrete ejections \citep{Corbel_2000, Gallo_2004} and could be taken as a rough lower limit on the size of the cavity\footnote{At least in the NE direction, as an asymmetric cavity has been reported for \xte{} \citep{Hao, Steiner_xte}.}. 
As argued by \cite{Hao}, the presence of an under-dense cavity could be a common characteristic of BH XRBs environments \citep{Heinz_2002}, and its existence could be strictly required for the jet to travel such a long distance \citep{Hao}. 

This scenario is also supported by the flux evolution of RK1, which is shown in Figure \ref{fig:twojets_lc} and consists for the most part of two optically thin decay phases. The first decay, between MJD 58588 and 58621, has an e-folding time of $\sim$12 days (see Section \ref{sec:RK1}), during which the quickly decaying radio emission is likely produced by the ejecta radiating while in free adiabatic expansion (e.g. \citealt{vanderlaan}). The second decay (MJD 58782--58837) is significantly slower than the first, with an e-folding time of $\sim$41 days, and for which the radio emission is likely produced by particles re-accelerated in external shocks between the jet and the ISM, possibly at the cavity wall.
This is similar to what was observed in \xte{} \citep{Corbel2002_xte, Tomsick_2003, Migliori2017} and \hh{} \citep{Corbel2005_h17}. Decelerating ejecta have also been observed in the 2017/2018 outburst of \maxififth{} \citep{Russell_1535}, and in the 2018 outburst of \maxieight{} by \cite{Bright}, which reported a $\sim$6 days quick first decay, immediately followed by two slower $\sim$50 and $\sim$20 days decay phases. However, while jet-ISM interaction may take place, \maxieight{}{} is unlikely to be located in a low density cavity \citep{Bright}, and this could explain the lack of late-time re-brightenings of the discrete ejecta.

There are other possible scenarios for the late time behaviour of RK1. The knot was unresolved in all observations during the second part of the motion, and, as it becomes fainter, we notice some scatter in its position between MJD 58810 and 58840 (see Figure \ref{fig:first_jet_angsep}). This may be due either to a partial jet fragmentation, or to the presence of residual emission from cooling particles, or also to the presence of a receding jet tail, if the radio emission is connected to reverse shocks \citep{Migliori2017}. Internal shocks between unseen shells of plasma have also been invoked to explain the re-brightening of discrete ejections \citep{Fomalont_2001, Fender_belloni_gallo, Motta_sco}. 

Given the observation of possibly two radio flares at the hard-to-soft state transition, we could speculate that another possible explanation for the late re-brightening of RK1 is that the radio emission observed after MJD 58775 was produced by the collision of two separate, approaching discrete ejections, launched previously at different times on the same trajectory. We assume that the first component was RK1, inferred to be launched on MJD 58518.9 and tracked until MJD 58621, and the second one was ejected on the second radio flare reported to peak on MJD 58527 (see Figure \ref{fig:flare_zoom}). The second component cannot be RK2, as this would have required an acceleration of the discrete ejecta. Using Equation \ref{eq:theta_beta} and assuming constant speed, we obtain that, for the second component to catch up with RK1 on MJD 58775 at an angular distance from the core of $\sim$26 arcsec, we should have a difference in speed between the two jets of $\Delta\beta = \beta_2 - \beta_1$ between $10^{-2}$ and $10^{-3}$, considering the whole range of acceptable values of inclination angles $\theta$ inferred for RK1 (see discussion in Section \ref{sec:transient_jets} and Figure \ref{fig:theta_beta}). While in in principle such values of $\Delta\beta$ could be possible, the difference in speed between the two components appears to be rather small, especially if we compare it with what obtained for multiple ejections in other sources (e.g. \citealt{Tetarenko2017}). Therefore, some \textit{fine-tuning} seems to be required for this scenario to match the observational constraints. However, we mention that we could have missed the exact time of the re-brightening of RK1, and a different (earlier) re-appearance of RK1 could alleviate the requirement on $\Delta\beta$.

\subsubsection{Jet opening angle and expansion speed}
\label{sec:Jet_opening_angle}

RK1 remained unresolved in the direction perpendicular to its motion for the entire monitoring. Hence, we can put constraints on the jet opening angle and on the speed of its transverse expansion. Our best constraints can be obtained with the last detection of RK1 with ATCA, on MJD 58831. Using the $\sim$2.5 arcsec ATCA beam at 9 GHz and an angular separation from the core of 27 arcsec, we conclude that the RK1 opening angle is $\leq$6\degree  (half-opening angle $\leq$3\degree) and the transverse expansion speed must be $\leq$0.05 $c$. These values are consistent with what is derived for other BH LMXBs (e.g. \citealt{Miller-jones2006, Russell_1535}). In particular, our values are consistent in terms of transverse expansion speed with the multiple ejections (except the two fastest components) observed and modeled in V404 Cyg \citep{Tetarenko2017}, although we find a smaller jet opening angle for RK1. Our results are in agreement with the small jet expansion speeds calculated in \cite{Fender_2019_equipartition}. Similar computations are not informative for RK2, given that the second knot was detected at a maximum separation of only $\sim$5 arcsec from the core position.

\subsubsection{Considerations on RK2}
\label{sec:Considerations on RK2}

For the second discrete ejections RK2, we are only able to constrain the first part of its motion, as RK2 quickly faded below our detection limits, roughly 1 month after its inferred ejection date. However, as RK2 was ejected in the same direction of RK1 and with a very similar initial proper motion, we would expect a similar behaviour between the two discrete components. With an e-folding time of $\sim$28 days, the decay of RK2 (between MJD 58588 and 58628, see Figure \ref{fig:twojets_lc}) is slower than the first decay of RK1, although the emission is more likely to be produced by the adiabatically expanding ejecta rather than particle re-acceleration in external shocks with the ISM, as RK2 was only detected during the early phase of its motion. If that is the case, this implies that RK2 has a smaller fractional expansion speed in comparison to RK1.

In the context of a cavity surrounding \maxithirt{}, if RK2 had the same energetic content of RK1, it should have hit the wall at the same angular distance as RK1 did before. It is thus possible that the detection of radio emission $\sim$27 arcsec from the core on MJD 58887, which we attributed to RK1, is in fact due to RK2 reaching the same ISM density bump, hence undergoing deceleration and subsequent re-brightening. If we assume that RK2 kept the same initial proper motion of $\sim$100 mas day$^{-1}$, it would have taken roughly 270 days for it to get $\sim$27 arcsec far from the core, which is broadly consistent with the last point attributed to RK1. It becomes even more consistent if we include deceleration in this scenario. However, the flux of RK1 on its last detection is roughly consistent with the overall decay in its re-brightening, and RK2 was never detected at late times at the $\sim$mJy levels the re-brightening of RK1. Hence it is possible that RK2 had less material to interact with, after RK1 swept up most of the material first.

\subsection{Transient jet energetics}
\label{sec:energetics}

We constrain the minimum energy and the corresponding magnetic field required to produce the synchrotron luminosity of the first, strong radio flare (e.g. \citealt{Longair, Fender2006}). If we make the standard assumption of equipartition between electrons and magnetic fields in the jet plasma, we can compute the minimum energy as following:
\begin{equation}
        E_{\tu{min}} \simeq 8 \times 10^{6} \eta^{\frac{4}{7}} \bigg(\frac{V}{\tu{cm}^3}\bigg)^{\frac{3}{7}} \bigg(\frac{\nu}{\tu{GHz}}\bigg)^{\frac{2}{7}} \bigg(\frac{L_{\nu}}{\tu{erg} \ \tu{s}^{-1} \tu{Hz}^{-1}}\bigg)^{\frac{4}{7}} \tu{erg}
        \label{eq:equipartition}
\end{equation}
where $\eta$ is a parameter linked to the ratio of energy in protons to that in electrons. Here we assume that synchrotron emission from protons is negligible, so that $\eta=1$.
Since the knot cannot expand faster than light, we can estimate an upper limit on the volume of the emission region $V$ at launch\footnote{However, \cite{Fender_2019_equipartition} suggest that these ejections expand at only at small fractions (0.01$\sim$0.1) of $c$.}: $V = (4/3)\pi(c\Delta t)^3 $. The conservative rise time $\Delta t$ of the RK1 ejection can be obtained by the interval between the last optically thick ATCA detection of the \maxithirt{} core on MJD 58521.9 and the high $\sim$0.5 Jy peak observed with MeerKAT on MJD 58523.2, for which optically thin emission is detected with ATCA on the same day, likely produced by the adiabatic expansion of the plasma knot. We then convert the peak flux of the flare observed with MeerKAT at 1.3 GHz to the monochromatic radio luminosity $L_{\nu} = 4\pi D^2 S_{\nu}$, where $S_{\nu}$ is the measured flux density. 
Neglecting for the moment any Doppler effects, we obtain a minimum energy $E_{\tu{min}}\sim$ 10$^{42}$ erg, a minimum average power $P_{\tu{min}} \sim$ 10$^{37}$ erg s$^{-1}$ ($\sim$10\% of the simultaneous X-ray luminosity) and a corresponding equipartition magnetic field $B_{\tu{eq}} \sim$ 10 mGauss. The Lorentz factor of electrons radiating around 1.3 GHz is $\sim$5000. 
Those numbers are consistent with what has been found for other discrete ejections \citep{Fender1999, Curran, Russell_1535, Bright}, but the magnetic field is two orders of magnitude higher than what was obtained for \maxieight{} \citep{Espinasse_xray}.

Given the constraints on the ejecta speed and inclination angle provided in Section \ref{sec:transient_jets} (see Figure \ref{fig:theta_beta}), it is possible to estimate the effect of Doppler boosting on the inferred jet energetics. Assuming the bulk jet Lorentz factor $\Gamma_{\tu{j}}$ to be in the range between $\sim$1.4 (from the constraint $\beta_1 \gtrsim 0.7$) and 5, with a corresponding inclination angle $1\degree \lesssim \theta \lesssim 44\degree$ from Equation \ref{eq:theta_beta}, we obtain the Doppler factor of an approaching component $\delta_{\tu{app}} = \Gamma_{\tu{j}}^{-1}(1 - \beta \cos{\theta})^{-1}$ to be in the range between $\sim$1.4 and $\sim$9.8. The minimum energy in the ejecta rest frame can be computed as $E_{\tu{min, RF}} \simeq \Gamma_{\tu{j}} E_{\tu{min}} \delta^{(4\alpha -5)/7}$ (including the jet kinetic energy, \citealt{Fender2006}), where $E_{\tu{min}}$ comes from Equation \ref{eq:equipartition} and $\alpha$ is the radio spectral index of the ejecta. We take $\alpha = -0.5$ (measured value on MJD 58523, see Table \ref{tab:core_flux_table}) and we obtain a $E_{\tu{min, RF}}$ in the range $4\sim8 \times 10^{41}$ erg. We can also derive the minimum intrinsic power and magnetic field as, respectively $P_{\tu{min, RF}} \simeq \Gamma_{\tu{j}} P_{\tu{min}} \delta^{(4\alpha -12)/7}$ and $B_{\tu{eq, RF}} =  B_{\tu{eq}} \delta^{(2\alpha -13)/7}$, obtaining $P_{\tu{min, RF}}$ to range between $\sim$0.3 and $\sim$5 $\times 10^{36}$ erg s$^{-1}$ and $B_{\tu{eq, RF}}$ to range between $\sim$0.1 and $\sim$6 mGauss. In the rest frame of the ejecta, the minimum energy, power and equipartion magnetic field are lower than the ones derived ignoring Doppler effects. We then conclude that the inclusion of Doppler boosting reduces the energetics constraints obtained with the equipartition assumption for the acceptable combinations of $\Gamma_{\tu{j}}$ and $\theta$.

Considering the major flare, an independent estimation of the ejecta expansion speed and on the minimum energy can be obtained by assuming that the flare was due to synchrothron self-absorption, that it was initially optically thick and that at the peak the optical depth was $\tau \simeq 1$, as outlined in \cite{Fender_2019_equipartition}. From Equations 27 and 28 of \cite{Fender_2019_equipartition}, and ignoring Doppler effects, we obtain an expansion speed of $\beta_{\tu{exp}} \simeq 0.015$ and a minimum energy of $E_{\tu{min}}\sim$ 10$^{39}$ erg (less constraining than the previous estimation). This expansion speed leads to a size of the emitting region that is much smaller than what was previously estimated and is consistent with what discussed in Section \ref{sec:Jet_opening_angle} and with previous results for other sources \citep{Miller-jones2006, Tetarenko2017, Russell_1535, Fender_2019_equipartition}. The minimum energy limit is less constraining, but both the values inferred for $E_{\tu{min}}$ in \maxithirt{} in this Section are likely to be largely conservative, as it appears that a dominant fraction of the ejecta energy is not radiated, but is instead transferred as kinetic energy to the surrounding environment \citep{Bright}.
If we include again Doppler boosting in our estimation, we can obtain in the rest frame of the ejecta the minimum energy $E_{\tu{min, RF}} \simeq \Gamma_{\tu{j}} E_{\tu{min}} \delta^{-97/34}$ and an ejecta expansion speed $\beta_{\tu{RF, exp}} = \beta_{\tu{exp}} \delta^{-49/34}$, as outlined in \cite{Fender_2019_equipartition}. Assuming the same range as above for $\Gamma_{\tu{j}}$ and $\theta$, we obtain $E_{\tu{min, RF}}$ in the range between  $0.09\sim6 \times 10^{38}$ erg and an expansion speed $\beta_{\tu{RF, exp}}$ to range between $\sim$6$\times 10^{-4}$ and $\sim$0.01. Also in this case, taking into account Doppler corrections leads to a lower minimum energy for the synchrotron flare, and yields  an even smaller jet expansion speed, in agreement with \cite{Fender_2019_equipartition}. For RK2, similar estimations are not possible, given that we do not have a good constraint on the rise time of the radio flare at the moment of its ejection.

\section{CONCLUSIONS} 
\label{sec:conclusion}
\maxithirt{} is a new black hole candidate in a LMXB system that fits very well in the general picture adopted for X-ray binaries. In this work we have presented the X-ray and radio monitoring of  \maxithirt{} during its 2019/2020, discovery outburst. With our X-ray monitoring we have been able to follow the whole outburst, in which the source displayed a rather typical X-ray evolution in the first part, completing a whole cycle in the HID, and then exhibited a complex sequence of hard-state-only re-brightenings in the second part. Thanks to our radio observations, we monitored the rise, quenching, and re-activation of compact jets in different phases of the outburst, specifically constraining the jet quenching factor to 3.5 orders of magnitude in the soft state.
Two subsequent, single-sided, approaching discrete ejections have been tracked as they moved away from the core in the same direction, with the highest proper motion ($\gtrsim$ 100 mas day$^{-1}$) observed so far among XRBs ejecta.
We were able to put similar constraints for the two knots on the intrinsic speed $\beta \geq 0.7$ and the inclination angle of the jet axis $\theta \leq 45\degree$. Since the first knot was followed for 11 months, we were able to constrain its opening angle to $\leq6\degree$. We precisely characterise its motion, composed of a first part of ballistic motion followed by a strong deceleration happening after the knot traveled at least 0.3 pc. This deceleration is in agreement with the scenario of microquasars embedded in low density, parsec-scale ISM cavities \citep{Hao}.
From the modeling of the motion of the two knots, we can infer that the ejection of the first knot happened before a strong radio flare, while the system was in the IMS. Moreover, the constraints on the minimum energy and power required to produce the observed radio flare are in agreement with the values generally measured for discrete ejections. The second jet was likely ejected during a short excursion of system from the soft to the intermediate state, a behaviour already observed in several XRBs. In this context, high sensitivity and high cadence radio monitorings are crucial to improve our understanding on the jets launching mechanism and on the environment surrounding microquasars.

\section*{Data availability}
The un-calibrated MeerKAT and ATCA visibility data are publicly available at the SARAO and ATNF archives, respectively at \url{https://archive.sarao.ac.za} and \url{https://atoa.atnf.csiro.au}. The \textit{Swift}/XRT data are instead available from the \textit{Swift} archive:
\url{https://www.swift.ac.uk/swift_portal}, while the MAXI data can be downloaded from \url{http://maxi.riken.jp/mxondem}.

\section*{Acknowledgements}

We thank the anonymous referee for the careful reading of the manuscript and for providing valuable comments. We thank the staff at the South African Radio Astronomy Observatory (SARAO) for scheduling these observations. The MeerKAT telescope is operated by the South African Radio Astronomy Observatory, which is a facility of the National Research Foundation, an agency of the Department of Science and Innovation. This work was carried out in part using facilities and data processing pipelines developed at the Inter-University Institute for Data Intensive Astronomy (IDIA). IDIA is a partnership of the Universities of Cape Town, of the Western Cape and of Pretoria. FC, SC and TR thank Jamie Stevens and staff from the Australia Telescope National Facility (ATNF) for scheduling the ATCA radio observations. The Australia Telescope Compact Array is part of the Australia Telescope National Facility which is funded by the Australian Government for operation as a National Facility managed by CSIRO. We also thank \emph{Swift} for the scheduling of the X-ray observations. FC thanks Jerome Rodriguez for useful discussions regarding the \emph{Swift}/XRT data analysis. This research has made use of the XRT Data Analysis Software (XRTDAS) developed under the responsibility of the ASI Science Data Center (ASDC), Italy. FC and SC thank Diego Altamirano and Liang Zhang for useful discussions on the NICER data. We acknowledge the use of data obtained from the High Energy Astrophysics Science Archive Research Center (HEASARC), provided by NASA's Goddard Space Flight Center.  
FC acknowledges support from the project Initiative d’Excellence (IdEx) of Universit\'{e} de Paris (ANR-18-IDEX-0001).
We acknowledge the use of the Nan\c cay Data Center, hosted by the Nan\c cay Radio Observatory (Observatoire de Paris-PSL, CNRS, Universit\'{e} d'Orl\'{e}ans), and also supported by Region Centre-Val de Loire.
This research has made use of MAXI data provided by RIKEN, JAXA and the MAXI team. GRS acknowledges support from Natural Sciences and Engineering Research Council of Canada (NSERC) Discovery Grants (RGPIN-06569-2016). AH acknowledges support by the I-Core Program of the Planning and Budgeting Committee and the Israel Science Foundation, support by the ISF grant 647/18, and support from from the United States - Israel Binational Science Foundation (BSF). This research was supported by a Grant from the GIF, the German-Israeli Foundation for Scientific Research and Development.




\bibliographystyle{mnras}

\bibliography{mnras_maxi1348}

\begin{thebibliography}{}
\makeatletter
\relax
\def\mn@urlcharsother{\let\do\@makeother \do\$\do\&\do\#\do\^\do\_\do\%\do\~}
\def\mn@doi{\begingroup\mn@urlcharsother \@ifnextchar [ {\mn@doi@}
  {\mn@doi@[]}}
\def\mn@doi@[#1]#2{\def\@tempa{#1}\ifx\@tempa\@empty \href
  {http://dx.doi.org/#2} {doi:#2}\else \href {http://dx.doi.org/#2} {#1}\fi
  \endgroup}
\def\mn@eprint#1#2{\mn@eprint@#1:#2::\@nil}
\def\mn@eprint@arXiv#1{\href {http://arxiv.org/abs/#1} {{\tt arXiv:#1}}}
\def\mn@eprint@dblp#1{\href {http://dblp.uni-trier.de/rec/bibtex/#1.xml}
  {dblp:#1}}
\def\mn@eprint@#1:#2:#3:#4\@nil{\def\@tempa {#1}\def\@tempb {#2}\def\@tempc
  {#3}\ifx \@tempc \@empty \let \@tempc \@tempb \let \@tempb \@tempa \fi \ifx
  \@tempb \@empty \def\@tempb {arXiv}\fi \@ifundefined
  {mn@eprint@\@tempb}{\@tempb:\@tempc}{\expandafter \expandafter \csname
  mn@eprint@\@tempb\endcsname \expandafter{\@tempc}}}

\bibitem[\protect\citeauthoryear{{Al Yazeedi}, {Russell}, {Lewis}, {Baglio},
  {Bramich}  \& {Saikia}}{{Al Yazeedi} et~al.}{2019}]{Yazeedi_atel_2}
{Al Yazeedi} A.,  {Russell} D.~M.,  {Lewis} F.,  {Baglio} M.~C.,  {Bramich}
  D.~M.,   {Saikia} P.,  2019, The Astronomer's Telegram, \href
  {https://ui.adsabs.harvard.edu/abs/2019ATel13188....1A} {13188, 1}

\bibitem[\protect\citeauthoryear{{Arnaud}}{{Arnaud}}{1996}]{Arnaud_xspec}
{Arnaud} K.~A.,  1996, in {Jacoby} G.~H.,  {Barnes} J.,  eds,  Astronomical
  Society of the Pacific Conference Series Vol. 101, Astronomical Data Analysis
  Software and Systems V. p.~17

\bibitem[\protect\citeauthoryear{{Belloni}}{{Belloni}}{2010}]{Belloni_2010}
{Belloni} T.~M.,  2010, {States and Transitions in Black Hole Binaries}.
Berlin Springer Verlag, p.~53, \mn@doi{10.1007/978-3-540-76937-8_3}

\bibitem[\protect\citeauthoryear{{Belloni} \& {Motta}}{{Belloni} \&
  {Motta}}{2016}]{Belloni_Motta2016}
{Belloni} T.~M.,  {Motta} S.~E.,  2016, {Transient Black Hole Binaries}.
Springer, p.~61, \mn@doi{10.1007/978-3-662-52859-4_2}

\bibitem[\protect\citeauthoryear{{Belloni}, {Homan}, {Casella}, {van der Klis},
  {Nespoli}, {Lewin}, {Miller}  \& {M{\'e}ndez}}{{Belloni}
  et~al.}{2005}]{Belloni_2005}
{Belloni} T.,  {Homan} J.,  {Casella} P.,  {van der Klis} M.,  {Nespoli} E.,
  {Lewin} W.~H.~G.,  {Miller} J.~M.,   {M{\'e}ndez} M.,  2005, \mn@doi [\aap]
  {10.1051/0004-6361:20042457}, \href
  {https://ui.adsabs.harvard.edu/abs/2005A&A...440..207B} {440, 207}

\bibitem[\protect\citeauthoryear{Belloni, Zhang, Kylafis, Reig  \&
  Altamirano}{Belloni et~al.}{2020}]{Belloni_1348}
Belloni T.~M.,  Zhang L.,  Kylafis N.~D.,  Reig P.,   Altamirano D.,  2020,
  \mn@doi [\mnras] {10.1093/mnras/staa1843}, 496, 4366

\bibitem[\protect\citeauthoryear{{Blandford} \& {K{\"o}nigl}}{{Blandford} \&
  {K{\"o}nigl}}{1979}]{Blandford_Konigl}
{Blandford} R.~D.,  {K{\"o}nigl} A.,  1979, \mn@doi [\apj] {10.1086/157262},
  \href {https://ui.adsabs.harvard.edu/abs/1979ApJ...232...34B} {232, 34}

\bibitem[\protect\citeauthoryear{Bright et~al.,}{Bright et~al.}{2020}]{Bright}
Bright J.,  et~al., 2020, \mn@doi [Nature Astronomy]
  {10.1038/s41550-020-1023-5}, 4, 1

\bibitem[\protect\citeauthoryear{Brocksopp, Jonker, Fender, Groot, van~der Klis
   \& Tingay}{Brocksopp et~al.}{2001}]{Brocksopp_2001}
Brocksopp C.,  Jonker P.,  Fender R.~P.,  Groot P.,  van~der Klis M.,   Tingay
  S.,  2001, \mn@doi [Astrophys. Space Sci.] {10.1023/A:1011698018918}, 276,
  117

\bibitem[\protect\citeauthoryear{{Brocksopp} et~al.,}{{Brocksopp}
  et~al.}{2002}]{Brocksopp_2002}
{Brocksopp} C.,  et~al., 2002, \mn@doi [\mnras]
  {10.1046/j.1365-8711.2002.05230.x}, \href
  {https://ui.adsabs.harvard.edu/abs/2002MNRAS.331..765B} {331, 765}

\bibitem[\protect\citeauthoryear{{Brocksopp}, {Miller-Jones}, {Fender}  \&
  {Stappers}}{{Brocksopp} et~al.}{2007}]{Brocksopp_2007}
{Brocksopp} C.,  {Miller-Jones} J.~C.~A.,  {Fender} R.~P.,   {Stappers} B.~W.,
  2007, \mn@doi [\mnras] {10.1111/j.1365-2966.2007.11846.x}, \href
  {https://ui.adsabs.harvard.edu/abs/2007MNRAS.378.1111B} {378, 1111}

\bibitem[\protect\citeauthoryear{{Brocksopp}, {Corbel}, {Tzioumis},
  {Broderick}, {Rodriguez}, {Yang}, {Fender}  \& {Paragi}}{{Brocksopp}
  et~al.}{2013}]{Brocksopp}
{Brocksopp} C.,  {Corbel} S.,  {Tzioumis} A.,  {Broderick} J.~W.,  {Rodriguez}
  J.,  {Yang} J.,  {Fender} R.~P.,   {Paragi} Z.,  2013, \mn@doi [\mnras]
  {10.1093/mnras/stt493}, \href
  {https://ui.adsabs.harvard.edu/abs/2013MNRAS.432..931B} {432, 931}

\bibitem[\protect\citeauthoryear{{Burrows} et~al.,}{{Burrows}
  et~al.}{2005}]{Burrows_xrt}
{Burrows} D.~N.,  et~al., 2005, \mn@doi [\ssr] {10.1007/s11214-005-5097-2},
  \href {https://ui.adsabs.harvard.edu/abs/2005SSRv..120..165B} {120, 165}

\bibitem[\protect\citeauthoryear{{Camilo} et~al.,}{{Camilo}
  et~al.}{2018}]{Camilo2018}
{Camilo} F.,  et~al., 2018, \mn@doi [\apj] {10.3847/1538-4357/aab35a}, \href
  {https://ui.adsabs.harvard.edu/abs/2018ApJ...856..180C} {856, 180}

\bibitem[\protect\citeauthoryear{{Carotenuto}, {Tremou}, {Corbel}, {Fender},
  {Woudt}  \& {Miller-Jones}}{{Carotenuto} et~al.}{2019}]{Carotenuto_atel}
{Carotenuto} F.,  {Tremou} E.,  {Corbel} S.,  {Fender} R.,  {Woudt} P.,
  {Miller-Jones} J.,  2019, The Astronomer's Telegram, \href
  {https://ui.adsabs.harvard.edu/abs/2019ATel12497....1C} {12497, 1}

\bibitem[\protect\citeauthoryear{{Carotenuto}, {Corbel}, {Fender}, {Woudt}  \&
  {Miller-Jones}}{{Carotenuto} et~al.}{2020}]{Carotenuto_atel_2}
{Carotenuto} F.,  {Corbel} S.,  {Fender} R.,  {Woudt} P.,   {Miller-Jones} J.,
  2020, The Astronomer's Telegram, \href
  {https://ui.adsabs.harvard.edu/abs/2020ATel13467....1C} {13467, 1}

\bibitem[\protect\citeauthoryear{{Cash}}{{Cash}}{1979}]{cstat}
{Cash} W.,  1979, \mn@doi [\apj] {10.1086/156922}, \href
  {https://ui.adsabs.harvard.edu/abs/1979ApJ...228..939C} {228, 939}

\bibitem[\protect\citeauthoryear{{Chauhan} et~al.,}{{Chauhan}
  et~al.}{2021}]{Chauhan2020}
{Chauhan} J.,  et~al., 2021, \mn@doi [\mnras] {10.1093/mnrasl/slaa195}, \href
  {https://ui.adsabs.harvard.edu/abs/2021MNRAS.501L..60C} {501, L60}

\bibitem[\protect\citeauthoryear{{Chen}, {Shrader}  \& {Livio}}{{Chen}
  et~al.}{1997}]{Chen_1997}
{Chen} W.,  {Shrader} C.~R.,   {Livio} M.,  1997, \mn@doi [\apj]
  {10.1086/304921}, \href
  {https://ui.adsabs.harvard.edu/abs/1997ApJ...491..312C} {491, 312}

\bibitem[\protect\citeauthoryear{{Chen} et~al.,}{{Chen}
  et~al.}{2019}]{HXMT_atel}
{Chen} Y.~P.,  et~al., 2019, The Astronomer's Telegram, \href
  {https://ui.adsabs.harvard.edu/abs/2019ATel12470....1C} {12470, 1}

\bibitem[\protect\citeauthoryear{{Corbel} \& {Fender}}{{Corbel} \&
  {Fender}}{2002}]{Corbel2002}
{Corbel} S.,  {Fender} R.~P.,  2002, \mn@doi [\apjl] {10.1086/341870}, \href
  {https://ui.adsabs.harvard.edu/abs/2002ApJ...573L..35C} {573, L35}

\bibitem[\protect\citeauthoryear{{Corbel}, {Fender}, {Tzioumis}, {Nowak},
  {McIntyre}, {Durouchoux}  \& {Sood}}{{Corbel} et~al.}{2000}]{Corbel_2000}
{Corbel} S.,  {Fender} R.~P.,  {Tzioumis} A.~K.,  {Nowak} M.,  {McIntyre} V.,
  {Durouchoux} P.,   {Sood} R.,  2000, \aap, \href
  {https://ui.adsabs.harvard.edu/abs/2000A&A...359..251C} {359, 251}

\bibitem[\protect\citeauthoryear{{Corbel}, {Fender}, {Tzioumis}, {Tomsick},
  {Orosz}, {Miller}, {Wijnand s}  \& {Kaaret}}{{Corbel}
  et~al.}{2002}]{Corbel2002_xte}
{Corbel} S.,  {Fender} R.~P.,  {Tzioumis} A.~K.,  {Tomsick} J.~A.,  {Orosz}
  J.~A.,  {Miller} J.~M.,  {Wijnand s} R.,   {Kaaret} P.,  2002, \mn@doi
  [Science] {10.1126/science.1075857}, \href
  {https://ui.adsabs.harvard.edu/abs/2002Sci...298..196C} {298, 196}

\bibitem[\protect\citeauthoryear{{Corbel}, {Fender}, {Tomsick}, {Tzioumis}  \&
  {Tingay}}{{Corbel} et~al.}{2004}]{Corbel2004}
{Corbel} S.,  {Fender} R.~P.,  {Tomsick} J.~A.,  {Tzioumis} A.~K.,   {Tingay}
  S.,  2004, \mn@doi [\apj] {10.1086/425650}, \href
  {https://ui.adsabs.harvard.edu/abs/2004ApJ...617.1272C} {617, 1272}

\bibitem[\protect\citeauthoryear{{Corbel}, {Kaaret}, {Fender}, {Tzioumis},
  {Tomsick}  \& {Orosz}}{{Corbel} et~al.}{2005}]{Corbel2005_h17}
{Corbel} S.,  {Kaaret} P.,  {Fender} R.~P.,  {Tzioumis} A.~K.,  {Tomsick}
  J.~A.,   {Orosz} J.~A.,  2005, \mn@doi [\apj] {10.1086/432499}, \href
  {https://ui.adsabs.harvard.edu/abs/2005ApJ...632..504C} {632, 504}

\bibitem[\protect\citeauthoryear{{Corbel}, {Tomsick}  \& {Kaaret}}{{Corbel}
  et~al.}{2006}]{Corbel2006}
{Corbel} S.,  {Tomsick} J.~A.,   {Kaaret} P.,  2006, \mn@doi [\apj]
  {10.1086/498230}, \href
  {https://ui.adsabs.harvard.edu/abs/2006ApJ...636..971C} {636, 971}

\bibitem[\protect\citeauthoryear{{Corbel}, {Koerding}  \& {Kaaret}}{{Corbel}
  et~al.}{2008}]{Corbel2008}
{Corbel} S.,  {Koerding} E.,   {Kaaret} P.,  2008, \mn@doi [\mnras]
  {10.1111/j.1365-2966.2008.13542.x}, \href
  {https://ui.adsabs.harvard.edu/abs/2008MNRAS.389.1697C} {389, 1697}

\bibitem[\protect\citeauthoryear{{Corbel} et~al.,}{{Corbel}
  et~al.}{2013}]{Corbel2013_IR}
{Corbel} S.,  et~al., 2013, \mn@doi [\mnras] {10.1093/mnrasl/slt018}, \href
  {https://ui.adsabs.harvard.edu/abs/2013MNRAS.431L.107C} {431, L107}

\bibitem[\protect\citeauthoryear{{Coriat} et~al.,}{{Coriat}
  et~al.}{2011}]{Coriat}
{Coriat} M.,  et~al., 2011, \mn@doi [\mnras]
  {10.1111/j.1365-2966.2011.18433.x}, \href
  {https://ui.adsabs.harvard.edu/abs/2011MNRAS.414..677C} {414, 677}

\bibitem[\protect\citeauthoryear{{Corral-Santana}, {Casares},
  {Mu{\~n}oz-Darias}, {Bauer}, {Mart{\'\i}nez-Pais}  \&
  {Russell}}{{Corral-Santana} et~al.}{2016}]{Corral_santana}
{Corral-Santana} J.~M.,  {Casares} J.,  {Mu{\~n}oz-Darias} T.,  {Bauer} F.~E.,
  {Mart{\'\i}nez-Pais} I.~G.,   {Russell} D.~M.,  2016, \mn@doi [\aap]
  {10.1051/0004-6361/201527130}, \href
  {https://ui.adsabs.harvard.edu/abs/2016A&A...587A..61C} {587, A61}

\bibitem[\protect\citeauthoryear{{Curran} et~al.,}{{Curran}
  et~al.}{2014}]{Curran}
{Curran} P.~A.,  et~al., 2014, \mn@doi [\mnras] {10.1093/mnras/stt2125}, \href
  {https://ui.adsabs.harvard.edu/abs/2014MNRAS.437.3265C} {437, 3265}

\bibitem[\protect\citeauthoryear{{Denisenko} et~al.,}{{Denisenko}
  et~al.}{2019}]{Denisenko_atel}
{Denisenko} D.,  et~al., 2019, The Astronomer's Telegram, \href
  {https://ui.adsabs.harvard.edu/abs/2019ATel12430....1D} {12430, 1}

\bibitem[\protect\citeauthoryear{{Dubus}, {Hameury}  \& {Lasota}}{{Dubus}
  et~al.}{2001}]{Dubus2001}
{Dubus} G.,  {Hameury} J.~M.,   {Lasota} J.~P.,  2001, \mn@doi [\aap]
  {10.1051/0004-6361:20010632}, \href
  {https://ui.adsabs.harvard.edu/abs/2001A&A...373..251D} {373, 251}

\bibitem[\protect\citeauthoryear{{Dunn}, {Fender}, {K{\"o}rding}, {Belloni}  \&
  {Cabanac}}{{Dunn} et~al.}{2010}]{Dunn_2010}
{Dunn} R.~J.~H.,  {Fender} R.~P.,  {K{\"o}rding} E.~G.,  {Belloni} T.,
  {Cabanac} C.,  2010, \mn@doi [\mnras] {10.1111/j.1365-2966.2010.16114.x},
  \href {https://ui.adsabs.harvard.edu/abs/2010MNRAS.403...61D} {403, 61}

\bibitem[\protect\citeauthoryear{{Esin}, {McClintock}  \& {Narayan}}{{Esin}
  et~al.}{1997}]{Esin}
{Esin} A.~A.,  {McClintock} J.~E.,   {Narayan} R.,  1997, \mn@doi [\apj]
  {10.1086/304829}, \href
  {https://ui.adsabs.harvard.edu/abs/1997ApJ...489..865E} {489, 865}

\bibitem[\protect\citeauthoryear{{Espinasse} et~al.,}{{Espinasse}
  et~al.}{2020}]{Espinasse_xray}
{Espinasse} M.,  et~al., 2020, \mn@doi [\apjl] {10.3847/2041-8213/ab88b6},
  \href {https://ui.adsabs.harvard.edu/abs/2020ApJ...895L..31E} {895, L31}

\bibitem[\protect\citeauthoryear{{Falcke} \& {Biermann}}{{Falcke} \&
  {Biermann}}{1996}]{Falcke_Biermann_1996}
{Falcke} H.,  {Biermann} P.~L.,  1996, \aap, \href
  {https://ui.adsabs.harvard.edu/abs/1996A&A...308..321F} {308, 321}

\bibitem[\protect\citeauthoryear{{Fender}}{{Fender}}{2001}]{Fender_2001}
{Fender} R.~P.,  2001, \mn@doi [\mnras] {10.1046/j.1365-8711.2001.04080.x},
  \href {https://ui.adsabs.harvard.edu/abs/2001MNRAS.322...31F} {322, 31}

\bibitem[\protect\citeauthoryear{{Fender}}{{Fender}}{2006}]{Fender2006}
{Fender} R.,  2006, in Compact stellar X-ray sources. pp 381--419

\bibitem[\protect\citeauthoryear{{Fender} \& {Bright}}{{Fender} \&
  {Bright}}{2019}]{Fender_2019_equipartition}
{Fender} R.,  {Bright} J.,  2019, \mn@doi [\mnras] {10.1093/mnras/stz2000},
  \href {https://ui.adsabs.harvard.edu/abs/2019MNRAS.489.4836F} {489, 4836}

\bibitem[\protect\citeauthoryear{Fender \& Mu{\~n}oz-Darias}{Fender \&
  Mu{\~n}oz-Darias}{2016}]{Fender_balance}
Fender R.,  Mu{\~n}oz-Darias T.,  2016, \mn@doi [Lect. Notes Phys.]
  {10.1007/978-3-319-19416-5\_3}, 905, 65

\bibitem[\protect\citeauthoryear{{Fender}, {Bell Burnell}, {Waltman}, {Pooley},
  {Ghigo}  \& {Foster}}{{Fender} et~al.}{1997}]{Fender1997}
{Fender} R.~P.,  {Bell Burnell} S.~J.,  {Waltman} E.~B.,  {Pooley} G.~G.,
  {Ghigo} F.~D.,   {Foster} R.~S.,  1997, \mn@doi [\mnras]
  {10.1093/mnras/288.4.849}, \href
  {https://ui.adsabs.harvard.edu/abs/1997MNRAS.288..849F} {288, 849}

\bibitem[\protect\citeauthoryear{Fender, Garrington, McKay, Muxlow, Pooley,
  Spencer, Stirling  \& Waltman}{Fender et~al.}{1999}]{Fender1999}
Fender R.~P.,  Garrington S.~T.,  McKay D.~J.,  Muxlow T. W.~B.,  Pooley G.~G.,
   Spencer R.~E.,  Stirling A.~M.,   Waltman E.~B.,  1999, \mn@doi [\mnras]
  {10.1046/j.1365-8711.1999.02364.x}, 304, 865

\bibitem[\protect\citeauthoryear{{Fender}, {Belloni}  \& {Gallo}}{{Fender}
  et~al.}{2004}]{Fender_belloni_gallo}
{Fender} R.~P.,  {Belloni} T.~M.,   {Gallo} E.,  2004, \mn@doi [\mnras]
  {10.1111/j.1365-2966.2004.08384.x}, \href
  {https://ui.adsabs.harvard.edu/abs/2004MNRAS.355.1105F} {355, 1105}

\bibitem[\protect\citeauthoryear{{Fender}, {Homan}  \& {Belloni}}{{Fender}
  et~al.}{2009}]{Fender_2009}
{Fender} R.~P.,  {Homan} J.,   {Belloni} T.~M.,  2009, \mn@doi [\mnras]
  {10.1111/j.1365-2966.2009.14841.x}, \href
  {https://ui.adsabs.harvard.edu/abs/2009MNRAS.396.1370F} {396, 1370}

\bibitem[\protect\citeauthoryear{{Fender} et~al.,}{{Fender}
  et~al.}{2017}]{ThunderKAT}
{Fender} R.,  et~al., 2017, arXiv e-prints, \href
  {https://ui.adsabs.harvard.edu/abs/2017arXiv171104132F} {p. arXiv:1711.04132}

\bibitem[\protect\citeauthoryear{{Fomalont}, {Geldzahler}  \&
  {Bradshaw}}{{Fomalont} et~al.}{2001}]{Fomalont_2001}
{Fomalont} E.~B.,  {Geldzahler} B.~J.,   {Bradshaw} C.~F.,  2001, \mn@doi
  [\apjl] {10.1086/320490}, \href
  {https://ui.adsabs.harvard.edu/abs/2001ApJ...553L..27F} {553, L27}

\bibitem[\protect\citeauthoryear{{Gaensler}, {Green}  \&
  {Manchester}}{{Gaensler} et~al.}{1998}]{SNR}
{Gaensler} B.~M.,  {Green} A.~J.,   {Manchester} R.~N.,  1998, \mn@doi [\mnras]
  {10.1046/j.1365-8711.1998.01814.x}, \href
  {https://ui.adsabs.harvard.edu/abs/1998MNRAS.299..812G} {299, 812}

\bibitem[\protect\citeauthoryear{{Gallo}, {Corbel}, {Fender}, {Maccarone}  \&
  {Tzioumis}}{{Gallo} et~al.}{2004}]{Gallo_2004}
{Gallo} E.,  {Corbel} S.,  {Fender} R.~P.,  {Maccarone} T.~J.,   {Tzioumis}
  A.~K.,  2004, \mn@doi [\mnras] {10.1111/j.1365-2966.2004.07435.x}, \href
  {https://ui.adsabs.harvard.edu/abs/2004MNRAS.347L..52G} {347, L52}

\bibitem[\protect\citeauthoryear{{Gehrels} et~al.,}{{Gehrels}
  et~al.}{2004}]{Gehrels}
{Gehrels} N.,  et~al., 2004, \mn@doi [\apj] {10.1086/422091}, \href
  {https://ui.adsabs.harvard.edu/abs/2004ApJ...611.1005G} {611, 1005}

\bibitem[\protect\citeauthoryear{{Hameury}, {Lasota}  \& {Warner}}{{Hameury}
  et~al.}{2000}]{Hameury2000}
{Hameury} J.-M.,  {Lasota} J.-P.,   {Warner} B.,  2000, \aap, \href
  {https://ui.adsabs.harvard.edu/abs/2000A&A...353..244H} {353, 244}

\bibitem[\protect\citeauthoryear{{Hao} \& {Zhang}}{{Hao} \&
  {Zhang}}{2009}]{Hao}
{Hao} J.~F.,  {Zhang} S.~N.,  2009, \mn@doi [\apj]
  {10.1088/0004-637X/702/2/1648}, \href
  {https://ui.adsabs.harvard.edu/abs/2009ApJ...702.1648H} {702, 1648}

\bibitem[\protect\citeauthoryear{{Heinz}}{{Heinz}}{2002}]{Heinz_2002}
{Heinz} S.,  2002, \mn@doi [\aap] {10.1051/0004-6361:20020402}, \href
  {https://ui.adsabs.harvard.edu/abs/2002A&A...388L..40H} {388, L40}

\bibitem[\protect\citeauthoryear{{Heinz} \& {Sunyaev}}{{Heinz} \&
  {Sunyaev}}{2003}]{Heinz_Sunyaev_2003}
{Heinz} S.,  {Sunyaev} R.~A.,  2003, \mn@doi [\mnras]
  {10.1046/j.1365-8711.2003.06918.x}, \href
  {https://ui.adsabs.harvard.edu/abs/2003MNRAS.343L..59H} {343, L59}

\bibitem[\protect\citeauthoryear{{Hjellming} \& {Rupen}}{{Hjellming} \&
  {Rupen}}{1995}]{GRO1655}
{Hjellming} R.~M.,  {Rupen} M.~P.,  1995, \mn@doi [\nat] {10.1038/375464a0},
  \href {https://ui.adsabs.harvard.edu/abs/1995Natur.375..464H} {375, 464}

\bibitem[\protect\citeauthoryear{{Homan} \& {Belloni}}{{Homan} \&
  {Belloni}}{2005}]{Homan_belloni}
{Homan} J.,  {Belloni} T.,  2005, \mn@doi [\apss] {10.1007/s10509-005-1197-4},
  \href {https://ui.adsabs.harvard.edu/abs/2005Ap&SS.300..107H} {300, 107}

\bibitem[\protect\citeauthoryear{{Homan}, {Wijnands}, {van der Klis},
  {Belloni}, {van Paradijs}, {Klein-Wolt}, {Fender}  \& {M{\'e}ndez}}{{Homan}
  et~al.}{2001}]{Homan_2001}
{Homan} J.,  {Wijnands} R.,  {van der Klis} M.,  {Belloni} T.,  {van Paradijs}
  J.,  {Klein-Wolt} M.,  {Fender} R.,   {M{\'e}ndez} M.,  2001, \mn@doi [\apjs]
  {10.1086/318954}, \href
  {https://ui.adsabs.harvard.edu/abs/2001ApJS..132..377H} {132, 377}

\bibitem[\protect\citeauthoryear{{Homan}, {Fridriksson}, {Jonker}, {Russell},
  {Gallo}, {Kuulkers}, {Rea}  \& {Altamirano}}{{Homan}
  et~al.}{2013}]{Homan_2013}
{Homan} J.,  {Fridriksson} J.~K.,  {Jonker} P.~G.,  {Russell} D.~M.,  {Gallo}
  E.,  {Kuulkers} E.,  {Rea} N.,   {Altamirano} D.,  2013, \mn@doi [\apj]
  {10.1088/0004-637X/775/1/9}, \href
  {https://ui.adsabs.harvard.edu/abs/2013ApJ...775....9H} {775, 9}

\bibitem[\protect\citeauthoryear{{Homan} et~al.,}{{Homan}
  et~al.}{2020}]{Homan_qpo}
{Homan} J.,  et~al., 2020, \mn@doi [\apjl] {10.3847/2041-8213/ab7932}, \href
  {https://ui.adsabs.harvard.edu/abs/2020ApJ...891L..29H} {891, L29}

\bibitem[\protect\citeauthoryear{{Ingram} \& {Motta}}{{Ingram} \&
  {Motta}}{2019}]{Ingram_2019}
{Ingram} A.~R.,  {Motta} S.~E.,  2019, \mn@doi [\nar]
  {10.1016/j.newar.2020.101524}, \href
  {https://ui.adsabs.harvard.edu/abs/2019NewAR..8501524I} {85, 101524}

\bibitem[\protect\citeauthoryear{{Jonas} \& {MeerKAT Team}}{{Jonas} \& {MeerKAT
  Team}}{2016}]{Jonas2016}
{Jonas} J.,  {MeerKAT Team} 2016, in MeerKAT Science: On the Pathway to the
  SKA. p.~1

\bibitem[\protect\citeauthoryear{{Kaaret}, {Corbel}, {Tomsick}, {Fender},
  {Miller}, {Orosz}, {Tzioumis}  \& {Wijnand s}}{{Kaaret}
  et~al.}{2003}]{Kaaret_2003}
{Kaaret} P.,  {Corbel} S.,  {Tomsick} J.~A.,  {Fender} R.,  {Miller} J.~M.,
  {Orosz} J.~A.,  {Tzioumis} A.~K.,   {Wijnand s} R.,  2003, \mn@doi [\apj]
  {10.1086/344540}, \href
  {https://ui.adsabs.harvard.edu/abs/2003ApJ...582..945K} {582, 945}

\bibitem[\protect\citeauthoryear{{Kaiser}, {Sunyaev}  \& {Spruit}}{{Kaiser}
  et~al.}{2000}]{Kaiser}
{Kaiser} C.~R.,  {Sunyaev} R.,   {Spruit} H.~C.,  2000, \aap, \href
  {https://ui.adsabs.harvard.edu/abs/2000A&A...356..975K} {356, 975}

\bibitem[\protect\citeauthoryear{{Kalemci}, {Din{\c{c}}er}, {Tomsick},
  {Buxton}, {Bailyn}  \& {Chun}}{{Kalemci} et~al.}{2013}]{Kalemci}
{Kalemci} E.,  {Din{\c{c}}er} T.,  {Tomsick} J.~A.,  {Buxton} M.~M.,  {Bailyn}
  C.~D.,   {Chun} Y.~Y.,  2013, \mn@doi [\apj] {10.1088/0004-637X/779/2/95},
  \href {https://ui.adsabs.harvard.edu/abs/2013ApJ...779...95K} {779, 95}

\bibitem[\protect\citeauthoryear{{Kennea} \& {Negoro}}{{Kennea} \&
  {Negoro}}{2019}]{Kennea_atel}
{Kennea} J.~A.,  {Negoro} H.,  2019, The Astronomer's Telegram, \href
  {https://ui.adsabs.harvard.edu/abs/2019ATel12434....1K} {12434, 1}

\bibitem[\protect\citeauthoryear{{K{\"o}rding} \& {Falcke}}{{K{\"o}rding} \&
  {Falcke}}{2005}]{Kording_2005}
{K{\"o}rding} E.,  {Falcke} H.,  2005, \mn@doi [\apss]
  {10.1007/s10509-005-1193-8}, \href
  {https://ui.adsabs.harvard.edu/abs/2005Ap&SS.300..211K} {300, 211}

\bibitem[\protect\citeauthoryear{{Lasota}}{{Lasota}}{2001}]{Lasota2001}
{Lasota} J.-P.,  2001, \mn@doi [\nar] {10.1016/S1387-6473(01)00112-9}, \href
  {https://ui.adsabs.harvard.edu/abs/2001NewAR..45..449L} {45, 449}

\bibitem[\protect\citeauthoryear{{Liu}, {Dong}, {Weng}  \& {Wu}}{{Liu}
  et~al.}{2019}]{Liu_index}
{Liu} H.,  {Dong} A.,  {Weng} S.,   {Wu} Q.,  2019, \mn@doi [\mnras]
  {10.1093/mnras/stz1622}, \href
  {https://ui.adsabs.harvard.edu/abs/2019MNRAS.487.5335L} {487, 5335}

\bibitem[\protect\citeauthoryear{Longair}{Longair}{2011}]{Longair}
Longair M.,  2011, High Energy Astrophysics.
Cambridge University Press

\bibitem[\protect\citeauthoryear{{Maccarone}}{{Maccarone}}{2002}]{Maccarone_2002}
{Maccarone} T.~J.,  2002, \mn@doi [\mnras] {10.1046/j.1365-8711.2002.05876.x},
  \href {https://ui.adsabs.harvard.edu/abs/2002MNRAS.336.1371M} {336, 1371}

\bibitem[\protect\citeauthoryear{{Maccarone}}{{Maccarone}}{2003}]{Maccarone}
{Maccarone} T.~J.,  2003, \mn@doi [\aap] {10.1051/0004-6361:20031146}, \href
  {https://ui.adsabs.harvard.edu/abs/2003A&A...409..697M} {409, 697}

\bibitem[\protect\citeauthoryear{{Maccarone}, {Osler}, {Miller-Jones}, {Atri},
  {Russell}, {Meier}, {McHardy}  \& {Longa-Pe{\~n}a}}{{Maccarone}
  et~al.}{2020}]{Maccarone_2020}
{Maccarone} T.~J.,  {Osler} A.,  {Miller-Jones} J. C.~A.,  {Atri} P.,
  {Russell} D.~M.,  {Meier} D.~L.,  {McHardy} I.~M.,   {Longa-Pe{\~n}a} P.~A.,
  2020, \mn@doi [\mnras] {10.1093/mnrasl/slaa120}, \href
  {https://ui.adsabs.harvard.edu/abs/2020MNRAS.498L..40M} {498, L40}

\bibitem[\protect\citeauthoryear{{Markoff}, {Falcke}  \& {Fender}}{{Markoff}
  et~al.}{2001}]{Markoff_2001}
{Markoff} S.,  {Falcke} H.,   {Fender} R.,  2001, \mn@doi [\aap]
  {10.1051/0004-6361:20010420}, \href
  {https://ui.adsabs.harvard.edu/abs/2001A&A...372L..25M} {372, L25}

\bibitem[\protect\citeauthoryear{{Markoff}, {Nowak}  \& {Wilms}}{{Markoff}
  et~al.}{2005}]{Markoff_corona}
{Markoff} S.,  {Nowak} M.~A.,   {Wilms} J.,  2005, \mn@doi [\apj]
  {10.1086/497628}, \href
  {https://ui.adsabs.harvard.edu/abs/2005ApJ...635.1203M} {635, 1203}

\bibitem[\protect\citeauthoryear{{Mart{\'\i}-Vidal}, {Vlemmings}, {Muller}  \&
  {Casey}}{{Mart{\'\i}-Vidal} et~al.}{2014}]{uvmultifit}
{Mart{\'\i}-Vidal} I.,  {Vlemmings} W.~H.~T.,  {Muller} S.,   {Casey} S.,
  2014, \mn@doi [\aap] {10.1051/0004-6361/201322633}, \href
  {https://ui.adsabs.harvard.edu/abs/2014A&A...563A.136M} {563, A136}

\bibitem[\protect\citeauthoryear{{Matsuoka} et~al.,}{{Matsuoka}
  et~al.}{2009}]{Matsuoka_maxi}
{Matsuoka} M.,  et~al., 2009, \mn@doi [\pasj] {10.1093/pasj/61.5.999}, \href
  {https://ui.adsabs.harvard.edu/abs/2009PASJ...61..999M} {61, 999}

\bibitem[\protect\citeauthoryear{McClintock, Remillard, Rupen, Torres, Steeghs,
  Levine  \& Orosz}{McClintock et~al.}{2009}]{McClintock_2009}
McClintock J.~E.,  Remillard R.~A.,  Rupen M.~P.,  Torres M. A.~P.,  Steeghs
  D.,  Levine A.~M.,   Orosz J.~A.,  2009, \mn@doi [The Astrophysical Journal]
  {10.1088/0004-637x/698/2/1398}, 698, 1398

\bibitem[\protect\citeauthoryear{{McMullin}, {Waters}, {Schiebel}, {Young}  \&
  {Golap}}{{McMullin} et~al.}{2007}]{CASA}
{McMullin} J.~P.,  {Waters} B.,  {Schiebel} D.,  {Young} W.,   {Golap} K.,
  2007, in {Shaw} R.~A.,  {Hill} F.,   {Bell} D.~J.,  eds,  Astronomical
  Society of the Pacific Conference Series Vol. 376, Astronomical Data Analysis
  Software and Systems XVI. p.~127

\bibitem[\protect\citeauthoryear{{M{\'e}sz{\'a}ros} \&
  {Rees}}{{M{\'e}sz{\'a}ros} \& {Rees}}{1997}]{Meszaros_1997}
{M{\'e}sz{\'a}ros} P.,  {Rees} M.~J.,  1997, \mn@doi [\apj] {10.1086/303625},
  \href {https://ui.adsabs.harvard.edu/abs/1997ApJ...476..232M} {476, 232}

\bibitem[\protect\citeauthoryear{{Migliori}, {Corbel}, {Tomsick}, {Kaaret},
  {Fender}, {Tzioumis}, {Coriat}  \& {Orosz}}{{Migliori}
  et~al.}{2017}]{Migliori2017}
{Migliori} G.,  {Corbel} S.,  {Tomsick} J.~A.,  {Kaaret} P.,  {Fender} R.~P.,
  {Tzioumis} A.~K.,  {Coriat} M.,   {Orosz} J.~A.,  2017, \mn@doi [\mnras]
  {10.1093/mnras/stx1864}, \href
  {https://ui.adsabs.harvard.edu/abs/2017MNRAS.472..141M} {472, 141}

\bibitem[\protect\citeauthoryear{{Miller-Jones}, {Fender}  \&
  {Nakar}}{{Miller-Jones} et~al.}{2006}]{Miller-jones2006}
{Miller-Jones} J.~C.~A.,  {Fender} R.~P.,   {Nakar} E.,  2006, \mn@doi [\mnras]
  {10.1111/j.1365-2966.2006.10092.x}, \href
  {https://ui.adsabs.harvard.edu/abs/2006MNRAS.367.1432M} {367, 1432}

\bibitem[\protect\citeauthoryear{{Miller-Jones}, {Jonker}, {Ratti}, {Torres},
  {Brocksopp}, {Yang}  \& {Morrell}}{{Miller-Jones}
  et~al.}{2011}]{Miller_jones_sedov}
{Miller-Jones} J.~C.~A.,  {Jonker} P.~G.,  {Ratti} E.~M.,  {Torres} M.~A.~P.,
  {Brocksopp} C.,  {Yang} J.,   {Morrell} N.~I.,  2011, \mn@doi [\mnras]
  {10.1111/j.1365-2966.2011.18704.x}, \href
  {https://ui.adsabs.harvard.edu/abs/2011MNRAS.415..306M} {415, 306}

\bibitem[\protect\citeauthoryear{{Miller-Jones} et~al.,}{{Miller-Jones}
  et~al.}{2012}]{Miller-Jones_h1743}
{Miller-Jones} J.~C.~A.,  et~al., 2012, \mn@doi [\mnras]
  {10.1111/j.1365-2966.2011.20326.x}, \href
  {https://ui.adsabs.harvard.edu/abs/2012MNRAS.421..468M} {421, 468}

\bibitem[\protect\citeauthoryear{{Miller-Jones} et~al.,}{{Miller-Jones}
  et~al.}{2019}]{Miller-Jones2019}
{Miller-Jones} J. C.~A.,  et~al., 2019, \mn@doi [\nat]
  {10.1038/s41586-019-1152-0}, \href
  {https://ui.adsabs.harvard.edu/abs/2019Natur.569..374M} {569, 374}

\bibitem[\protect\citeauthoryear{{Mioduszewski}, {Rupen}, {Hjellming}, {Pooley}
   \& {Waltman}}{{Mioduszewski} et~al.}{2001}]{Mioduszewski}
{Mioduszewski} A.~J.,  {Rupen} M.~P.,  {Hjellming} R.~M.,  {Pooley} G.~G.,
  {Waltman} E.~B.,  2001, \mn@doi [\apj] {10.1086/320965}, \href
  {https://ui.adsabs.harvard.edu/abs/2001ApJ...553..766M} {553, 766}

\bibitem[\protect\citeauthoryear{{Mirabel} \& {Rodr{\'\i}guez}}{{Mirabel} \&
  {Rodr{\'\i}guez}}{1994}]{Mirabel1994}
{Mirabel} I.~F.,  {Rodr{\'\i}guez} L.~F.,  1994, \mn@doi [\nat]
  {10.1038/371046a0}, \href
  {https://ui.adsabs.harvard.edu/abs/1994Natur.371...46M} {371, 46}

\bibitem[\protect\citeauthoryear{{Motta} \& {Fender}}{{Motta} \&
  {Fender}}{2019}]{Motta_sco}
{Motta} S.~E.,  {Fender} R.~P.,  2019, \mn@doi [\mnras]
  {10.1093/mnras/sty3331}, \href
  {https://ui.adsabs.harvard.edu/abs/2019MNRAS.483.3686M} {483, 3686}

\bibitem[\protect\citeauthoryear{{Negoro} et~al.,}{{Negoro}
  et~al.}{2019}]{Negoro_atel}
{Negoro} H.,  et~al., 2019, The Astronomer's Telegram, \href
  {https://ui.adsabs.harvard.edu/abs/2019ATel12838....1N} {12838, 1}

\bibitem[\protect\citeauthoryear{{Nesci} \& {Fiocchi}}{{Nesci} \&
  {Fiocchi}}{2019}]{Nesci_atel}
{Nesci} R.,  {Fiocchi} M.,  2019, The Astronomer's Telegram, \href
  {https://ui.adsabs.harvard.edu/abs/2019ATel12448....1N} {12448, 1}

\bibitem[\protect\citeauthoryear{{Offringa}}{{Offringa}}{2010}]{Offringa}
{Offringa} A.~R.,  2010, {AOFlagger: RFI Software} (\mn@eprint {ascl}
  {1010.017})

\bibitem[\protect\citeauthoryear{{Offringa} et~al.,}{{Offringa}
  et~al.}{2014}]{WSCLEAN}
{Offringa} A.~R.,  et~al., 2014, \mn@doi [\mnras] {10.1093/mnras/stu1368},
  \href {https://ui.adsabs.harvard.edu/abs/2014MNRAS.444..606O} {444, 606}

\bibitem[\protect\citeauthoryear{{Parikh}, {Russell}, {Wijnands},
  {Miller-Jones}, {Sivakoff}  \& {Tetarenko}}{{Parikh} et~al.}{2019}]{Parikh}
{Parikh} A.~S.,  {Russell} T.~D.,  {Wijnands} R.,  {Miller-Jones} J.~C.~A.,
  {Sivakoff} G.~R.,   {Tetarenko} A.~J.,  2019, \mn@doi [\apjl]
  {10.3847/2041-8213/ab2636}, \href
  {https://ui.adsabs.harvard.edu/abs/2019ApJ...878L..28P} {878, L28}

\bibitem[\protect\citeauthoryear{{P{\'e}ault} et~al.,}{{P{\'e}ault}
  et~al.}{2019}]{Peault_2019}
{P{\'e}ault} M.,  et~al., 2019, \mn@doi [\mnras] {10.1093/mnras/sty2796}, \href
  {https://ui.adsabs.harvard.edu/abs/2019MNRAS.482.2447P} {482, 2447}

\bibitem[\protect\citeauthoryear{{Pirbhoy}, {Baglio}, {Russell}, {Bramich},
  {Saikia}, {Yazeedi}  \& {Lewis}}{{Pirbhoy} et~al.}{2020}]{Pirbhoy_atel}
{Pirbhoy} S.~F.,  {Baglio} M.~C.,  {Russell} D.~M.,  {Bramich} D.~M.,  {Saikia}
  P.,  {Yazeedi} A.~A.,   {Lewis} F.,  2020, The Astronomer's Telegram, \href
  {https://ui.adsabs.harvard.edu/abs/2020ATel13451....1P} {13451, 1}

\bibitem[\protect\citeauthoryear{{Plotkin}, {Gallo}  \& {Jonker}}{{Plotkin}
  et~al.}{2013}]{Plotkin2013}
{Plotkin} R.~M.,  {Gallo} E.,   {Jonker} P.~G.,  2013, \mn@doi [\apj]
  {10.1088/0004-637X/773/1/59}, \href
  {https://ui.adsabs.harvard.edu/abs/2013ApJ...773...59P} {773, 59}

\bibitem[\protect\citeauthoryear{{Plotkin}, {Gallo}, {Markoff}, {Homan},
  {Jonker}, {Miller-Jones}, {Russell}  \& {Drappeau}}{{Plotkin}
  et~al.}{2015}]{Plotkin2015}
{Plotkin} R.~M.,  {Gallo} E.,  {Markoff} S.,  {Homan} J.,  {Jonker} P.~G.,
  {Miller-Jones} J. C.~A.,  {Russell} D.~M.,   {Drappeau} S.,  2015, \mn@doi
  [\mnras] {10.1093/mnras/stu2385}, \href
  {https://ui.adsabs.harvard.edu/abs/2015MNRAS.446.4098P} {446, 4098}

\bibitem[\protect\citeauthoryear{{Plotkin} et~al.,}{{Plotkin}
  et~al.}{2017}]{Plotkin_2017_v404}
{Plotkin} R.~M.,  et~al., 2017, \mn@doi [\apj] {10.3847/1538-4357/834/2/104},
  \href {https://ui.adsabs.harvard.edu/abs/2017ApJ...834..104P} {834, 104}

\bibitem[\protect\citeauthoryear{{Poutanen}, {Veledina}  \&
  {Revnivtsev}}{{Poutanen} et~al.}{2014}]{Poutanen_2014}
{Poutanen} J.,  {Veledina} A.,   {Revnivtsev} M.~G.,  2014, \mn@doi [\mnras]
  {10.1093/mnras/stu1989}, \href
  {https://ui.adsabs.harvard.edu/abs/2014MNRAS.445.3987P} {445, 3987}

\bibitem[\protect\citeauthoryear{{Rees}}{{Rees}}{1966}]{Rees_1966}
{Rees} M.~J.,  1966, \mn@doi [\nat] {10.1038/211468a0}, \href
  {https://ui.adsabs.harvard.edu/abs/1966Natur.211..468R} {211, 468}

\bibitem[\protect\citeauthoryear{{Rees} \& {Meszaros}}{{Rees} \&
  {Meszaros}}{1992}]{Rees_1992}
{Rees} M.~J.,  {Meszaros} P.,  1992, \mn@doi [\mnras]
  {10.1093/mnras/258.1.41P}, \href
  {https://ui.adsabs.harvard.edu/abs/1992MNRAS.258P..41R} {258, 41}

\bibitem[\protect\citeauthoryear{{Remillard} \& {McClintock}}{{Remillard} \&
  {McClintock}}{2006}]{Remillard_xrb}
{Remillard} R.~A.,  {McClintock} J.~E.,  2006, \mn@doi [\araa]
  {10.1146/annurev.astro.44.051905.092532}, \href
  {https://ui.adsabs.harvard.edu/abs/2006ARA&A..44...49R} {44, 49}

\bibitem[\protect\citeauthoryear{{Rib{\'o}}, {Dhawan}  \& {Mirabel}}{{Rib{\'o}}
  et~al.}{2004}]{Ribo}
{Rib{\'o}} M.,  {Dhawan} V.,   {Mirabel} I.~F.,  2004, in European VLBI Network
  on New Developments in VLBI Science and Technology. pp 111--112 (\mn@eprint
  {arXiv} {astro-ph/0412657})

\bibitem[\protect\citeauthoryear{Rodriguez, Corbel  \& Tomsick}{Rodriguez
  et~al.}{2003}]{Rodriguez_2003}
Rodriguez J.,  Corbel S.,   Tomsick J.~A.,  2003, \mn@doi [The Astrophysical
  Journal] {10.1086/377478}, 595, 1032

\bibitem[\protect\citeauthoryear{{Rushton} et~al.,}{{Rushton}
  et~al.}{2017}]{Rushton}
{Rushton} A.~P.,  et~al., 2017, \mn@doi [\mnras] {10.1093/mnras/stx526}, \href
  {https://ui.adsabs.harvard.edu/abs/2017MNRAS.468.2788R} {468, 2788}

\bibitem[\protect\citeauthoryear{{Russell} et~al.,}{{Russell}
  et~al.}{2013a}]{Russell2013}
{Russell} D.~M.,  et~al., 2013a, \mn@doi [\mnras] {10.1093/mnras/sts377}, \href
  {https://ui.adsabs.harvard.edu/abs/2013MNRAS.429..815R} {429, 815}

\bibitem[\protect\citeauthoryear{{Russell} et~al.,}{{Russell}
  et~al.}{2013b}]{Russell_2013b}
{Russell} D.~M.,  et~al., 2013b, \mn@doi [\apjl] {10.1088/2041-8205/768/2/L35},
  \href {https://ui.adsabs.harvard.edu/abs/2013ApJ...768L..35R} {768, L35}

\bibitem[\protect\citeauthoryear{{Russell}, {Soria}, {Miller-Jones}, {Curran},
  {Markoff}, {Russell}  \& {Sivakoff}}{{Russell} et~al.}{2014}]{Russell_2014}
{Russell} T.~D.,  {Soria} R.,  {Miller-Jones} J.~C.~A.,  {Curran} P.~A.,
  {Markoff} S.,  {Russell} D.~M.,   {Sivakoff} G.~R.,  2014, \mn@doi [\mnras]
  {10.1093/mnras/stt2498}, \href
  {https://ui.adsabs.harvard.edu/abs/2014MNRAS.439.1390R} {439, 1390}

\bibitem[\protect\citeauthoryear{{Russell} et~al.,}{{Russell}
  et~al.}{2015}]{Russell_2015}
{Russell} T.~D.,  et~al., 2015, \mn@doi [\mnras] {10.1093/mnras/stv723}, \href
  {https://ui.adsabs.harvard.edu/abs/2015MNRAS.450.1745R} {450, 1745}

\bibitem[\protect\citeauthoryear{{Russell} et~al.,}{{Russell}
  et~al.}{2019a}]{Russell_1535}
{Russell} T.~D.,  et~al., 2019a, \mn@doi [\apj] {10.3847/1538-4357/ab3d36},
  \href {https://ui.adsabs.harvard.edu/abs/2019ApJ...883..198R} {883, 198}

\bibitem[\protect\citeauthoryear{{Russell}, {Baglio}  \& {Lewis}}{{Russell}
  et~al.}{2019b}]{DRussell_atel_opt_1}
{Russell} D.~M.,  {Baglio} C.~M.,   {Lewis} F.,  2019b, The Astronomer's
  Telegram, \href {https://ui.adsabs.harvard.edu/abs/2019ATel12439....1R}
  {12439, 1}

\bibitem[\protect\citeauthoryear{{Russell}, {Anderson}, {Miller-Jones},
  {Degenaar}, {Eijnden}, {Sivakoff}  \& {Tetarenko}}{{Russell}
  et~al.}{2019c}]{TRussell_atel}
{Russell} T.,  {Anderson} G.,  {Miller-Jones} J.,  {Degenaar} N.,  {Eijnden} J.
  v.~d.,  {Sivakoff} G.~R.,   {Tetarenko} A.,  2019c, The Astronomer's
  Telegram, \href {https://ui.adsabs.harvard.edu/abs/2019ATel12456....1R}
  {12456, 1}

\bibitem[\protect\citeauthoryear{{Russell}, {Al Yazeedi}, {Bramich}, {Baglio}
  \& {Lewis}}{{Russell} et~al.}{2019d}]{DRussell_atel_opt_2}
{Russell} D.~M.,  {Al Yazeedi} A.,  {Bramich} D.~M.,  {Baglio} M.~C.,   {Lewis}
  F.,  2019d, The Astronomer's Telegram, \href
  {https://ui.adsabs.harvard.edu/abs/2019ATel12829....1R} {12829, 1}

\bibitem[\protect\citeauthoryear{{Russell} et~al.,}{{Russell}
  et~al.}{2020a}]{Russell_2020_break_frequency}
{Russell} T.~D.,  et~al., 2020a, \mn@doi [\mnras] {10.1093/mnras/staa2650},
  \href {https://ui.adsabs.harvard.edu/abs/2020MNRAS.498.5772R} {498, 5772}

\bibitem[\protect\citeauthoryear{Russell et~al.,}{Russell
  et~al.}{2020b}]{Russell2020}
Russell T.~D.,  et~al., 2020b, \mn@doi [\mnras] {10.1093/mnras/staa2650}, 498,
  5772

\bibitem[\protect\citeauthoryear{{Saikia}, {Russell}, {Bramich},
  {Miller-Jones}, {Baglio}  \& {Degenaar}}{{Saikia} et~al.}{2019}]{Saikia}
{Saikia} P.,  {Russell} D.~M.,  {Bramich} D.~M.,  {Miller-Jones} J. C.~A.,
  {Baglio} M.~C.,   {Degenaar} N.,  2019, \mn@doi [\apj]
  {10.3847/1538-4357/ab4a09}, \href
  {https://ui.adsabs.harvard.edu/abs/2019ApJ...887...21S} {887, 21}

\bibitem[\protect\citeauthoryear{{Sanna} et~al.,}{{Sanna}
  et~al.}{2019}]{Nicer_atel}
{Sanna} A.,  et~al., 2019, The Astronomer's Telegram, \href
  {https://ui.adsabs.harvard.edu/abs/2019ATel12447....1S} {12447, 1}

\bibitem[\protect\citeauthoryear{{Shimomukai} et~al.,}{{Shimomukai}
  et~al.}{2020}]{Shimomukai_atel}
{Shimomukai} R.,  et~al., 2020, The Astronomer's Telegram, \href
  {https://ui.adsabs.harvard.edu/abs/2020ATel13459....1S} {13459, 1}

\bibitem[\protect\citeauthoryear{{Sobczak}, {McClintock}, {Remillard}, {Cui},
  {Levine}, {Morgan}, {Orosz}  \& {Bailyn}}{{Sobczak}
  et~al.}{2000}]{Sobczak_2000}
{Sobczak} G.~J.,  {McClintock} J.~E.,  {Remillard} R.~A.,  {Cui} W.,  {Levine}
  A.~M.,  {Morgan} E.~H.,  {Orosz} J.~A.,   {Bailyn} C.~D.,  2000, \mn@doi
  [\apj] {10.1086/317229}, \href
  {https://ui.adsabs.harvard.edu/abs/2000ApJ...544..993S} {544, 993}

\bibitem[\protect\citeauthoryear{Soleri, Belloni  \& Casella}{Soleri
  et~al.}{2008}]{Soleri2008}
Soleri P.,  Belloni T.,   Casella P.,  2008, \mn@doi [\mnras]
  {10.1111/j.1365-2966.2007.12596.x}, 383, 1089

\bibitem[\protect\citeauthoryear{{Steiner} \& {McClintock}}{{Steiner} \&
  {McClintock}}{2012}]{Steiner_xte}
{Steiner} J.~F.,  {McClintock} J.~E.,  2012, \mn@doi [\apj]
  {10.1088/0004-637X/745/2/136}, \href
  {https://ui.adsabs.harvard.edu/abs/2012ApJ...745..136S} {745, 136}

\bibitem[\protect\citeauthoryear{{Steiner}, {McClintock}, {Remillard}, {Gou},
  {Yamada}  \& {Narayan}}{{Steiner} et~al.}{2010}]{Steiner_2010}
{Steiner} J.~F.,  {McClintock} J.~E.,  {Remillard} R.~A.,  {Gou} L.,  {Yamada}
  S.,   {Narayan} R.,  2010, \mn@doi [\apjl] {10.1088/2041-8205/718/2/L117},
  \href {https://ui.adsabs.harvard.edu/abs/2010ApJ...718L.117S} {718, L117}

\bibitem[\protect\citeauthoryear{{Steiner}, {McClintock}  \& {Reid}}{{Steiner}
  et~al.}{2012}]{Steiner_h17}
{Steiner} J.~F.,  {McClintock} J.~E.,   {Reid} M.~J.,  2012, \mn@doi [\apjl]
  {10.1088/2041-8205/745/1/L7}, \href
  {https://ui.adsabs.harvard.edu/abs/2012ApJ...745L...7S} {745, L7}

\bibitem[\protect\citeauthoryear{{Tasse} et~al.,}{{Tasse} et~al.}{2018}]{Tasse}
{Tasse} C.,  et~al., 2018, \mn@doi [\aap] {10.1051/0004-6361/201731474}, \href
  {https://ui.adsabs.harvard.edu/abs/2018A&A...611A..87T} {611, A87}

\bibitem[\protect\citeauthoryear{{Tetarenko}, {Sivakoff}, {Heinke}  \&
  {Gladstone}}{{Tetarenko} et~al.}{2016}]{Watchdog}
{Tetarenko} B.~E.,  {Sivakoff} G.~R.,  {Heinke} C.~O.,   {Gladstone} J.~C.,
  2016, \mn@doi [\apjs] {10.3847/0067-0049/222/2/15}, \href
  {https://ui.adsabs.harvard.edu/abs/2016ApJS..222...15T} {222, 15}

\bibitem[\protect\citeauthoryear{{Tetarenko} et~al.,}{{Tetarenko}
  et~al.}{2017}]{Tetarenko2017}
{Tetarenko} A.~J.,  et~al., 2017, \mn@doi [\mnras] {10.1093/mnras/stx1048},
  \href {https://ui.adsabs.harvard.edu/abs/2017MNRAS.469.3141T} {469, 3141}

\bibitem[\protect\citeauthoryear{{Tetarenko}, {Dubus}, {Lasota}, {Heinke}  \&
  {Sivakoff}}{{Tetarenko} et~al.}{2018}]{Tetarenko_B_2018}
{Tetarenko} B.~E.,  {Dubus} G.,  {Lasota} J.~P.,  {Heinke} C.~O.,   {Sivakoff}
  G.~R.,  2018, \mn@doi [\mnras] {10.1093/mnras/sty1798}, \href
  {https://ui.adsabs.harvard.edu/abs/2018MNRAS.480....2T} {480, 2}

\bibitem[\protect\citeauthoryear{{Tetarenko}, {Casella}, {Miller-Jones},
  {Sivakoff}, {Tetarenko}, {Maccarone}, {Gand hi}  \& {Eikenberry}}{{Tetarenko}
  et~al.}{2019}]{Tetarenko_2019}
{Tetarenko} A.~J.,  {Casella} P.,  {Miller-Jones} J.~C.~A.,  {Sivakoff} G.~R.,
  {Tetarenko} B.~E.,  {Maccarone} T.~J.,  {Gand hi} P.,   {Eikenberry} S.,
  2019, \mn@doi [\mnras] {10.1093/mnras/stz165}, \href
  {https://ui.adsabs.harvard.edu/abs/2019MNRAS.484.2987T} {484, 2987}

\bibitem[\protect\citeauthoryear{{Tetarenko}, {Dubus}, {Marcel}, {Done}  \&
  {Clavel}}{{Tetarenko} et~al.}{2020}]{Tetarenko_B_2020}
{Tetarenko} B.~E.,  {Dubus} G.,  {Marcel} G.,  {Done} C.,   {Clavel} M.,  2020,
  \mn@doi [\mnras] {10.1093/mnras/staa1367}, \href
  {https://ui.adsabs.harvard.edu/abs/2020MNRAS.495.3666T} {495, 3666}

\bibitem[\protect\citeauthoryear{{Tominaga} et~al.,}{{Tominaga}
  et~al.}{2020}]{Tominaga_1348}
{Tominaga} M.,  et~al., 2020, \mn@doi [\apjl] {10.3847/2041-8213/abaaaa}, \href
  {https://ui.adsabs.harvard.edu/abs/2020ApJ...899L..20T} {899, L20}

\bibitem[\protect\citeauthoryear{Tomsick, Corbel, Fender, Miller, Orosz,
  Tzioumis, Wijnands  \& Kaaret}{Tomsick et~al.}{2003}]{Tomsick_2003}
Tomsick J.~A.,  Corbel S.,  Fender R.,  Miller J.~M.,  Orosz J.~A.,  Tzioumis
  T.,  Wijnands R.,   Kaaret P.,  2003, \mn@doi [The Astrophysical Journal]
  {10.1086/344703}, 582, 933

\bibitem[\protect\citeauthoryear{{Tomsick}, {Kalemci}  \& {Kaaret}}{{Tomsick}
  et~al.}{2004}]{Tomsick_2004}
{Tomsick} J.~A.,  {Kalemci} E.,   {Kaaret} P.,  2004, \mn@doi [\apj]
  {10.1086/380484}, \href
  {https://ui.adsabs.harvard.edu/abs/2004ApJ...601..439T} {601, 439}

\bibitem[\protect\citeauthoryear{{Tomsick} et~al.,}{{Tomsick}
  et~al.}{2008}]{Tomsick_2008}
{Tomsick} J.~A.,  et~al., 2008, \mn@doi [\apj] {10.1086/587797}, \href
  {https://ui.adsabs.harvard.edu/abs/2008ApJ...680..593T} {680, 593}

\bibitem[\protect\citeauthoryear{{Vadawale}, {Rao}, {Naik}, {Yadav},
  {Ishwara-Chandra}, {Pramesh Rao}  \& {Pooley}}{{Vadawale}
  et~al.}{2003}]{Vadawale_2003}
{Vadawale} S.~V.,  {Rao} A.~R.,  {Naik} S.,  {Yadav} J.~S.,  {Ishwara-Chandra}
  C.~H.,  {Pramesh Rao} A.,   {Pooley} G.~G.,  2003, \mn@doi [\apj]
  {10.1086/378672}, \href
  {https://ui.adsabs.harvard.edu/abs/2003ApJ...597.1023V} {597, 1023}

\bibitem[\protect\citeauthoryear{Vahdat~Motlagh, Kalemci  \&
  Maccarone}{Vahdat~Motlagh et~al.}{2019}]{Vahdat}
Vahdat~Motlagh A.,  Kalemci E.,   Maccarone T.~J.,  2019, \mn@doi [\mnras]
  {10.1093/mnras/stz569}, 485, 2744

\bibitem[\protect\citeauthoryear{{Verner}, {Ferland}, {Korista}  \&
  {Yakovlev}}{{Verner} et~al.}{1996}]{Verner1996}
{Verner} D.~A.,  {Ferland} G.~J.,  {Korista} K.~T.,   {Yakovlev} D.~G.,  1996,
  \mn@doi [\apj] {10.1086/177435}, \href
  {https://ui.adsabs.harvard.edu/abs/1996ApJ...465..487V} {465, 487}

\bibitem[\protect\citeauthoryear{{Wang}, {Dai}  \& {Lu}}{{Wang}
  et~al.}{2003}]{Wang_model}
{Wang} X.~Y.,  {Dai} Z.~G.,   {Lu} T.,  2003, \mn@doi [\apj] {10.1086/375638},
  \href {https://ui.adsabs.harvard.edu/abs/2003ApJ...592..347W} {592, 347}

\bibitem[\protect\citeauthoryear{{Wilms}, {Allen}  \& {McCray}}{{Wilms}
  et~al.}{2000}]{Wilms}
{Wilms} J.,  {Allen} A.,   {McCray} R.,  2000, \mn@doi [\apj] {10.1086/317016},
  \href {https://ui.adsabs.harvard.edu/abs/2000ApJ...542..914W} {542, 914}

\bibitem[\protect\citeauthoryear{Yang, Brocksopp, Corbel, Paragi, Tzioumis  \&
  Fender}{Yang et~al.}{2010}]{Yang2010}
Yang J.,  Brocksopp C.,  Corbel S.,  Paragi Z.,  Tzioumis T.,   Fender R.~P.,
  2010, \mn@doi [\mnras] {10.1111/j.1745-3933.2010.00948.x}, 409, L64

\bibitem[\protect\citeauthoryear{{Yatabe} et~al.,}{{Yatabe}
  et~al.}{2019}]{Yatabe2019}
{Yatabe} F.,  et~al., 2019, The Astronomer's Telegram, \href
  {https://ui.adsabs.harvard.edu/abs/2019ATel12425....1Y} {12425, 1}

\bibitem[\protect\citeauthoryear{Yuan \& Cui}{Yuan \&
  Cui}{2005}]{yuan_cui_2005}
Yuan F.,  Cui W.,  2005, \mn@doi [The Astrophysical Journal] {10.1086/431453},
  629, 408

\bibitem[\protect\citeauthoryear{{Zdziarski} \& {Gierli{\'n}ski}}{{Zdziarski}
  \& {Gierli{\'n}ski}}{2004}]{Zdziarski_corona}
{Zdziarski} A.~A.,  {Gierli{\'n}ski} M.,  2004, \mn@doi [Progress of
  Theoretical Physics Supplement] {10.1143/PTPS.155.99}, \href
  {https://ui.adsabs.harvard.edu/abs/2004PThPS.155...99Z} {155, 99}

\bibitem[\protect\citeauthoryear{{Zhang} et~al.,}{{Zhang}
  et~al.}{2020a}]{Zhang2020}
{Zhang} L.,  et~al., 2020a, \mn@doi [\mnras] {10.1093/mnras/staa2842}, \href
  {https://ui.adsabs.harvard.edu/abs/2020MNRAS.499..851Z} {499, 851}

\bibitem[\protect\citeauthoryear{{Zhang} et~al.,}{{Zhang}
  et~al.}{2020b}]{Zhang_atel}
{Zhang} L.,  et~al., 2020b, The Astronomer's Telegram, \href
  {https://ui.adsabs.harvard.edu/abs/2020ATel13465....1Z} {13465, 1}

\bibitem[\protect\citeauthoryear{{van der Laan}}{{van der
  Laan}}{1966}]{vanderlaan}
{van der Laan} H.,  1966, \mn@doi [\nat] {10.1038/2111131a0}, \href
  {https://ui.adsabs.harvard.edu/abs/1966Natur.211.1131V} {211, 1131}

\makeatother
\end{thebibliography}


\appendix
\onecolumn

\section{Radio data}

\begin{ThreePartTable}

\begin{TableNotes}
  \item[a] {\footnotesize Combination of the MJD 58830 and 58831 ATCA epochs to obtain a higher significance detection of RK1 at both 5.5 and 9 GHz.}
\end{TableNotes}

\LTcapwidth=\textwidth
\begin{center}
\setlength{\extrarowheight}{.1em}
\setlength{\tabcolsep}{3.5pt}
\begin{longtable}{*{8}{c}}
\caption{Radio flux densities $S_{\nu}$ of \maxithirt{} and of the first radio knot RK1. Observation MJDs represent the middle of the observation, where errors represent the observation duration (time on source). For the ATCA observations we report the array configuration as a subscript. The spectral index $\alpha$ of each component ($S_{\nu} \propto \nu^{\alpha}$) is computed only for simultaneous ATCA multi frequency observations. Upper limits are reported at the 3$\sigma$ level, and in the case of RK1 we only report upper limits for epochs after the first detection.}\\
\hhline{========}
Calendar date & MJD & Telescope & Central frequency & $S_{\nu, \tu{core}}$ & $\alpha_{\tu{core}}$ & $S_{\nu, \tu{RK1}}$ & $\alpha_{\tu{RK1}}$\\
$[$UT$]$ & & & [GHz] & [mJy] & & [mJy] & \\
\hline
\endfirsthead

\caption{Continued from previous page. Radio flux densities $S_{\nu}$ of \maxithirt{} and of the first radio knot RK1.}\\
\hhline{========}
Calendar date & MJD & Telescope & Central frequency & $S_{\nu, \tu{core}}$ & $\alpha_{\tu{core}}$ & $S_{\nu, \tu{RK1}}$ & $\alpha_{\tu{RK1}}$\\
$[$UT$]$ & & & [GHz] & [mJy] & & [mJy] & \\
\hline
\endhead

\hline \multicolumn{8}{r}{{Continued on next page}}\\
\endfoot

\insertTableNotes
\endlastfoot
2019-01-26 & 58509.93 $\pm$ 0.09   & ATCA$_{\tu{H}75}$  & 5.5  & 3.44  $\pm$ 0.11 & 0.02 $\pm$ 0.09   & &  \\
           &                       &                    & 9    & 3.48  $\pm$ 0.12 &                 & &  \\      
2019-01-28 & 58511.00 $\pm$ 0.13   & ATCA$_{\tu{H}75}$  & 5.5  & 6.19  $\pm$ 0.42 & 0.1 $\pm$ 0.2   & &  \\
           &              		   &  				    & 9    & 6.49  $\pm$ 0.47 &  			      & &  \\
2019-01-29 & 58512.029 $\pm$ 0.007 & MeerKAT            & 1.3  & 13.70 $\pm$ 0.05 &                 & &  \\
2019-01-31 & 58514.01 $\pm$ 0.09   & ATCA$_{\tu{H}75}$  & 5.5  & 21.9  $\pm$ 0.8  & 0.18 $\pm$ 0.02 & &  \\
           &                       &                    & 9    & 24.0  $\pm$ 0.5  &                 & & \\                                                                        
           &                       &                    & 16.7 & 27.76 $\pm$ 0.20 &                  & & \\
           &                       &                    & 21.2 & 28.49 $\pm$ 0.22 &                  & & \\
2019-02-01 & 58515.161 $\pm$ 0.005 & MeerKAT 		    & 1.3  & 28.57 $\pm$ 0.12 &                 & &\\
2019-02-01 & 58515.90 $\pm$ 0.21   & ATCA$_{\tu{H}75}$  & 8    & 34.4  $\pm$ 1.4  &  & &  \\
2019-02-05 & 58519.860 $\pm$ 0.003 & ATCA$_{\tu{H}75}$  & 5.5  & 135.3 $\pm$ 0.9  & 0.155 $\pm$ 0.003  & &\\
           &   					   &                    & 9    & 155.4 $\pm$ 0.4  &  & & \\                                      
           & 58519.870 $\pm$ 0.003 &                    & 16.7 & 165.2 $\pm$ 0.4  &  & &\\
           &  					   &                    & 21.2 & 176.3 $\pm$ 0.4  &  & &\\
2019-02-07 & 58521.97 $\pm$ 0.09   & ATCA$_{\tu{H}75}$  & 5.5  & 52.27 $\pm$ 0.06 & $-$0.004 $\pm$ 0.003  & &\\
           &                       &                    & 9    & 52.13 $\pm$ 0.08 &  & &\\
           & 58521.99 $\pm$ 0.07   &                    & 16.7 & 52.8  $\pm$ 0.6  &  & &\\
           &  					   &                    & 21.2 & 51.68 $\pm$ 0.6  &  & &\\
2019-02-09 & 58523.219 $\pm$ 0.005 & MeerKAT            & 1.3  & 485.6 $\pm$ 1.6  &  & &\\
2019-02-09 & 58523.93 $\pm$ 0.06   & ATCA$_{\tu{H}75}$  & 5.5  & 129.8 $\pm$ 1.5  & $-$0.508 $\pm$ 0.009  & &\\
           & 					   &                    & 9    & 105   $\pm$ 1    &    & &\\                                                                                                                     
           & 58523.94 $\pm$ 0.03   &                    & 16.7 & 74.5  $\pm$ 0.4  &  & &\\                                            
           & 					   &                    & 21.2 & 66.7  $\pm$ 0.5  &  & & \\
2019-02-11 & 58525.861 $\pm$ 0.003 & ATCA$_{\tu{H}75}$  & 5.5  & 223.0 $\pm$ 0.7  & $-$0.377 $\pm$ 0.003  & &\\
		   & 					   &                    & 9    & 191.5 $\pm$ 0.5  &  & &\\
		   & 58525.871 $\pm$ 0.003 &                    & 16.7 & 146.3 $\pm$ 0.3  &  & &\\
		   & 					   &                    & 21.2 & 140.0 $\pm$ 0.5  &  & &\\
2019-02-13 & 58527.9 $\pm$ 0.1     & ATCA$_{\tu{H}75}$  & 5.5  & 60.0  $\pm$ 0.5  & $-$0.46 $\pm$ 0.01  & &\\
           & 				       &                    & 9    & 45.2  $\pm$ 1.4  &  & & \\             
           & 58527.96 $\pm$ 0.07   &                    & 16.7 & 36.7  $\pm$ 0.4  &  & &\\
           &  					   &                    & 21.2 & 31.5  $\pm$ 0.5  &  & &\\
2019-02-15 & 58529.85 $\pm$ 0.02   & ATCA$_{\tu{H}75}$  & 5.5  & 6.27  $\pm$ 0.04 & $-$0.46 $\pm$ 0.04  & &\\
           &   				       &                    & 9    & 4.54  $\pm$ 0.03 &  & &\\
2019-02-16 & 58530.114 $\pm$ 0.005 & MeerKAT            & 1.3  & 16.0  $\pm$ 0.2  & & &\\                        
2019-02-17 & 58531.85 $\pm$ 0.02   & ATCA$_{\tu{H}75}$  & 5.5  & 5.17  $\pm$ 0.03 & $-$0.29 $\pm$ 0.03  & &\\
           &                       &                    & 9    & 3.61  $\pm$ 0.03 & & & \\
2019-02-23 & 58537.092 $\pm$ 0.005 & MeerKAT            & 1.3  & 2.06  $\pm$ 0.04 & & &\\
2019-03-01 & 58543.076 $\pm$ 0.005 & MeerKAT            & 1.3  & 1.51  $\pm$ 0.04 & & & \\
2019-03-09 & 58551.100 $\pm$ 0.005 & MeerKAT            & 1.3  & $<$0.12            & &    2.72 $\pm$ 0.04 &\\
2019-03-18 & 58560.075 $\pm$ 0.005 & MeerKAT            & 1.3  & $<$0.12            & &    6.75 $\pm$ 0.06   &\\
2019-03-25 & 58567.074 $\pm$ 0.005 & MeerKAT            & 1.3  & $<$0.12            & &    2.52 $\pm$ 0.04  &\\ 
2019-03-31 & 58573.76  $\pm$ 0.11  & ATCA$_{6\tu{A}}$   & 5.5  & 8.48  $\pm$ 0.12 & $-$0.37 $\pm$ 0.04 & 0.65 $\pm$ 0.09 & $-$1.1 $\pm$ 0.3\\
           &                       &                    & 9    & 7.06  $\pm$ 0.11 &  				  & 0.38 $\pm$ 0.03 &\\                
2019-04-01 & 58574.061 $\pm$ 0.005 & MeerKAT 		    & 1.3  & 8.09  $\pm$ 0.07 &                  & 1.78 $\pm$ 0.07 &\\   
2019-04-08 & 58581.72  $\pm$ 0.18  & ATCA$_{\tu{H}75}$  & 5.5  & 9.36  $\pm$ 0.13 & $-$0.37 $\pm$ 0.03 & 1.12 $\pm$ 0.01 & $-$1.27 $\pm$ 0.03\\
           &                       &      			    & 9    & 7.78  $\pm$ 0.08 &   				  & 0.64 $\pm$ 0.05 & \\
2019-04-09 & 58582.053 $\pm$ 0.005 & MeerKAT            & 1.3  & 9.35  $\pm$ 0.09 &                  & 3.50 $\pm$ 0.09 & \\ 
2019-04-15 & 58588.053 $\pm$ 0.005 & MeerKAT            & 1.3  & <0.14 			  &                  & 4.48 $\pm$ 0.06 &\\    
2019-04-16 & 58589.81  $\pm$ 0.11  & ATCA$_{750\tu{C}}$ & 5.5  & <0.12            &                           & 1.30 $\pm$ 0.01 & $-$0.77 $\pm$ 0.05\\     
           &                       &                    & 9    & <0.045           &                           & 0.89 $\pm$ 0.02 &\\         
2019-04-20 & 58593.074 $\pm$ 0.005 & MeerKAT            & 1.3  & <0.12            &                  & 2.80 $\pm$ 0.07 &\\
2019-04-29 & 58602.144 $\pm$ 0.005 & MeerKAT            & 1.3  & 4.73  $\pm$ 0.04 &                  & 1.32 $\pm$ 0.05 &\\
2019-04-30 & 58603.31  $\pm$ 0.08  & ATCA$_{750\tu{C}}$ & 5.5  & 8.68  $\pm$ 0.14 & 0.09 $\pm$ 0.05  & 0.60 $\pm$ 0.05 & $-$0.9 $\pm$ 0.3\\
           &                       &                    & 9    & 9.09  $\pm$ 0.18 &                  & 0.40 $\pm$ 0.06 & \\
2019-05-04 & 58607.908 $\pm$ 0.005 & MeerKAT 			& 1.3  & 3.43  $\pm$ 0.04 & 				   & 0.92 $\pm$ 0.04 &\\
2019-05-05 & 58608.68  $\pm$ 0.21  & ATCA$_{1.5\tu{B}}$ & 8.4  & 2.75  $\pm$ 0.05 &                  & <0.18 &\\
2019-05-09 & 58612.44  $\pm$ 0.04  & ATCA$_{1.5\tu{B}}$ & 5.5  & 0.77  $\pm$ 0.02 & 0.21 $\pm$ 0.02  & <0.08 &\\
           &                       &                    & 9    & 0.75  $\pm$ 0.11 &                  & <0.08 &\\
           &                       &                    & 17   & 0.95  $\pm$ 0.02 &                  & <0.12 &\\
           &                       &                    & 19   & 0.96  $\pm$ 0.03 &                  & <0.18 &\\           
2019-05-11 & 58614.909 $\pm$ 0.005 & MeerKAT			& 1.3  & 0.32  $\pm$ 0.04 &                  & 0.40 $\pm$ 0.04 &\\
2019-05-14 & 58617.47  $\pm$ 0.06  & ATCA$_{1.5\tu{B}}$ & 5.5  & 0.71  $\pm$ 0.02 & 0.27 $\pm$ 0.03  & <0.08 &\\
           &                       &                    & 9    & 0.88  $\pm$ 0.11 &                  & <0.08 &               \\
           &                       &                    & 17   & 0.96  $\pm$ 0.04 &                  & <0.18 &            \\
           &                       &                    & 19   & 1.03  $\pm$ 0.03 &                  & <0.18 &            \\
2019-05-18 & 58621.888 $\pm$ 0.005 & MeerKAT            & 1.3  & 0.27  $\pm$ 0.04 & 				   & 0.22 $\pm$ 0.04 &\\
2019-05-25 & 58628.019 $\pm$ 0.005 & MeerKAT            & 1.3  & 0.13  $\pm$ 0.04 & 				   & <0.12           &\\
2019-05-26 & 58629.340 $\pm$ 0.045 & ATCA$_{6\tu{A}}$   & 5.5  & 0.153 $\pm$ 0.011 & 0.0 $\pm$ 0.2  & <0.03 & \\
           &                       &       			    & 9    & 0.150 $\pm$ 0.010 &                & <0.03 &             \\
2019-05-31 & 58634.896 $\pm$ 0.005 & MeerKAT            & 1.3  & 0.79  $\pm$ 0.06 &  		           & <0.12 &\\
2019-06-08 & 58642.878 $\pm$ 0.005 & MeerKAT            & 1.3  & 3.99  $\pm$ 0.05 & 				   & <0.15 &\\
2019-06-09 & 58643.46  $\pm$ 0.10  & ATCA$_{6\tu{A}}$   & 5.5  & 4.09  $\pm$ 0.02 & $-$0.03 $\pm$ 0.02 & <0.03 &\\    
           &                       &     			    & 9    & 4.02  $\pm$ 0.03 &        		   & <0.03 &        \\
2019-06-16 & 58650.881 $\pm$ 0.005 & MeerKAT 			& 1.3  & 6.25  $\pm$ 0.09 &                  & <0.15 &\\
2019-06-24 & 58658.797 $\pm$ 0.005 & MeerKAT 			& 1.3  & 4.60  $\pm$ 0.04 &                  & <0.15 &\\
2019-06-26 & 58660.27  $\pm$ 0.10  & ATCA$_{6\tu{A}}$   & 5.5  & 4.18  $\pm$ 0.04 & 0.05 $\pm$ 0.03  & <0.05 &\\
           &                       &    			    & 9    & 4.28  $\pm$ 0.05 &                  & <0.05 &         \\
2019-06-30 & 58664.724 $\pm$ 0.005 & MeerKAT            & 1.3  & 4.95  $\pm$ 0.05 &                  & <0.15 &\\
2019-07-07 & 58671.861 $\pm$ 0.005 & MeerKAT            & 1.3  & 2.29  $\pm$ 0.03 &                  & <0.15 &\\
2019-07-14 & 58678.912 $\pm$ 0.005 & MeerKAT            & 1.3  & 1.66  $\pm$ 0.03 &                  & <0.15 &\\
2019-07-22 & 58686.860 $\pm$ 0.005 & MeerKAT            & 1.3  & 1.11  $\pm$ 0.03 &                  & <0.15 &\\
2019-07-25 & 58689.44   $\pm$ 0.22 & ATCA$_{750\tu{C}}$ & 5.5  & 1.19  $\pm$ 0.01 & $-$0.02 $\pm$ 0.03 & <0.03 &\\
           &                       &     				& 9    & 1.18  $\pm$ 0.02 &                  & <0.03 &     \\
2019-07-27 & 58691.750 $\pm$ 0.005 & MeerKAT            & 1.3  & 0.70  $\pm$ 0.03 &                  & <0.15 &\\
2019-08-04 & 58699.713 $\pm$ 0.005 & MeerKAT            & 1.3  & 0.44  $\pm$ 0.04 &                  & <0.12 &\\
2019-08-10 & 58705.765 $\pm$ 0.005 & MeerKAT            & 1.3  & 0.14  $\pm$ 0.04 &                  & <0.12 &\\
2019-08-16 & 58711.867 $\pm$ 0.005 & MeerKAT            & 1.3  & <0.14           &                  & <0.14 &\\                      
2019-08-21 & 58716.22  $\pm$ 0.10  & ATCA$_{750\tu{C}}$ & 5.5  & <0.033          &                  & <0.03 &\\
           &                       &                    & 9    & <0.033          &                  & <0.03 &\\
2019-08-23 & 58718.658 $\pm$ 0.005 & MeerKAT            & 1.3  & <0.11           &                  & <0.12 &\\		
2019-08-31 & 58726.741 $\pm$ 0.005 & MeerKAT            & 1.3  & <0.12           &                  & <0.12 &\\
2019-09-01 & 58727.08  $\pm$ 0.13  & ATCA$_{6\tu{A}}$   & 5.5  & <0.033          & 				   & <0.03 &\\
           &                       &     			    & 9    & <0.024			&                  & <0.024 &\\
2019-10-19 & 58775.616 $\pm$ 0.005 & MeerKAT            & 1.3  & <0.15           &                  &  0.98 $\pm$ 0.04 &\\		
2019-10-26 & 58782.597 $\pm$ 0.005 & MeerKAT            & 1.3  & <0.15           &  				   &  0.64 $\pm$ 0.03 &\\		
2019-11-01 & 58788.670 $\pm$ 0.005 & MeerKAT            & 1.3  & <0.18           &  				   &  0.60 $\pm$ 0.04 &\\
2019-11-02 & 58789.183 $\pm$ 0.045 & ATCA$_{750\tu{C}}$ & 5.5  &  <0.09          &                  & <0.09 &\\
           &                       &                    & 9    &  <0.09          &                  & <0.09 &\\
2019-11-10 & 58797.408 $\pm$ 0.005 & MeerKAT            & 1.3  &  <0.12          &                  & 0.31 $\pm$ 0.04 &\\		   
2019-11-18 & 58805.403 $\pm$ 0.005 & MeerKAT            & 1.3  &  <0.12          &                  & 0.35 $\pm$ 0.03 &\\		
2019-11-24 & 58811.345 $\pm$ 0.005 & MeerKAT            & 1.3  & 0.50  $\pm$ 0.03 &                  & 0.29 $\pm$ 0.03 &\\ 
2019-11-27 & 58814.72  $\pm$ 0.09  & ATCA$_{1.5\tu{C}}$ & 5.5  & 1.07  $\pm$ 0.01 & 0.31 $\pm$ 0.04  & 0.060 $\pm$ 0.009 &\\
           &                       &       			    & 9    & 1.24  $\pm$ 0.02 &   			   & <0.04 &             \\
2019-11-30 & 58817.449 $\pm$ 0.005 & MeerKAT            & 1.3  & 0.47  $\pm$ 0.03 &                  & 0.29 $\pm$ 0.03 &\\
2019-12-03 & 58820.73  $\pm$ 0.13  & ATCA$_{1.5\tu{C}}$ & 5.5  & 0.56  $\pm$ 0.01 & 0.19 $\pm$ 0.06  & 0.44 $\pm$ 0.09 &\\
           &                       &					& 9    & 0.61  $\pm$ 0.01 &                  & <0.03 &           \\               
2019-12-07 & 58824.396 $\pm$ 0.005 & MeerKAT            & 1.3  & 0.16  $\pm$ 0.03 &                  & 0.25 $\pm$ 0.03 &\\
2019-12-10 & 58827.75  $\pm$ 0.16  & ATCA$_{1.5\tu{C}}$ & 5.5  & 0.15  $\pm$ 0.01 & 0.9 $\pm$ 0.2    & 0.067 $\pm$ 0.007 &\\
           &                       &                    & 9    & 0.21  $\pm$ 0.01 &                  & <0.02 &            \\
2019-12-13 & 58830.82  $\pm$ 0.24  & ATCA$_{1.5\tu{C}}$ & 5.5  & 0.053 $\pm$ 0.007 & 0.4 $\pm$ 0.3  &                   & 			\\
           &                       &                    & 9    & 0.065 $\pm$ 0.005 &                &                   &            \\
2019-12-14 & 58831.82  $\pm$ 0.24  & ATCA$_{1.5\tu{C}}$ & 5.5  & 0.054 $\pm$ 0.006 & $-$0.3 $\pm$ 0.3  &                   & 				\\
           &                       &                    & 9    & 0.046 $\pm$ 0.005 &                &                   &            \\
2019-12-14\tnote{a} & 58831.31  $\pm$ 0.73  & ATCA$_{1.5\tu{C}}$ & 5.5  &                   &                & 0.043\tnote{a} $\pm$ 0.005 & $-$1.4 $\pm$ 0.6\\
           &                       &                    & 9    &                   &                & 0.018\tnote{a} $\pm$ 0.003 &            \\
2019-12-15 & 58832.387 $\pm$ 0.005 & MeerKAT            & 1.3  & <0.12           &                  & <0.12 &\\ 		
2019-12-20 & 58837.428 $\pm$ 0.005 & MeerKAT            & 1.3  & <0.12           &                  & 0.20 $\pm$ 0.04 &\\ 		
2019-12-28 & 58845.272 $\pm$ 0.005 & MeerKAT            & 1.3  & <0.15           &                  & <0.15 &\\ 		
2020-01-03 & 58851.397 $\pm$ 0.005 & MeerKAT            & 1.3  & <0.18           &                  & <0.18 &\\ 		      
2020-02-08 & 58887.185 $\pm$ 0.005 & MeerKAT            & 1.3  & 0.38  $\pm$ 0.03 &                  & 0.18 $\pm$ 0.2 &\\
2020-02-15 & 58894.147 $\pm$ 0.005 & MeerKAT            & 1.3  & 0.22  $\pm$ 0.02 &                  & <0.12 &\\
2020-02-21 & 58900.236 $\pm$ 0.005 & MeerKAT            & 1.3  & <0.12           &                  & <0.12 &\\		
2020-03-02 & 58910.147 $\pm$ 0.005 & MeerKAT            & 1.3  & <0.12           &                  & <0.12 &\\		                                                                                
\hline
\label{tab:core_flux_table}
\end{longtable}
\end{center}
\end{ThreePartTable}

\begin{ThreePartTable}

\begin{TableNotes}
  \item[a] {\footnotesize Combination of the MJD 58830 and 58831 ATCA epochs to obtain a higher significance detection of RK1 at both 5.5 and 9 GHz.}
\end{TableNotes}
\LTcapwidth=\textwidth
\begin{center}
\setlength{\extrarowheight}{.3em}
\begin{longtable}{*{5}{c}}
\caption{Measured positions of RK1 and its angular separation from \maxithirt{}. The positions are corrected using a bright background point source close to the target. The errors reported for the coordinates are only the statistical ones from source fitting. The angular separation is computed as the great circle distance between the radio knot and the \maxithirt{} position, either the one obtained from the fit in case of core detection in the same epoch, either the reference one reported in Section \ref{sec:radio_core_results}, in case of core non-detection. This allows to ignore global systematics in the error associated to the angular separation for the epochs in which both the RK1 and the core are detected.}\\
\hhline{=====}
Calendar date & MJD & Right Ascension & Declination & Angular separation\\
$[$UT$]$ & & &  & [arcsec]\\
\hline
\endfirsthead

\caption{Continued from previous page. Measured positions of RK1 and its angular separation from \maxithirt{}.}\\
\hhline{=====}
Calendar date & MJD & Right Ascension & Declination & Angular separation\\
$[$UT$]$ & & &  & [arcsec]\\
\hline
\endhead

\hline \multicolumn{5}{r}{{Continued on next page}}\\
\endfoot

\insertTableNotes
\endlastfoot
2019-03-09 & 58551.100 & 13$^{\textup{h}}$48$^{\textup{h}}$13.03$^{\textup{s}}$ $\pm$ 0.09\arcsec & $-63\degree$16\arcmin26.26\arcsec $\pm$ 0.20\arcsec & 2.80  $\pm$ 0.46\\    
2019-03-18 & 58560.075 & 13$^{\textup{h}}$48$^{\textup{h}}$13.11$^{\textup{s}}$ $\pm$ 0.02\arcsec & $-63\degree$16\arcmin25.31\arcsec $\pm$ 0.04\arcsec & 3.89  $\pm$ 0.44\\
2019-03-25 & 58567.074 & 13$^{\textup{h}}$48$^{\textup{h}}$13.19$^{\textup{s}}$ $\pm$ 0.04\arcsec & $-63\degree$16\arcmin24.45\arcsec $\pm$ 0.08\arcsec & 4.92  $\pm$ 0.44\\ 
2019-03-31 & 58573.759 & 13$^{\textup{h}}$48$^{\textup{h}}$13.29$^{\textup{s}}$ $\pm$ 0.19\arcsec & $-63\degree$16\arcmin23.57\arcsec $\pm$ 0.23\arcsec & 6.00  $\pm$ 0.20\\ 
2019-04-01 & 58574.061 & 13$^{\textup{h}}$48$^{\textup{h}}$13.23$^{\textup{s}}$ $\pm$ 0.42\arcsec & $-63\degree$16\arcmin23.83\arcsec $\pm$ 0.52\arcsec & 5.58  $\pm$ 0.49\\
2019-04-08 & 58581.715 & 13$^{\textup{h}}$48$^{\textup{h}}$13.363$^{\textup{s}}$ $\pm$ 0.005\arcsec & $-63\degree$16\arcmin22.77\arcsec $\pm$ 0.01\arcsec & 6.78  $\pm$ 0.02\\
2019-04-09 & 58582.053 & 13$^{\textup{h}}$48$^{\textup{h}}$13.31$^{\textup{s}}$ $\pm$ 0.41\arcsec & $-63\degree$16\arcmin23.51\arcsec $\pm$ 0.13\arcsec & 6.15  $\pm$ 0.48\\
2019-04-15 & 58588.053 & 13$^{\textup{h}}$48$^{\textup{h}}$13.36$^{\textup{s}}$ $\pm$ 0.04\arcsec & $-63\degree$16\arcmin22.43\arcsec $\pm$ 0.05\arcsec & 7.23  $\pm$ 0.44\\
2019-04-16 & 58589.806 & 13$^{\textup{h}}$48$^{\textup{h}}$13.40$^{\textup{s}}$ $\pm$ 0.02\arcsec & $-63\degree$16\arcmin22.13\arcsec $\pm$ 0.01\arcsec & 7.64  $\pm$ 0.25\\
2019-04-20 & 58593.074 & 13$^{\textup{h}}$48$^{\textup{h}}$13.44$^{\textup{s}}$ $\pm$ 0.08\arcsec & $-63\degree$16\arcmin21.97\arcsec $\pm$ 0.07\arcsec & 7.90  $\pm$ 0.44\\                                                                             
2019-04-29 & 58602.144 & 13$^{\textup{h}}$48$^{\textup{h}}$13.72$^{\textup{s}}$ $\pm$ 0.75\arcsec & $-63\degree$16\arcmin20.88\arcsec $\pm$ 0.15\arcsec & 9.39  $\pm$ 0.25\\
2019-04-30 & 58603.306 & 13$^{\textup{h}}$48$^{\textup{h}}$13.58$^{\textup{s}}$ $\pm$ 0.21\arcsec & $-63\degree$16\arcmin21.05\arcsec $\pm$ 0.04\arcsec & 9.14  $\pm$ 0.18\\
2019-05-04 & 58607.908 & 13$^{\textup{h}}$48$^{\textup{h}}$13.52$^{\textup{s}}$ $\pm$ 0.08\arcsec & $-63\degree$16\arcmin20.64\arcsec $\pm$ 0.28\arcsec & 9.20  $\pm$ 0.25\\
2019-05-11 & 58614.909 & 13$^{\textup{h}}$48$^{\textup{h}}$13.53$^{\textup{s}}$ $\pm$ 0.15\arcsec & $-63\degree$16\arcmin20.40\arcsec $\pm$ 0.37\arcsec & 9.69  $\pm$ 0.31\\
2019-05-18 & 58621.888 & 13$^{\textup{h}}$48$^{\textup{h}}$13.69$^{\textup{s}}$ $\pm$ 0.31\arcsec & $-63\degree$16\arcmin18.88\arcsec $\pm$ 0.81\arcsec & 11.61 $\pm$ 0.68\\
2019-10-19 & 58775.616 & 13$^{\textup{h}}$48$^{\textup{h}}$14.99$^{\textup{s}}$ $\pm$ 0.28\arcsec & $-63\degree$16\arcmin06.31\arcsec $\pm$ 0.15\arcsec & 26.74 $\pm$ 0.46\\         
2019-10-26 & 58782.597 & 13$^{\textup{h}}$48$^{\textup{h}}$14.94$^{\textup{s}}$ $\pm$ 0.25\arcsec & $-63\degree$16\arcmin06.28\arcsec $\pm$ 0.15\arcsec & 26.58 $\pm$ 0.46\\            
2019-11-01 & 58788.670 & 13$^{\textup{h}}$48$^{\textup{h}}$14.84$^{\textup{s}}$ $\pm$ 0.58\arcsec & $-63\degree$16\arcmin05.68\arcsec $\pm$ 0.14\arcsec & 26.73 $\pm$ 0.48\\           
2019-11-10 & 58797.408 & 13$^{\textup{h}}$48$^{\textup{h}}$14.99$^{\textup{s}}$ $\pm$ 0.28\arcsec & $-63\degree$16\arcmin05.47\arcsec $\pm$ 0.67\arcsec & 27.45 $\pm$ 0.71\\          
2019-11-18 & 58805.403 & 13$^{\textup{h}}$48$^{\textup{h}}$14.99$^{\textup{s}}$ $\pm$ 0.13\arcsec & $-63\degree$16\arcmin05.50\arcsec $\pm$ 0.36\arcsec & 27.41 $\pm$ 0.53\\          
2019-11-24 & 58811.345 & 13$^{\textup{h}}$48$^{\textup{h}}$14.97$^{\textup{s}}$ $\pm$ 0.18\arcsec & $-63\degree$16\arcmin06.92\arcsec $\pm$ 0.39\arcsec & 26.63 $\pm$ 0.41\\
2019-11-27 & 58814.722 & 13$^{\textup{h}}$48$^{\textup{h}}$14.82$^{\textup{s}}$ $\pm$ 0.78\arcsec & $-63\degree$16\arcmin06.05\arcsec $\pm$ 0.05\arcsec & 26.36 $\pm$ 0.26\\
2019-11-30 & 58817.449 & 13$^{\textup{h}}$48$^{\textup{h}}$14.90$^{\textup{s}}$ $\pm$ 0.25\arcsec & $-63\degree$16\arcmin06.27\arcsec $\pm$ 0.27\arcsec & 27.19 $\pm$ 0.31\\
2019-12-03 & 58820.734 & 13$^{\textup{h}}$48$^{\textup{h}}$15.16$^{\textup{s}}$ $\pm$ 0.79\arcsec & $-63\degree$16\arcmin06.74\arcsec $\pm$ 0.41\arcsec & 27.03 $\pm$ 0.43\\
2019-12-07 & 58824.396 & 13$^{\textup{h}}$48$^{\textup{h}}$14.85$^{\textup{s}}$ $\pm$ 0.50\arcsec & $-63\degree$16\arcmin07.20\arcsec $\pm$ 0.65\arcsec & 26.40 $\pm$ 1.36\\
2019-12-10 & 58827.754 & 13$^{\textup{h}}$48$^{\textup{h}}$14.93$^{\textup{s}}$ $\pm$ 0.17\arcsec & $-63\degree$16\arcmin05.48\arcsec $\pm$ 0.15\arcsec & 27.33 $\pm$ 0.24\\
2019-12-14\tnote{a} & 58831.316 & 13$^{\textup{h}}$48$^{\textup{h}}$15.05$^{\textup{s}}$ $\pm$ 0.16\arcsec & $-63\degree$16\arcmin05.88\arcsec $\pm$ 0.28\arcsec & 27.44 $\pm$ 0.33\\
2019-12-20 & 58837.428 & 13$^{\textup{h}}$48$^{\textup{h}}$14.93$^{\textup{s}}$ $\pm$ 1.19\arcsec & $-63\degree$16\arcmin05.28\arcsec $\pm$ 0.53\arcsec & 27.38 $\pm$ 0.69\\
2020-02-08 & 58887.185 & 13$^{\textup{h}}$48$^{\textup{h}}$15.03$^{\textup{s}}$ $\pm$ 0.23\arcsec & $-63\degree$16\arcmin05.23\arcsec $\pm$ 0.59\arcsec & 27.94 $\pm$ 0.56\\
\hline
\label{tab:first_jet_angsep}
\end{longtable}
\end{center}
\end{ThreePartTable}

\LTcapwidth=\textwidth
\begin{center}
\setlength{\extrarowheight}{.1em}
\begin{longtable}{*{6}{c}}
\caption{Radio flux densities $S_{\nu}$ of the second radio knot RK2. Observation MJDs represent the middle of the observation, where errors represent the observation duration (time on source). For the ATCA observations we report the array configuration as a subscript. The spectral index $\alpha$ of each component ($S_{\nu} \propto \nu^{\alpha}$) is computed only for simultaneous ATCA multi frequency observations. 
For brevity, we only report the 6 detections of RK2.}\\
\hhline{======}
Calendar date & MJD & Telescope & Central frequency & $S_{\nu, \tu{RK2}}$ & $\alpha_{\tu{RK2}}$\\
$[$UT$]$ & & & [GHz] & [mJy] & \\
\hline
\endfirsthead

\caption{Continued from previous page}\\
\hhline{======}
Calendar date & MJD & Telescope & Central frequency & $S_{\nu, \tu{RK2}}$ & $\alpha_{\tu{RK2}}$\\
$[$UT$]$ & & & [GHz] & [mJy] & \\
\hline
\endhead

\hline \multicolumn{6}{r}{{Continued on next page}}
\endfoot

\endlastfoot
2019-04-15 & 58588.053 $\pm$ 0.005 & MeerKAT & 1.3 & 1.41 $\pm$ 0.06\\
2019-04-16 & 58589.81 $\pm$ 0.11  & ATCA$_{750\tu{C}}$ & 5.5 & 0.39 $\pm$ 0.01 & $-$0.6 $\pm$ 0.1\\
 &  &  & 9 & 0.29 $\pm$ 0.02 &\\
2019-04-20 & 58593.074 $\pm$ 0.005 & MeerKAT & 1.3 & 1.12 $\pm$ 0.07\\
2019-05-11 & 58614.909 $\pm$ 0.005 & MeerKAT & 1.3 & 0.46 $\pm$ 0.04\\
2019-05-18 & 58621.888 $\pm$ 0.005 & MeerKAT & 1.3 & 0.52 $\pm$ 0.04\\
2019-05-25 & 58628.019 $\pm$ 0.005 & MeerKAT & 1.3 & 0.31 $\pm$ 0.04\\
\hline
\label{tab:RK2_flux_table}
\end{longtable}
\end{center}

\LTcapwidth=\textwidth
\begin{center}
\setlength{\extrarowheight}{.3em}
\begin{longtable}{*{5}{c}}
\caption{Measured positions of RK2 and its angular separation from \maxithirt{}. The positions are corrected using a bright background point source close to the target. The errors reported for the coordinates are only the statistical ones from source fitting. The angular separation is computed as the great circle distance between the radio knot and the \maxithirt{} position, either the one obtained from the fit in case of core detection in the same epoch, either the reference one reported in Section \ref{sec:radio_core_results}, in case of core non-detection. This allows to ignore global systematics in the error associated to the angular separation for the epochs in which both the RK2 and the core are detected.}\\
\hhline{=====}
Calendar date & MJD & Right Ascension & Declination & Angular separation\\
$[$UT$]$ & & &  & [arcsec]\\
\hline
\endfirsthead

\caption{Continued from previous page. Measured positions of RK2 and its angular separation from \maxithirt{}.}\\
\hhline{=====}
Calendar date & MJD & Right Ascension & Declination & Angular separation\\
$[$UT$]$ & & &  & [arcsec]\\
\hline
\endhead

\hline \multicolumn{5}{r}{{Continued on next page}}\\
\endfoot

\endlastfoot
2019-04-15 & 58588.053 & 13$^{\textup{h}}$48$^{\textup{h}}$12.81$^{\textup{s}}$ $\pm$ 0.11\arcsec & $-63\degree$16\arcmin27.87\arcsec $\pm$ 0.16\arcsec & 0.70 $\pm$ 0.52\\
2019-04-16 & 58589.806 & 13$^{\textup{h}}$48$^{\textup{h}}$12.86$^{\textup{s}}$ $\pm$ 0.07\arcsec & $-63\degree$16\arcmin28.01\arcsec $\pm$ 0.04\arcsec & 0.69 $\pm$ 0.24\\
2019-04-20 & 58593.074 & 13$^{\textup{h}}$48$^{\textup{h}}$12.92$^{\textup{s}}$ $\pm$ 0.21\arcsec & $-63\degree$16\arcmin27.57\arcsec $\pm$ 0.18\arcsec & 1.31 $\pm$ 0.43\\
2019-05-11 & 58614.909 & 13$^{\textup{h}}$48$^{\textup{h}}$13.07$^{\textup{s}}$ $\pm$ 0.12\arcsec & $-63\degree$16\arcmin25.90\arcsec $\pm$ 0.31\arcsec & 3.23 $\pm$ 0.50\\
2019-05-18 & 58621.888 & 13$^{\textup{h}}$48$^{\textup{h}}$13.15$^{\textup{s}}$ $\pm$ 0.12\arcsec & $-63\degree$16\arcmin25.52\arcsec $\pm$ 0.32\arcsec & 4.15 $\pm$ 0.23\\
2019-05-25 & 58628.019 & 13$^{\textup{h}}$48$^{\textup{h}}$13.18$^{\textup{s}}$ $\pm$ 0.49\arcsec & $-63\degree$16\arcmin24.25\arcsec $\pm$ 0.59\arcsec & 4.96 $\pm$ 0.52\\
\hline
\label{tab:second_jet_angsep}
\end{longtable}
\end{center}

\section{X-ray data}

\begin{ThreePartTable}

\begin{TableNotes}
  \item[a] {\footnotesize Starting date of the observation.} 
  \item[b] {\footnotesize In units of 10$^{-10}$ erg cm$^{-2}$ s$^{-1}$.}
  \item[c] {\footnotesize Another observations was taken on MJD 58509 (ObsID 00885807001), but we discard it since it was erroneously performed in PC mode.}
  \item[d] {\footnotesize We discard the observation taken one day before (ObsID 00011107046) due to a satellite pointing error.}
  \item[e] {\footnotesize We do not consider the observation taken one day before, on MJD 58727 (ObsID 00011107053), as it was taken in WT mode, providing a less constraining upper limit.}
  \item[f] {\footnotesize Epoch combined with ObsID 00011107075, as it was performed less than 3 hours after ObsID 00011107074.}
  \item[g] {\footnotesize Due to the shorter exposure time, a larger circular region was needed to have enough basckground counts for the upper limit estimation.}
\end{TableNotes}

\LTcapwidth=\textwidth
\begin{center}
\setlength{\extrarowheight}{.4em}
\setlength{\tabcolsep}{3.5pt}
\begin{longtable}{*{10}{c}}
\caption{X-ray spectral parameters of \maxithirt{} obtained from \emph{Swift}/XRT observations. The models used are discussed in Section \ref{sec:Swift/XRT observations} and interstellar absorption is accounted for adopting $N_{\textup{H}} = 0.86\times10^{22}$ cm$^{-2}$ . We report the annulus inner and outer radii of the spectrum extraction region in pixels (1 pixel $=$ 2.357 arcsec) and the source spectral state. We then show the obtained unabsorbed 1--10 keV flux in units of 10$^{-10}$ erg cm$^{-2}$ s$^{-1}$, the photon index $\Gamma$, the accretion disk inner temperature $k_{\textup{B}}T_{\tu{in}}$ in keV units (for the epochs in which the {\tt diskbb} component was needed) and the reduced $\chi^2$ over the degrees of freedom of the fit. Epochs for which the $\chi^2$ is not reported were fitted using Cash statistics due to low number of counts. Throughout this table, we quote 90\% confidence errors.}\\
\hhline{==========}
ObsID & Day & MJD\tnote{a} & Exposure & Extraction & Spectral & Unabsobed flux & $\Gamma$ & $k_{\textup{B}}T_{\tu{in}}$ & $\chi^2$/d.o.f\\
 & [UT] & & time [s] & region & state & 1--10 keV\tnote{b} & & [keV] & \\
\hline
\endfirsthead

\caption{Continued from previous page. X-ray spectral parameters of \maxithirt{} obtained from \emph{Swift}/XRT observations.}\\
\hhline{==========}
ObsID & Day & MJD\tnote{a} & Exposure & Extraction & Spectral & Unabsobed flux & $\Gamma$ & $k_{\textup{B}}T_{\tu{in}}$ & $\chi^2$/d.o.f\\
 & [UT] & & time [s] & region & state & 1--10 keV\tnote{b} & & [keV] & \\
\hline
\endhead

\hline \multicolumn{10}{r}{{Continued on next page}}\\
\endfoot
\insertTableNotes
\endlastfoot
00885807000\tnote{c}  & 2019-01-26  &  58509.438 &  1221  &  0$-$20   &  HS  &  $38.13^{+0.35}_{-0.35}$        &  $1.41^{+0.01}_{-0.01}$     &                             &  0.99/618\\  
00885960000  & 2019-01-27  &  58510.038 &  1453  &  3$-$30   &  HS  &  $95.57^{+0.72}_{-0.72}$        &  $1.52^{+0.01}_{-0.01}$     &                             &  1.09/646\\  
00886266000  & 2019-01-28  &  58511.576 &  1838  &  6$-$30   &  HS  &  $313.9^{+1.7}_{-1.7}$          &  $1.783^{+0.008}_{-0.008}$  &                             &  1.06/682\\  
00886496000  & 2019-01-29  &  58512.420 &  317   &  7$-$30   &  HS  &  $375.6^{+4.7}_{-4.6}$          &  $1.73^{+0.02}_{-0.02}$     &                             &  1.00/487\\  
00011107001  & 2019-01-30  &  58513.108 &  832   &  10$-$30  &  HS  &  $484.1^{+3.9}_{-3.9}$          &  $1.836^{+0.012}_{-0.012}$  &                             &  0.98/613\\  
00088843001  & 2019-02-01  &  58515.753 &  2019  &  12$-$30  &  HS  &  $625.3^{+3.4}_{-2.7}$          &  $1.898^{+0.008}_{-0.008}$  &                             &  1.14/673\\  
00011107002  & 2019-02-03  &  58517.672 &  868   &  12$-$30  &  IMS &  $707.0^{+4.6}_{-5.7}$          &  $2.10^{+0.01}_{-0.01}$     &                             &  1.02/585\\  
00011107003  & 2019-02-05  &  58519.335 &  1370  &  17$-$30  &  IMS &  $1143.0^{+9.5}_{-9.5}$         &  $2.25^{+0.06}_{-0.05}$     &  $0.63^{+0.02}_{-0.02}$     &  1.06/581\\  
00011107004  & 2019-02-07  &  58521.074 &  1750  &  31$-$50  &  IMS &  $2164^{+14}_{-14}$             &  2.4                        &  $0.743^{+0.009}_{-0.009}$  &  1.25/528\\  
00011107005  & 2019-02-09  &  58523.000 &  219   &  28$-$50  &  SS  &  $2594^{+44}_{-44}$             &  2.4                        &  $0.823^{+0.014}_{-0.014}$  &  1.16/296\\  
00011107006  & 2019-02-09  &  58523.450 &  1412  &  33$-$50  &  SS  &  $2444^{+16}_{-16}$             &  2.4                        &  $0.759^{+0.007}_{-0.007}$  &  1.11/525\\  
00011107007  & 2019-02-10  &  58524.526 &  760   &  33$-$50  &  SS  &  $2670^{+23}_{-23}$             &  2.4                        &  $0.729^{+0.009}_{-0.009}$  &  1.03/461\\  
00011107008  & 2019-02-11  &  58525.774 &  1472  &  33$-$50  &  SS  &  $2314^{+15}_{-15}$             &  2.4                        &  $0.719^{+0.007}_{-0.007}$  &  1.14/517\\  
00011107009  & 2019-02-12  &  58526.257 &  1120  &  32$-$50  &  SS  &  $2670^{+18}_{-18}$             &  2.4                        &  $0.735^{+0.007}_{-0.007}$  &  1.13/474\\  
00011107010  & 2019-02-13  &  58527.440 &  1462  &  32$-$50  &  SS  &  $2590^{+16}_{-16}$             &  2.4                        &  $0.763^{+0.007}_{-0.007}$  &  1.17/533\\  
00011107011  & 2019-02-16  &  58530.28  &  1462  &  32$-$50  &  SS  &  $2217^{+14}_{-14}$             &  2.4                        &  $0.682^{+0.005}_{-0.005}$  &  1.03/469\\  
00011107012  & 2019-02-17  &  58531.362 &  1457  &  31$-$50  &  SS  &  $2181^{+26}_{-25}$             &  2.4                        &  $0.674^{+0.005}_{-0.005}$  &  1.05/484\\  
00011107013  & 2019-02-19  &  58533.547 &  1023  &  30$-$50  &  SS  &  $1847^{+14}_{-14}$             &  2.4                        &  $0.664^{+0.007}_{-0.007}$  &  0.99/449\\  
00011107014  & 2019-02-21  &  58535.800 &  1128  &  30$-$50  &  SS  &  $1905^{+13}_{-13}$             &  2.4                        &  $0.657^{+0.006}_{-0.006}$  &  1.04/438\\  
00011107015  & 2019-03-02  &  58544.918 &  1100  &  26$-$50  &  SS  &  $1655^{+11}_{-11}$             &  2.4                        &  $0.636^{+0.006}_{-0.006}$  &  1.03/435\\  
00011107016  & 2019-03-05  &  58547.835 &  1003  &  25$-$50  &  SS  &  $1496^{+10}_{-10}$             &  2.4                        &  $0.645^{+0.006}_{-0.006}$  &  1.06/453\\  
00088843002  & 2019-03-08  &  58550.497 &  2081  &  22$-$50  &  SS  &  $1501^{+9}_{-9}$               &  2.4                        &  $0.642^{+0.005}_{-0.005}$  &  1.05/474\\  
00011107017  & 2019-03-11  &  58553.873 &  935   &  17$-$30  &  SS  &  $1233^{+7.7}_{-7.7}$           &  2.4                        &  $0.624^{+0.006}_{-0.006}$  &  1.06/424\\  
00011107019  & 2019-03-17  &  58559.134 &  1191  &  15$-$30  &  SS  &  $873.4^{+5.1}_{-5.1}$          &  2.4                        &  $0.624^{+0.004}_{-0.004}$  &  1.04/406\\  
00011107020  & 2019-03-20  &  58562.049 &  999   &  15$-$30  &  SS  &  $849.3^{+5.6}_{-5.6}$          &  2.4                        &  $0.629^{+0.006}_{-0.006}$  &  1.07/422\\  
00011107021  & 2019-03-23  &  58565.820 &  1102  &  13$-$30  &  SS  &  $643.5^{+3.7}_{-3.7}$          &  2.4                        &  $0.601^{+0.005}_{-0.005}$  &  1.03/401\\  
00011107022  & 2019-03-26  &  58568.025 &  1008  &  13$-$30  &  SS  &  $629.1^{+3.8}_{-3.8}$          &  2.4                        &  $0.600^{+0.005}_{-0.005}$  &  1.05/402\\  
00011107024  & 2019-04-01  &  58574.383 &  1023  &  12$-$30  &  SS  &  $569.1^{+3.6}_{-3.6}$          &  2.4                        &  $0.573^{+0.005}_{-0.005}$  &  1.06/422\\  
00011107025  & 2019-04-04  &  58577.239 &  1572  &  13$-$30  &  SS  &  $542.2^{+2.6}_{-2.6}$          &  2.4                        &  $0.544^{+0.003}_{-0.003}$  &  1.05/401\\  
00011107026  & 2019-04-10  &  58583.029 &  1224  &  11$-$30  &  SS  &  $415.5^{+2.4}_{-2.4}$          &  2.4                        &  $0.551^{+0.004}_{-0.004}$  &  1.00/384\\  
00011107027  & 2019-04-12  &  58585.209 &  977   &  11$-$30  &  SS  &  $442.4^{+2.8}_{-2.8}$          &  2.4                        &  $0.563^{+0.004}_{-0.004}$  &  1.17/362\\  
00011107028  & 2019-04-15  &  58588.727 &  1018  &  12$-$30  &  SS  &  $443.5^{+2.8}_{-2.8}$          &  2.4                        &  $0.518^{+0.004}_{-0.004}$  &  1.17/345\\  
00011107029  & 2019-04-24  &  58597.045 &  1013  &  10$-$30  &  IMS &  $326.7^{+2.1}_{-2.1}$          &  2.4                        &  $0.491^{+0.004}_{-0.004}$  &  0.95/314\\  
00011107032  & 2019-05-03  &  58606.396 &  529   &  2$-$30   &  HS  &  $46.4^{+0.5}_{-0.6}$           &  $2.28^{+0.03}_{-0.03}$     &                             &  1.11/391\\  
00011107033  & 2019-05-06  &  58609.474 &  875   &  0$-$20   &  HS  &  $21.5^{+0.3}_{-0.3}$           &  $1.86^{+0.02}_{-0.02}$     &                             &  1.10/435\\  
00011107034  & 2019-05-09  &  58612.438 &  975   &  0$-$20   &  HS  &  $9.11^{+0.26}_{-0.26}$         &  $1.66^{+0.04}_{-0.04}$     &                             &  1.17/256\\  
00011107035  & 2019-05-12  &  58615.972 &  604   &  0$-$20   &  HS  &  $2.80^{+0.13}_{-0.13}$         &  $1.76^{+0.07}_{-0.07}$     &                             &  0.88/107\\  
00011107036  & 2019-05-15  &  58618.746 &  1080  &  0$-$20   &  HS  &  $1.29^{+0.07}_{-0.07}$         &  $1.98^{+0.09}_{-0.09}$     &                             &  1.18/70 \\  
00011107037  & 2019-05-18  &  58621.397 &  913   &  0$-$20   &  HS  &  $0.59^{+0.02}_{-0.04}$         &  $1.96^{+0.11}_{-0.11}$     &                             &          \\  
00011107038  & 2019-05-21  &  58624.075 &  602   &  0$-$20   &  HS  &  $0.29^{+0.03}_{-0.03}$         &  $2.88^{+0.26}_{-0.25}$     &                             &          \\  
00011107039  & 2019-05-24  &  58627.445 &  993   &  0$-$20   &  HS  &  <0.59                        &                             &                             &          \\  
00011107040  & 2019-05-27  &  58630.699 &  986   &  0$-$20   &  HS  &  $0.52^{+0.04}_{-0.04}$         &  $1.87^{+014}_{-0.13}$      &                             &  0.89/31 \\  
00011107041  & 2019-05-30  &  58633.148 &  861   &  0$-$20   &  HS  &  $4.74^{+0.14}_{-0.12}$         &  $1.53^{+0.04}_{-0.04}$     &                             &  1.07/219\\  
00011107042  & 2019-06-05  &  58639.659 &  998   &  0$-$20   &  HS  &  $45.5^{+0.4}_{-0.4}$           &  $1.64^{+0.01}_{-0.01}$     &                             &  0.98/600\\  
00011107043  & 2019-06-13  &  58647.781 &  1070  &  0$-$20   &  HS  &  $128.2^{+1.9}_{-1.9}$          &  $1.87^{+0.02}_{-0.02}$     &                             &  1.08/468\\  
00011107044  & 2019-06-16  &  58650.831 &  998   &  0$-$20   &  HS  &  $99.3^{+1.5}_{-1.5}$           &  $1.81^{+0.03}_{-0.03}$     &                             &  1.04/415\\  
00011107045  & 2019-07-21  &  58685.017 &  935   &  0$-$20   &  HS  &  $27.2^{+0.3}_{-0.3}$           &  $1.54^{+0.02}_{-0.02}$     &                             &  0.88/547\\  
00011107047\tnote{d}  & 2019-07-26  &  58690.131 &  918   &  0$-$20   &  HS  &  $19.3^{+0.4}_{-0.4}$           &  $1.39^{+0.03}_{-0.03}$     &                             &  1.03/325\\  
00011107048  & 2019-08-04  &  58699.836 &  988   &  0$-$20   &  HS  &  $0.98^{+0.06}_{-0.05}$         &  $1.56^{+0.10}_{-0.10}$     &                             &  1.09/55 \\  
00011107049  & 2019-08-11  &  58706.413 &  898   &  0$-$20   &  HS  &  $0.071^{+0.012}_{-0.007}$      &  $2.64^{+0.35}_{-0.34}$     &                             &          \\  
00011107050  & 2019-08-18  &  58713.044 &  928   &  0$-$20   &  HS  &  <0.083      					&       					  &                             &          \\  
00011107051  & 2019-08-25  &  58720.155 &  673   &  0$-$20   &  HS  &  <0.048                       &    						  &                             &          \\  
00011107052  & 2019-08-31  &  58726.999 &  3024  &  0$-$20   &  HS  &  <0.002    			            &        					  &                             &          \\  
00011107054\tnote{e}  & 2019-09-02  &  58728.059 &  1115  &  0$-$20   &  HS  &  <0.005     			&  						      &                             &          \\  
00011107055  & 2019-09-08  &  58734.030 &  1702  &  0$-$20   &  HS  &  <0.003     			        &                             &                             &          \\  
00011107056  & 2019-10-19  &  58775.046 &  714   &  0$-$20   &  HS  &  $0.008^{+0.006}_{-0.004}$      &  2.2                        &                             &          \\  
00011107057  & 2019-10-26  &  58782.616 &  657   &  0$-$20   &  HS  &  <0.013                         &     	                      &                             &          \\  
00011107058  & 2019-11-02  &  58789.649 &  812   &  0$-$20   &  HS  &  <0.006                         &                             &                             &          \\  
00011107059  & 2019-11-29  &  58816.360 &  971   &  7$-$30   &  HS  &  $1.00^{+0.15}_{-0.14}$         &  $1.96^{+0.26}_{-0.25}$     &                             &  0.86/27 \\  
00011107060  & 2019-12-04  &  58821.802 &  1019  &  1$-$30   &  HS  &  $0.139^{+0.022}_{-0.019}$      &  $2.25^{+0.24}_{-0.23}$     &                             &          \\  
00011107061  & 2019-12-07  &  58824.050 &  932   &  0$-$20   &  HS  &  $0.061^{+0.017}_{-0.013}$      &  $2.01^{+0.37}_{-0.36}$     &                             &          \\  
00011107062  & 2019-12-09  &  58826.041 &  864   &  0$-$20   &  HS  &  $0.033^{+0.009}_{-0.008}$      &  2.2                        &                             &          \\  
00011107064  & 2019-12-11  &  58828.032 &  1623  &  0$-$20   &  HS  &  $0.014^{+0.004}_{-0.003}$      &  2.2                        &                             &          \\
00011107065  & 2019-12-14  &  58831.616 &  2707  &  0$-$20   &  HS  &  $0.0020^{+0.0018}_{-0.0012}$   &  2.2                        &                             &          \\
00011107066  & 2019-12-16  &  58833.673 &  944   &  0$-$20   &  HS  &  <0.007                         &                             &                             &          \\
00011107067  & 2019-12-18  &  58835.598 &  2947  &  0$-$20   &  HS  &  <0.004                         &                             &                             &          \\
00011107068  & 2019-12-21  &  58838.522 &  2166  &  0$-$20   &  HS  &  <0.006                         &                             &                             &          \\
00011107069  & 2019-12-23  &  58840.457 &  1447  &  0$-$20   &  HS  &  <0.014                         &                             &                             &          \\
00011107070  & 2019-12-29  &  58846.697 &  948   &  0$-$20   &  HS  &  <0.006                         &                             &                             &          \\
00011107071  & 2020-01-06  &  58854.210 &  757   &  0$-$20   &  HS  &  <0.006                         &                             &                             &          \\
00011107072  & 2020-01-13  &  58861.179 &  1096  &  0$-$20   &  HS  &  <0.004                         &                             &                             &          \\
00011107073  & 2020-02-10  &  58889.521 &  1186  &  4$-$30   &  HS  &  $0.98^{+0.10}_{-0.08}$         &  $1.84^{+0.15}_{-0.15}$     &                             &  0.93/19 \\
00011107074\tnote{f}  & 2020-02-18  &  58897.057 &  1274  &  0$-$20   &  HS  &  $0.08^{+0.01}_{-0.01}$         &  $1.72^{+0.30}_{-0.29}$     &                             &          \\
00011107076  & 2020-02-28  &  58907.132 &  267   &  0$-$30\tnote{g} &  HS  &  <0.022                  &                             &                             &          \\
00011107077  & 2020-03-03  &  58911.244 &  537   &  0$-$20   &  HS  &  <0.021                         &                             &                             &          \\
00011107078  & 2020-03-07  &  58915.621 &  1382  &  0$-$20   &  HS  &  <0.004                         &                             &                             &          \\
             
\label{tab:xray_1348}
\end{longtable}
\end{center}
\end{ThreePartTable}




\bsp	
\label{lastpage}
\end{document}